\documentclass[twocolumn,superscriptaddress,longbibliography]{revtex4-2}
\usepackage{graphicx}
\usepackage{amsmath,amsbsy,amssymb}
\usepackage{bm}
\usepackage{mathrsfs}
\usepackage{textgreek}
\usepackage{mleftright}
\usepackage{braket}
\usepackage{physics}
\usepackage{comment}
\usepackage{xcolor}

\usepackage{xurl}
\usepackage[colorlinks=true,breaklinks=true]{hyperref}
\hypersetup{linkcolor=blue,citecolor=blue,urlcolor=blue}
\graphicspath{{./figures/}}

\newcommand{\diff}{\mathrm{d}}

\newcommand{\imu}{\mathrm{i}}
\newcommand{\epn}{\mathrm{e}}

\newcommand{\ua}{\uparrow}
\newcommand{\da}{\downarrow}
\newcommand{\dg}{\dagger}
\newcommand{\la}{\langle}
\newcommand{\lla}{\langle\hspace{-0.5mm}\langle}
\newcommand{\ra}{\rangle}
\newcommand{\rra}{\rangle\hspace{-0.5mm}\rangle}
\newcommand{\al}{\alpha}
\newcommand{\sg}{\sigma}
\newcommand{\gm}{\gamma}
\newcommand{\ep}{\varepsilon}

\begin{document}

\title{
Spherical-tensor description of the Jahn--Teller--Hubbard molecule and \\
local electron--phonon entanglement
}

%%%%%%%%%%%%%%%%%%%%%%%%%%%%%%%%%%%%%
\author{Koichiro Takahashi}
%%%%%%%%%%%%%%%%%%%%%%%%%%%%%%%%%%%%%
\affiliation{Department of Physics, 
Saitama University, Saitama 338-8570, Japan}
\affiliation{Department of Physics, The Grainger College of Engineering,
University of Illinois Urbana-Champaign, Urbana, IL 61801, USA}

%%%%%%%%%%%%%%%%%%%%%%%%%%%%%%%%%%%%%
\author{Shuichiro Ebata}
%%%%%%%%%%%%%%%%%%%%%%%%%%%%%%%%%%%%%
\affiliation{Department of Physics, 
Saitama University, Saitama 338-8570, Japan}
%%%%%%%%%%%%%%%%%%%%%%%%%%%%%%%%%%%%%

%%%%%%%%%%%%%%%%%%%%%%%%%%%%%%%%%%%%%
\author{Naotaka Yoshinaga}
%%%%%%%%%%%%%%%%%%%%%%%%%%%%%%%%%%%%%
\affiliation{Department of Physics, 
Saitama University, Saitama 338-8570, Japan}
%%%%%%%%%%%%%%%%%%%%%%%%%%%%%%%%%%%%%

%%%%%%%%%%%%%%%%%%%%%%%%%%%%%%%%%%%%%
\author{Shintaro Hoshino}
%%%%%%%%%%%%%%%%%%%%%%%%%%%%%%%%%%%%%
\affiliation{Department of Physics, 
Chiba University, Chiba 263-8522, Japan}
\affiliation{Department of Physics, 
Saitama University, Saitama 338-8570, Japan}
%%%%%%%%%%%%%%%%%%%%%%%%%%%%%%%%%%%%%

\date{\today}

\begin{abstract}
We investigate the localized-electron character of the Mott-insulating phase in A$_3$C$_{60}$ using a single-site multiorbital electron model coupled to anisotropic molecular vibrations (Jahn--Teller phonons).
We apply the spherical-tensor formalism, a framework originally developed in nuclear physics, to analyze the electron--phonon-coupled ground-state multiplet.
Focusing on multipole moments, we find that both the conventional electronic quadrupole moment and the lattice displacement associated with the molecular vibrations vanish, even though the degenerate ground-state multiplet implies the presence of quadrupolar degrees of freedom.
By analyzing these degrees of freedom within the spherical-tensor framework, we introduce composite (two-body) quadrupole operators involving both electrons and phonons and study their parameter dependence numerically.
Furthermore, using quasispin selection rules, we demonstrate that the composite quadrupole does not couple to either the conventional quadrupole or lattice-displacement operators, thereby distinguishing it fundamentally from standard quadrupolar degrees of freedom.
In addition, we investigate the nature of the electron--phonon entanglement and characterize it from the viewpoint of angular momentum.
Analysis of the entanglement spectrum reveals that the ground state consists of superpositions of multi-phonon states with angular momenta $L_{\rm ph}=2$ and $L_{\rm ph}=3$, formed through coupling to three-electron states with $L=1$ and $L=2$.
\end{abstract}

\maketitle

\section{Introduction}

The multiorbital nature of strongly correlated electron systems in solids and molecules generates rich physical phenomena.
Such systems range from inorganic materials with $d$- and $f$-electrons to organic conductors with molecular orbitals.
In this paper, we focus on the alkali-doped fulleride A$_3$C$_{60}$, where 
the triply degenerate $t_{1u}$ orbital for each fullerene molecule is occupied by three electrons.
This material exhibits high-temperature superconductivity of $T_\mathrm{c}\sim 40$K,
and becomes a Mott insulator as the intermolecular distance increases~\cite{Gunnarsson97,Capone09,Takabayashi16,Nomura16}.

A striking feature of fulleride systems, compared with transition-metal-based multiorbital systems, is the nature of the Hund's coupling: the ferromagnetic Hund's coupling originating from the Coulomb interaction is small in fullerides due to the extended nature of the molecular orbital, and its sign is inverted by the coupling to molecular vibrations 
to become antiferromagnetic Hund's coupling~\cite{Fabrizio97}.
As a result, an unconventional type of orbital order is identified, in which the ordinary orbital order parameter is zero and the doublon orbital moment is regarded as a relevant order parameter~\cite{Hoshino17,Hoshino19}.
This anomalous metallic state is interpreted as a spontaneous orbital selective Mott state, and provides a candidate scenario for the experimentally observed Jahn--Teller metallic phase~\cite{Zadik15}.
The vanishing of the ordinary orbital moment persists in the ordered states of the Mott-insulating phase~\cite{Ishigaki18-2,Iwazaki21}.

Although previous theoretical studies have revealed characteristic behaviors arising from antiferromagnetic Hund's coupling in a purely electron model, such effective descriptions require justification through comparison with full electron--phonon models.
We refer to this electron–phonon coupled model as the ``Jahn--Teller--Hubbard'' model~\cite{Kaga22}, which can be viewed as a multiorbital extension of the Holstein--Hubbard model~\cite{Linden95}.
Phonons have been treated explicitly in earlier studies of superconductivity~\cite{Han00,Han03,Yamazaki14,Nomura15,Kaga22,Ishida25,Okada25, Okada_proc} as well as in investigations of single C$_{60}$ ions~\cite{Auerbach94,Manini94,Gunnarsson95,Gunnarsson95-2,Liu18-2,Matsuda18}.
Across these works, either Cartesian-coordinate or spherical-tensor representations have been employed, with the choice varying from study to study, which complicates direct comparison.
It is therefore desirable to clarify the correspondence between these two representations.
Furthermore, in the context of the unconventional orbital orders discussed above~\cite{Hoshino17,Ishigaki18-2,Werner17,Iwazaki21}, it is essential to examine the phonon properties associated with distortions of the fullerene cage, as they play a crucial role in experimentally characterizing unconventional ordered states.

In this paper, we consider a localized-electron model for the Mott-insulator phase coupled to anisotropic (Jahn--Teller) phonons.
While intersite interactions lead to various ordered states, as demonstrated for the multiorbital Hubbard model with antiferromagnetic Hund's coupling in Refs.~\cite{Ishigaki18-2,Iwazaki21}, here we focus on the single-molecule limit (Sec.~\ref{sec:model}) as a first step toward understanding electron--phonon--entangled Mott insulators.
Even in this simplified setting, the problem remains nontrivial because the Hilbert space in the molecular limit is infinite due to the phonon degrees of freedom.

To clarify the electron--phonon properties associated with unconventional orbital ordering, we introduce unconventional quadrupole tensors constructed from composite two-body operators of the form $c^\dagger c^\dagger c c$, rather than the standard one-body $c^\dagger c$ operators (where $c$ and $c^\dagger$ denote electron annihilation and creation operators, respectively).
We further examine how these composite quadrupoles couple to the standard quadrupole moments and lattice distortions using symmetry-based selection rules.
In addition, we investigate the nature of the entanglement induced by the electron--phonon coupling. 
By organizing the analysis from the viewpoint of angular momentum using the spherical-tensor formalism, the behavior of the reduced density matrix and the entanglement spectrum can be understood transparently.

To address this problem, we begin by discussing two types of representations: Cartesian tensors (Sec.~\ref{sec:Cart_rep_elec}) and spherical tensors (Sec.~\ref{sec:am_vs_cart_basis}).
The former is more intuitive and is naturally suited for applications in solid-state systems, whereas the latter, widely used in atomic and nuclear physics, provides a powerful and systematic framework for many-body calculations~\cite{Condon_book, Rose_book}.
Since the $t_{1u}$ system inherits an approximate spherical symmetry in the single-molecule limit, it is closely analogous to nuclear systems, in which fermions interact with bosons associated with collective excitations~\cite{Iachello_book,Talmi_book}.

A nucleus is an isolated quantum many-body system composed of nucleons in which the angular momentum is a good quantum number.
The spherical-tensor description provides a powerful 
theoretical
framework to construct operators that preserve 
rotational symmetry, enabling the derivation of selection rules and the evaluation of reduced matrix elements.
Historically, the coupling between collective and intrinsic nucleon motions has been studied since the early days of nuclear physics, beginning with investigations of nuclear fission phenomena~\cite{BohrKalckar1937} and the development of the Bohr--Mottelson collective model~\cite{BohrMottelson1953}.
This description was later systematized in an algebraic
framework by de Shalit~\cite{deShalit61}, leading to particle--boson coupling schemes such as the interacting boson--fermion model (IBFM)~\cite{Arima76}, as well as modern shell-model (SM) approaches~\cite{Mayer49,Jensen49}
with explicit phonon degrees of freedom~\cite{Otsuka78, Yoshinaga24}.

\begin{table}[t]
    \centering
    \begin{tabular}{c|c||c|c}
    \hline
       Research Field & Model &  $\ell_{\rm fermion}$ & $\ell_{\rm boson}$  \\
       \hline\hline
        Condensed Matter&Holstein--Hubbard & 0 & 0 \\
        Physics&Jahn--Teller--Hubbard & 1 & 0, 2 \\ \hline
        &de
        Shalit & 1 & $\forall$ \\
        Nuclear Physics&IBFM & 1 or 2 & 0, 2 \\
        &SM + phonon & model dep.  & 0, 2, 3 \\ \hline
        Quantum Optics & Rabi, JC & 
        1/2 & 0
        \\
    \hline
    \end{tabular}
\caption{
Family of Bohr--Mottelson models.
Note that spin and isospin degrees of freedom are omitted in the table.
In this paper, we restrict ourselves to the single-site case for the Holstein--Hubbard and Jahn--Teller--Hubbard models.
}
    \label{tab:my_label}
\end{table}

In this work, we draw on this mathematical structure and adapt it to an electron--phonon problem in condensed matter physics, without invoking any specific nuclear model, but rather exploiting the spherical-tensor formalism as a conceptual and algebraic bridge between the two fields.
The boson--fermion models, collectively referred to as the Bohr--Mottelson family of models, are summarized in Table~\ref{tab:my_label}.

From a different perspective, we briefly comment on a related class of systems. 
Although the context differs from the orbital-degenerate systems that constitute the main subject of this paper, 
the quantum Rabi model~\cite{Rabi36,Rabi37} and the Jaynes--Cummings (JC) model~\cite{Jaynes63} provide closely connected examples. 
These models describe a minimal light--matter interaction in which bosonic excitations are created and annihilated in association with transitions between two levels. 
Viewed from the standpoint of angular momentum, the model can be regarded as a pseudospin $\ell_{\rm fermion}=1/2$ coupled to a $\ell_{\rm boson}=0$ bosonic mode without internal angular momentum. 
In this sense, it may also be included in the family summarized in Table~\ref{tab:my_label}.
In these models, entanglement properties have been studied~\cite{Rossatto17,Liu17,Shi22}, and a similar theoretical perspective is incorporated into our model (Sec.~\ref{sec:numer_entangle}).

This paper includes a review section aimed at bridging the frameworks commonly used in condensed matter and nuclear physics.
Although some of the results presented here are not entirely new, we believe that a systematic comparison and unification of these two perspectives provides a useful guide to spherical tensors in condensed matter physics.
The original contribution of this work lies mainly in the analysis of composite quadrupoles in the presence of phonons and entanglement properties in terms of angular momentum representation (Secs.~\ref{sec:am_vs_cart_basis}, \ref{sec:numerical_results} and \ref{sec:quasispin}), which provide insight into the role of electron--phonon orbital degrees of freedom in fulleride compounds.

The rest of this paper is organized as follows.
Sec.~\ref{sec:model} introduces the model from a condensed-matter perspective 
using a Cartesian basis.
Sec.~\ref{sec:Cart_rep_elec} presents the Cartesian formulation for the wave functions and operators, while Sec.~\ref{sec:am_vs_cart_basis} develops the spherical-tensor approach.
Numerical analysis is given in Sec.~\ref{sec:numerical_results}, and the selection rules are discussed in Sec.~\ref{sec:quasispin}, followed by a summary in Sec.~\ref{sec:summary}.
Appendix~\ref{sec:Gell-Mann} lists the Gell-Mann matrices and the correspondence between Cartesian and spherical representations.
Additional numerical data and definitions of tensor operators are provided in Appendices~\ref{sec:nc_depend} and \ref{sec:spherical-tensor}, respectively.
Appendices~\ref{sec:spinorbital}--\ref{sec:Quasispin} describe the tensorial construction of wave functions and matrix elements based on methods developed in nuclear physics and provide a detailed and pedagogical account that may be useful for condensed-matter physicists interested in applying these methods to their systems.

\section{Model}
\label{sec:model}

\begin{table}[t]
    \centering
    \begin{tabular}{c|c}
    \hline
       Symbol  & Description \\ \hline
       $\sg=\ua,\da$ or $+\textstyle{1 \over 2},-\textstyle{1 \over 2}$   & spin index or spin value\\
       $\gm=x,y,z$  & $t_{1u}$ ($p$-orbital like) orbital index \\ 
 $\mu,\nu=x,y,z$& Cartesian coordinate index\\
       $\eta=xy,yz,\cdots$& Cartesian quadrupolar index\\
             $S,M_S$ & spin quantum numbers \\
       $\ell,m$; $L,M_L$ & angular momentum quantum numbers \\
       $K,M_K$ & quasispin quantum numbers \\ \hline
        $\hat \lambda^\eta$ & Gell-Mann matrices \\
        $\hat \ell^\mu$ & matrices for orbital angular momentum \\ \hline
       $\bm S$& spin angular momentum \\
      $\bm L$ & orbital angular momentum \\
      $\bm K$ & quasispin angular momentum\\ \hline
       $c_{\gm\sg}$&electron annihilation operator\\
       $D_\gm$& doublon annihilation operator\\
 $Q_\eta$ & quadrupole operator  \\
      $a_\eta$ & Jahn--Teller phonon annihilation operator\\
      $\phi_\eta$ & displacement operator \\ \hline
      $p_{m\sg}$ & (rank-1) $p$-fermion annihilation operator\\
      $d_M$ & (rank-2) $d$-boson annihilation operator\\
       $x_M$ & self-adjoint $d$-boson tensor\\ 
      $\mathscr L_M$ & rank-1 spherical tensor \\
      $\mathscr Q_M$ & rank-2 spherical tensor \\
      $A^{(L)}_M$
      & (rank-$L$) tensor with $M =  - L, \cdots ,L$ \\
      $[A^{(L_A)} \otimes B^{(L_B)}]^{(L)}$
      & (rank-$L$) tensor of $A^{(L_A)}$ and $B^{(L_B)}$\\
      $(A^{(L)}\cdot B^{(L)})$ &scalar product of $A^{(L)}$ and $B^{(L)}$ \\
      $D^{(L,S)}$ &double tensor of rank-$L$ and -$S$ \\
      $T^{(L,S,K)}$ &triple tensor of rank-$L$, -$S$ and -$K$  \\
      \hline
    \end{tabular}
    \caption{Notation for the indices and operators used in this paper. 
    For brevity, the magnetic quantum numbers $M_{S}$, $M_{L}$, and $M_{K}$ are denoted simply by $M$ in the main text when no confusion arises.}
    \label{tab:notation}
\end{table}

We begin with the definition of the electron-phonon-coupled model.
Because a large number of symbols are introduced in this paper, we summarize the notation in Table~\ref{tab:notation}.

\subsection{Hamiltonian}

We employ the Cartesian representation in this section. The standard representation of the Coulomb interaction is given by the Slater-Kanamori form~\cite{Georges13}:
\begin{align}
    &\mathscr H_{\rm C} = \frac{U}{2} \sum_{\gm\sg} n_{\gm \sg} n_{\gm \bar \sg}
    + \frac {U'} 2 \sum_{\gm\neq \gm', \sg} n_{\gm\sg} n_{\gm' \bar \sg}
    \nonumber \\
    &\hspace{2mm}
    + \frac{U'-J}{2} \sum_{\gm\neq \gm', \sg} n_{\gm\sg} n_{\gm'\sg} 
    -J \sum_{\gm\neq \gm'} S_\gm^+ S_{\gm'}^-
    + J \sum_{\gm\neq \gm'} D_\gm^\dg D_{\gm'},
    \label{eq:Hamiltonian}
\end{align}
where the orbital index is given by $\gm=x,y,z$, which has atomic $p$-orbital-like character.
The number operator is given by $n_{\gm\sg} = c^\dg_{\gm\sg} c_{\gm \sg}$.
The `barred' symbol for spin index is defined by $\displaystyle \bar \ua = \da$ and $\displaystyle \bar \da = \ua$.
We have also defined the spin and doublon-creation operators for each orbital by
\begin{align}
    \bm S_\gm &= \frac 1 2 \sum_{\sg\sg'} c_{\gm\sg}^\dg \bm \sg_{\sg\sg'} c_{\gm\sg'}
    , \\
    D_\gm^\dg &= c_{\gm\ua}^\dg c^\dg_{\gm \da},
\end{align}
where $\bm \sg$ is a Pauli matrix.
Eq.~\eqref{eq:Hamiltonian} contains five terms: 
The first three represent density-density Coulomb interactions corresponding to intra-orbital, inter-orbital (different spin), and inter-orbital (same spin) electrons, respectively.
The remaining two terms, proportional to $J$, describe the spin-flip and pair-hopping processes, which favor high-spin and low-spin states, respectively.

Since the molecule is nearly spherical, the $t_{1u}$ subspace approximately retains continuous rotational symmetry~\cite{Nomura16}.
In this case, there is the relation $U' = U-2J$, and one can rewrite the Coulomb interaction Hamiltonian as
\begin{align}
    \mathscr H_{\rm C} &= \frac{5J}{2} \sum_{\gm} D_\gm^\dg D_\gm
    - J \sum_{\gm\neq \gm'} (\bm S_\gm\cdot \bm S_{\gm'} - D_{\gm}^\dg D_{\gm'})
    \nonumber \\
    &\ \ \  + 
    \frac{2U-5J}{4}N(N-1),
\end{align}
where $N =\sum_{\gm\sg}n_{\gm\sg}$ is the total charge.
When we focus on the Mott insulator phase of A$_3$C$_{60}$, 
the occupation number of localized electrons in the $t_{1u}$  orbitals is three $(N=3)$.

Next we consider the Jahn-Teller phonons.
In general, there are $3\times 60-6=174$ vibrational modes for $C_{60}$, which are classified based on the molecular point group $I_h$.
Among them, two $A_g$ (one-dimensional) and eight $H_g$ (five-dimensional) modes are coupled to $t_{1u}$ orbital~\cite{Faber11}.
In constructing the model, 
we ignore the $A_{g}$ modes which is coupled to the charge $N$ (constant).
As for $H_g$, we single out representative one having largest coupling constant~\cite{Faber11,Kaga22}.
The $H_g$-type vibration consists of five degenerate modes labeled by $\eta=1,3,4,6,8$, each corresponding to a deformation described by the coordinate representations $xy$, $x^2-y^2$, $zx$, $yz$, and $x^2+y^2-2z^2$, respectively (see Appendix~\ref{sec:Gell-Mann}).
The phonon part of the Hamiltonian is thus given by
\begin{align}
    \mathscr H_{\rm p} = \omega_0 \sum_{\eta = 1,3,4,6,8} a_\eta^\dagger a_\eta,
\end{align}
where $a_\eta$ is the annihilation operator for the $\eta$-th mode and $\omega_0$ is the phonon frequency of the molecular vibration.

While the Hund's coupling $J$ is positive (ferromagnetic), such that the spin-flip term dominates over pair hopping, its effective sign is inverted due to the coupling to Jahn--Teller phonons~\cite{Fabrizio97}, which instead favors pair hopping.
In this paper, however, we do not adopt this effective description; instead, we treat the electron--phonon coupling explicitly, without invoking a sign reversal of the Hund's coupling:
\begin{align}
    \mathscr H_{\rm ep} &= g \sum_{\eta = 1,3,4,6,8} Q_\eta \phi_\eta,
\end{align}
where $g$ is the coupling constant and we have introduced the electric quadrupole operator (or orbital operator) by
\begin{align}
     Q_\eta &= \sum_{\gm\gm'\sg} c^\dg_{\gm\sg} \lambda^\eta _{\gm\gm'} c_{\gm'\sg},
\label{eq:quadrupole_definition}
\end{align}
which satisfies $Q_\eta^\dagger = Q_\eta$.
This Hermitian property is advantageous compared to the quadrupole tensor operator introduced later [cf.\ Eq.~\eqref{eq:QM_def}], and ensures that $Q_\eta$ represents a physical observable.
The Gell-Mann matrices $\lambda^\eta$ are defined in Appendix~\ref{sec:Gell-Mann}.
$\phi_\eta = a_\eta + a_\eta^\dg$ is the displacement operator associated with the fullerene molecule vibration. 
Physically, $Q_\eta$ corresponds to the anisotropy of the electronic spatial distribution, which couples to the molecular anisotropic (Jahn--Teller) distortion.
Once the orbital operator $Q_\eta$ is introduced, we can rewrite the Coulomb interaction term as~\cite{Iimura21}
\begin{align}
    \mathscr H_{\rm ee}&=\frac{J}{2} \sum_{\eta=1,3,4,6,8}:Q_\eta^2 :,
    \label{eq:H_ee}
\end{align}
where the colon symbol ($:$) denotes normal ordering, ensuring that the energy of the single-particle states is not counted.
The Hamiltonian $\mathscr H_{\rm ee}$ defined above represents the spin--orbital part of the electron--electron interaction, and
the explicit relation between $\mathscr H_{\rm ee}$ and $\mathscr H_{\rm C}$ is given by
\begin{align}
\mathscr H_{\rm C} &= \mathscr H_{\rm ee} + \left(\frac{U}{2} - \frac{2J}{3}\right) N(N-1),
\end{align}
where the term involving $N=\sum_{\gm\sg}c_{\gm\sg}^\dg c_{\gm\sg}=3$ gives a constant energy shift.

Collecting all the terms introduced above, the total Hamiltonian in the localized limit of the Mott-insulating phase is given by
\begin{align}
\mathscr H = \mathscr H_{\rm ee} + \mathscr H_{\rm p} + \mathscr H_{\rm ep},
\end{align}
which will be discussed in detail in the remainder of this paper.
We refer to this system as a ``Jahn--Teller--Hubbard molecule,'' in analogy with the single-orbital Hubbard model in its atomic limit, often called the ``Hubbard atom.''
In actual solids, intermolecular hopping is also present; however, in this paper we focus on the localized-electron limit of the model in the strong-interaction regime.
While similar single-site models have been considered in Refs.~\cite{Auerbach94,Chancey_book,Liu18}, the present work analyzes the model using the spherical-tensor formalism, enabling a detailed study of quadrupolar and entanglement properties.

In order to intuitively understand the effective interaction among electrons mediated by phonons, we rewrite the total interaction as
\begin{align}
    \mathscr H
    &= 
     \frac{J+J_{\rm ph}}{2} \sum_\eta :Q_\eta^2:
+ \, \omega_0 \sum_\eta :A_\eta^\dg A_\eta:,
\end{align}
where $\displaystyle J_{\rm ph} = - \frac{2g^2}{ \omega_0}$ ($<0$), and
\begin{align}
    A_\eta = a_\eta + \frac{g}{ \omega_0} Q_\eta
\end{align}
is a `shifted' boson operator.
If $A_\eta$ and $Q_\eta$ were independent physical degrees of freedom, the term proportional to $J_{\rm ph}$ would be regarded as a phonon-mediated static interaction, which modifies the Hund's coupling as $J_{\rm eff} = J + J_{\rm ph}$~\cite{Fabrizio97}.
This argument provides a simple interpretation of the effect of electron--phonon coupling.
Although $A_\eta$ behaves approximately as a canonical boson in the weak electron--phonon coupling limit ($g \to 0$), it is not a simple bosonic operator, as it satisfies nonstandard commutation relations:
\begin{equation}
\begin{aligned}
    [A_\eta, A_{\eta'}^\dg] &= \delta_{\eta\eta'} + \qty( \frac{g}{  \omega_0} )^2 [Q_\eta, Q_{\eta'}],
    \\
    [A_\eta, A_{\eta'}] &= [A_\eta^\dg, A_{\eta'}^\dg] = \qty( \frac{g}{  \omega_0} )^2 [Q_\eta, Q_{\eta'}].
\end{aligned}
\end{equation}
Thus, the noncommutativity of $Q_\eta$ complicates the multiorbital problem.
This situation contrasts with the single-orbital Holstein--Hubbard model~\cite{Linden95}, in which the shifted operator satisfies the canonical bosonic commutation relations~\cite{Lang62}.

\section{Cartesian representation of electrons}
\label{sec:Cart_rep_elec}

In this section, we describe the electronic degrees of freedom in terms of the Cartesian coordinate index $x,y,z$.
This representation is more intuitive for understanding the underlying degrees of freedom than the tensorial representation
discussed in the next section.

\subsection{Basis functions}

For electronic part with three electrons, there are $\displaystyle\binom{6}{3}=20$ states in total.
Here, to characterize these eigenstates, we utilize the total spin
\begin{align}
    \bm S_{\rm tot} &= \bm S_x + \bm S_y + \bm S_z
    ,
\end{align}
the number of doubly occupied orbitals (doublons)
\begin{align}
    N_D &= \sum_\gm D_\gm^\dg D_\gm
    ,
\end{align}
and the scalar spin chirality
\begin{align}
    C &= \bm S_x \cdot (\bm S_y\times \bm S_z)
    \\
    &= \frac{1}{6} 
    \sum_{\gm_1\gm_2\gm_3} \sum_{\mu_1\mu_2\mu_3} 
    \epsilon_{\mu_1\mu_2\mu_3}
    \epsilon_{\gm_1\gm_2\gm_3} S_{\gm_1}^{\mu_1}
    S_{\gm_2}^{\mu_2}
    S_{\gm_3}^{\mu_3}
    ,
\end{align}
where $\epsilon_{\mu_1\mu_2\mu_3}$ is the Levi-Civita symbol.
Note that the subscript $\gm=x,y,z$ labels the orbital component and the superscript $\mu =x,y,z$ represents a direction of the spin operator.
These operators, $\bm S_{\rm tot}$, $N_D$, and $C$, are respectively one-, two-, and three-body operators, and become good quantum numbers within the $N=3$ multiplet.
The total spin quantum number takes the values $S_{\rm tot}=\tfrac 3 2$ and $\tfrac 1 2$.

The eigenstates can be written explicitly as
\begin{align}
    &|\tfrac 3 2, 3\sg \ra=\frac{1}{\sqrt 6}
    \sum_{\gm\gm'\gm''} \epsilon_{\gm\gm'\gm''}
    c^\dg_{\gm \sg} c^\dg_{\gm'\sg} c^\dg_{\gm''\sg} |0\ra,
    \label{eq:three_half_2}
    \\
    &|\tfrac 3 2, \sg \ra=
    \frac{1}{\sqrt 6}
    \sum_{\gm\gm'\gm''} \epsilon_{\gm\gm'\gm''}
    c^\dg_{\gm \bar \sg} c^\dg_{\gm'\sg} c^\dg_{\gm''\sg} |0\ra,
    \label{eq:three_half_1}
    \\
    &|\tfrac 1 2,b,\gm,\sg \ra = 
    \frac{1}{\sqrt 8}\sum_{\gm'\gm''}|\epsilon_{\gm\gm'\gm''}|\ 
    c_{\gm\sg}^\dg (D_{\gm'}^\dg + D_{\gm''}^\dg)|0\ra,
    \label{eq:b_states}
    \\
    &|\tfrac 1 2,a,\gm,\sg \ra = 
    \frac{1}{\sqrt 8}
    \sum_{\gm'\gm''}\epsilon_{\gm\gm'\gm''} \  c_{\gm\sg}^\dg (D_{\gm'}^\dg - D_{\gm''}^\dg)|0\ra,
    \\
    &|\tfrac 1 2,\tau,\sg\ra=\frac{
    (c_{x\bar \sg}^\dg c_{y\sg}^\dg c_{z\sg}^\dg + \omega^{\tau}c_{x\sg}^\dg c_{y\bar\sg}^\dg c_{z\sg}^\dg+\omega^{-\tau}c_{x\sg}^\dg c_{y\sg}^\dg c_{z\bar\sg}^\dg) |0\ra}
    { 
    \sqrt{12}
    \, \sg},
    \label{eq:eigenstates_SKHam}
\end{align}
where $\omega = \epn^{2\pi \imu/3}$ is a primitive cubic root of unity.
The symbol $\sg$ takes $\sg=\ua$ (or $\sg=+1/2$) and $\sg=\da$ (or $\sg=-1/2$).
We have defined the sign of the eigenvalues of the scalar spin chirality: $\tau = {\rm sgn\,}C$.
The labels `$b$' and `$a$' indicate bonding and antibonding states of doublons.
The eigenenergies of $\mathscr H_{\rm ee}$ are $-5J$ for $|S_{\rm tot}=\tfrac 3 2, S_{\rm tot}^z\ra$, zero for $|\tfrac  1 2, b,\gm,\sg\ra$, and $-2J$ for the others ($|\tfrac 1 2 ,a,\gm,\sg \ra$ and $|\tfrac 1 2,\tau,\sg \ra$). The orbital part for $|S_{\rm tot}=\tfrac 3 2, S_{\rm tot}^z\ra$ forms a singlet structure as shown in Eqs.~\eqref{eq:three_half_1} and \eqref{eq:three_half_2}, which has the same structure as a color singlet of hadrons~\cite{Sakurai_book}.

In order to further characterize the basis functions,
we introduce the angular momentum operators,
\begin{equation}
\begin{aligned}
    L_\mu  &= \sum_{\gm\gm'\sg} c^\dg_{\gm\sg} \ell^\mu_{\gm\gm'} c_{\gm'\sg},
    \label{eq:first_def_of_L}
\end{aligned}
\end{equation}
for $\mu=x,y,z$,
where we have introduced the matrices $\hat \ell^x = \hat \lambda^7$, $\hat \ell^y = - \hat \lambda^5$ and $\hat \ell^z = \hat \lambda^2$ relevant to angular momentum operator.
The hat ($\hat \ $) symbol indicates the $3\times 3$ matrix.
In terms of spin and orbital angular momentum quantum number,
the above states in Eqs.~(\ref{eq:three_half_2}--\ref{eq:eigenstates_SKHam}) correspond to the total spin and angular momentum $(S_{\rm tot},L)=(\tfrac 3 2, 0)$, $(S_{\rm tot},L)=(\tfrac 1 2, 1)$ and $(S_{\rm tot},L)=(\tfrac 1 2, 2)$ states~\cite{Georges13}.

Note that spin $\sg$ is regarded as a good quantum number even when electron--phonon interaction is taken into account.
Hence we may focus on $\sg=+1/2$ (i.e., $\sg=\ua$) and can omit $\sg$.
In addition, $S_{\rm tot}=\tfrac 3 2$ states are not affected by the coupling to phonons.
Thus, we only have to consider eight $S_{\rm tot}=\tfrac 1 2$ states labeled as 
\begin{equation}
\begin{aligned}
\left.
\begin{matrix}
    |bx\ra&=&|\tfrac 1 2, b,\gm=x,\sg=\ua\ra\\[1mm]
    |by\ra&=&|\tfrac 1 2, b,\gm=y,\sg=\ua\ra\\[1mm]
    |bz\ra&=&|\tfrac 1 2, b,\gm=z,\sg=\ua\ra\\[1mm]
\end{matrix}
\ \ 
\right\} \ \ \ (L=1),
\\
\left.
\begin{matrix}
    |ax\ra&=&|\tfrac 1 2, a,\gm=x,\sg=\ua\ra\\[1mm]
    |ay\ra&=&|\tfrac 1 2, a,\gm=y,\sg=\ua\ra\\[1mm]
    |az\ra&=&|\tfrac 1 2, a,\gm=z,\sg=\ua\ra\\[1mm]
    |+\ra&=&|\tfrac 1 2, \tau=+,\sg=\ua\ra\\[1mm]
    |-\ra&=&|\tfrac 1 2, \tau=-,\sg=\ua\ra
\end{matrix}
\ \ 
\right\} \ \ \ (L=2),
\label{eq:shorthand_notation}
\end{aligned}
\end{equation}
where we have introduced a short-hand notation as shown in the left-hand side.
As for phonons, 
their basis functions are trivial: the eigenstates of $\mathscr H_{\mathrm{p}}$ are characterized by the boson numbers $n_\eta = a_\eta^\dg a_\eta$ for $\eta=1,3,4,6,8$.

\subsection{Composite quadrupole}
\label{sec:Cart_Comp_quadru}

\begin{figure}
    \centering
    \includegraphics[width=0.8\linewidth]{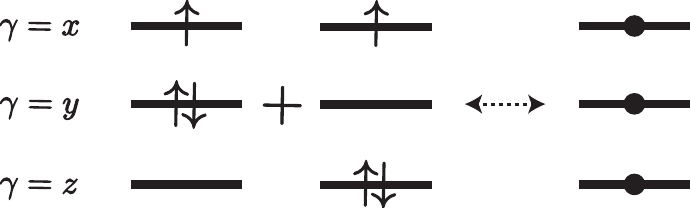}
    \caption{Schematic picture of the $|bx\ra = |\tfrac 1 2, b, x, \ua \ra$ state given in Eq.~\eqref{eq:b_states}. The right-hand side shows the mean occupation number for each orbital regardless of the spin, where the three orbitals are equally occupied in average.}
    \label{fig:schematic}
\end{figure}

We now discuss the degrees of freedom within the $|b\gamma\rangle$ ($L=1$) subspace defined in Eq.~\eqref{eq:shorthand_notation}, which becomes the ground state in the effectively antiferromagnetic Hund's coupling $J\to J_{\rm eff}<0$ [see the energies written below Eq.~\eqref{eq:eigenstates_SKHam}].
Because this subspace is labeled by a single orbital index $\gamma=x,y,z$, it is expected to exhibit properties analogous to those of a $p$ orbital.
However, the standard quadrupole operator is inactive~\cite{Anderson92, Ceulemans94,Chancey_book,Hoshino17}.
An intuitive explanation for the disappearance of the standard orbital moment in the $|b\gamma\rangle$ subspace is schematically illustrated in Fig.~\ref{fig:schematic}.
Focusing on the $|b x\rangle$ state, which exhibits anisotropy oriented along the $x$ direction, the occupation number of each orbital is given by
\begin{align}
    \langle b x | \sum_{\sigma} c_{\gamma\sigma}^\dagger c_{\gamma\sigma} | b x \rangle = 1
\end{align}
for $\gamma = x, y, z$.
Thus, no imbalance in the single-particle orbital occupations occurs, even though the wave function $|b x\rangle$ is anisotropic, as shown in Fig.~\ref{fig:schematic}.
By contrast, the doublon occupation numbers clearly capture the orbital imbalance:
\begin{equation}
\begin{aligned}
    &\langle b x | D_x^\dagger D_x | b x \rangle = 0 , \\
    &\langle b x | D_y^\dagger D_y | b x \rangle
      = \langle b x | D_z^\dagger D_z | b x \rangle = \tfrac{1}{2}.
\end{aligned}
\end{equation}
This indicates that doublon-based quadrupolar degrees of freedom are active in the $L=1$ subspace~\cite{Iwazaki21}.
Although this picture is most transparent in the Cartesian representation, it is less apparent 
in the spherical-tensor representation discussed in Sec.~\ref{sec:am_vs_cart_basis}; consequently, the Cartesian formulation provides a more intuitive physical interpretation.

As discussed above, the doublon captures orbital-symmetry breaking, which motivates the construction of doublon-based quadrupolar degrees of freedom.
A naive generic form of an active quadrupole operator may be written as
\begin{align}
    Q_{\eta}^{\rm C\,(tentative)}
    =
    \sum_{\gamma\gamma'}
    D_\gamma^\dagger \lambda^\eta_{\gamma\gamma'} D_{\gamma'},
\end{align}
which is analogous to Eq.~\eqref{eq:quadrupole_definition}.
Here, the superscript ``C'' denotes the composite nature of the operator.
The operator $Q_{\eta}^{\rm C\,(tentative)}$ indeed captures the orbital imbalance of the $L=1$ state, and its $\eta = 3,8$ components were discussed previously without explicit inclusion of phonon degrees of freedom~\cite{Hoshino17}.
However, this naively defined operator does not transform properly as a spherical tensor.
For example, the norm of the matrix elements
$\langle b\gamma \vert Q_{\eta}^{\rm C\,(tentative)} \vert b\gamma' \rangle$
depends on $\eta$, whereas that of a standard quadrupole operator is independent of $\eta$ under spherical symmetry.

The above observations lead to two questions:
(i) What is the proper definition of a composite quadrupole operator that behaves as a spherical tensor in the $L=1$ subspace?
(ii) How do the phonon degrees of freedom behave in relation to the electronic composite quadrupole?
The second question will be addressed in the next section (Sec.~\ref{sec:am_vs_cart_basis}).
Here, we focus on the first question (i) and construct a composite quadrupole operator that satisfies the required tensorial properties.
Specifically, we define the desired quadrupole operator as
\begin{align}
    Q^{\rm C}_\eta = \sum_{\mu\nu} \lambda_{\mu\nu}^\eta :L_\mu L_{\nu}:,
    \label{eq:def_QC_1}
\end{align}
where $\bm L$ is defined in Eq.~\eqref{eq:first_def_of_L}.
This form of composite operator is inspired by the following identity for the Gell-Mann matrices:
\begin{align}
\hat \lambda^\eta = - \sum_{\mu\nu} \lambda^\eta_{\mu\nu} 
\hat \ell^\mu \hat \ell^{\nu}
\end{align}
for $\eta=1,3,4,6,8$.
Namely, the combination of angular-momentum matrices ($\hat \ell^\mu$) leads to the quadrupole ($\hat \lambda^\eta$).
The operator $\bm L$ is constructed from a bilinear fermionic form, $c^\dagger c$, and therefore $Q^{\rm C}_{\eta}$ is a two-body operator.
Nonzero matrix elements of $\bm L$ in the $L=1$ subspace imply that $Q^{\rm C}_{\eta}$ is also finite in this subspace.
Indeed, by introducing the projection operator $P_L$ onto the electronic $L$ multiplet [see Eq.~\eqref{eq:shorthand_notation}], we obtain
\begin{align}
    P_{L=1} Q_{\eta}^{\rm C} P_{L=1}
    &= - \sum_{\sg}\sum_{\gm\gm'}
    \left|\tfrac 1 2,b,\gm,\sg\right\ra 
    \lambda^\eta_{\gm\gm'}
    \left\la \tfrac 1 2, b,\gm',\sg \right|,
\end{align}
which is the desired property as a spherical tensor, since it has the same structure as Eq.~\eqref{eq:quadrupole_definition}.

We can also introduce another form of two-body quadrupoles by using Eq.~\eqref{eq:def_composite_q}:
\begin{equation}
\begin{aligned}
    \overline{Q}^{\rm C}_{1} &= \sqrt 3 \, \overline{Q_8 Q_1} + \frac{3}{2} \, \overline{Q_4 Q_6},
    \\
    \overline{Q}^{\rm C}_{4} &= \frac{\sqrt 3}{2} (-\, \overline{Q_8 Q_4} + \sqrt 3 \, \overline{Q_3 Q_4}) + \frac 3 2 \, \overline{Q_1 Q_6},
    \\
    \overline{Q}^{\rm C}_{6} &= \frac{\sqrt 3}{2} (-\, \overline{Q_8 Q_6} - \sqrt 3 \, \overline{Q_3 Q_6}) + \frac 3 2 \, \overline{Q_1 Q_4},
    \\
 \overline{Q}^{\rm C}_{3} &= \sqrt 3 \, \overline{Q_8 Q_3} + \frac{3}{4} (Q_4^2-Q_6^2),
    \\
    \overline{Q}^{\rm C}_8
    &= \frac{\sqrt 3}{2} \qty[ Q_1^2+Q_3^2- Q_8^2  - \frac 1 2 (Q_4^2 + Q_6^2) ],
\end{aligned}
    \label{eq:QQdef_Cartesian}
\end{equation}
where $\overline{AB} = (AB+ BA)/ 2!$ is the symmetrized product.
This expression yields
\begin{align}
    P_{L=1} \overline{Q}^{\rm C}_{\eta} P_{L=1}
    &= -\frac{7}{4} P_{L=1} Q^{\rm C}_{\eta} P_{L=1}.
    \label{eq:rel_QCs}
\end{align}
Namely, there are two ways to define the composite (two-body) quadrupole operators [Eqs.~(\ref{eq:def_QC_1},\ref{eq:QQdef_Cartesian})], and, once projected onto the $L=1$ subspace, they reduce to the same expression up to a constant factor as in Eq.~\eqref{eq:rel_QCs}.

In this way, the electric quadrupole necessitates consideration of composite operators within the $L=1$ multiplet, whereas the magnetic degrees of freedom, whose lowest-order contributions are angular momentum (magnetic dipole) represented by one-body operators, have nonzero matrix elements within this restricted Hilbert space.
Although composite quadrupoles involving phonons can be constructed in a manner similar to the formulation above, we instead employ the spherical tensor representation, since its compatibility with the quasispin formulation (see Sec.~\ref{sec:am_vs_cart_basis}) is particularly convenient for interpreting the results.

We remark that, while the composite quadrupole considered here arises in the context of fulleride materials, composite physical quantities have also been studied in condensed matter physics as order parameters associated with symmetry breaking~\cite{Andreev84,Emery92,Balatsky93,Coleman95,Nourafkan08,Hoshino11,Hoshino14,Fernandes19,Geilhufe25,Miki26,Kuniyoshi26}. 
In nuclear physics, similar concepts have mainly been discussed in terms of intrinsic collective degrees of freedom and excitation modes \cite{Teng25}, due to the limited controllability of external fields.

\section{Spherical tensor representations}
\label{sec:am_vs_cart_basis}

While the Cartesian representations used in the previous sections are intuitive, a systematic formulation is more naturally achieved using tensorial algebra based on the Wigner--Eckart theorem.
We first introduce the spherical basis states (Sec.~\ref{sec:am_vs_cart_basis_A}), 
which are followed by the formulation of operators (Secs.~\ref{sec:am_vs_cart_basis_B}--\ref{sec:am_vs_cart_basis_E}).
We can construct the analytic solution for the case with one boson number as explained in Sec.~\ref{sec:am_vs_cart_basis_F}.

\subsection{Relation between spherical basis and Cartesian basis}
\label{sec:am_vs_cart_basis_A}

The eigenstate of $\mathscr H_{\rm ee}$ is denoted as $|LM\ra$
(labeled by eigenvalues of $\bm L^2$ and $L_z$). Namely, we write the eigenequation
\begin{align}
\big(\mathscr H_{\rm ee}+2J \big)
\  |LM\ra = E_F(L) \ 
|LM\ra ,
\end{align}
where the eigenenergy $E_F(L)$ is given by
\begin{equation}
\begin{aligned}
E_F(L = 0) &= -3J, \\
E_F(L = 1) &= 2J, \\
E_F(L = 2) &= 0.
\end{aligned}
\label{eq:fermi_system_energy}
\end{equation}
Note that the subscript $F$ emphasizes the energy of the fermionic systems, and we have shifted the energy by $2J$ to make the analytic equation simple [cf. Eq.~\eqref{eq:2by2mat}].
The wave function $|LM\ra$ is related to the previously introduced Cartesian basis representation [see Eq.~\eqref{eq:shorthand_notation}] as
\begin{equation}
\begin{aligned}
    |1,\pm 1\ra &= \mp \qty( |bx\ra \pm \imu  |by\ra)/\sqrt 2,
    \\
    |1,0\ra &= |bz\ra,
\end{aligned}
\label{eq:L1_def}
\end{equation}
for $L=1$, and
\begin{equation}
\begin{aligned}
|2,\pm 2\ra &= \qty(\omega |+\ra - \omega^* |-\ra \mp \sqrt 2 |az\ra)/2,
\\
    |2,\pm 1\ra &= \qty( |ax\ra  \mp \imu |ay\ra)/\sqrt 2,
    \\
    |2,0\ra &= \imu \qty( \omega |+\ra + \omega^* |-\ra )/\sqrt 2,
\end{aligned}
\label{eq:L2_def}
\end{equation}
for $L=2$.
Note that we have focused on the total spin $S=1/2$ sector which couples to Jahn-Teller phonons, and have not
considered the $S=3/2$ state ($L=0$) whose wave function has been given in Eq.~\eqref{eq:three_half_2}.

\subsection{Rank-1 electron tensors}
\label{sec:am_vs_cart_basis_B}

Next we consider the tensorial representation of operators.
The $p$-fermion creation operators are defined by
\begin{equation}
\begin{aligned}
    p_{\pm 1,\sg}^\dg &= \mp \qty( c_{x\sg}^\dg \pm \imu c_{y\sg}^\dg ) /\sqrt 2 ,
    \\
    p_{0,\sg}^\dg &= c_{z\sg}^\dg ,
    \label{eq:def_pfermion}
\end{aligned}
\end{equation}
which defines the unitary transformation $c^\dg_{\gm\sg} \to p^\dg_{m\sg}$, 
satisfying standard anticommutation relation.
Here, the label `$p$' is used to emphasize the character of the angular momentum $\ell =1$ (called $p$-fermion).
It is also worth noting that the $p$-fermions are regarded as a rank-1 spherical tensor:
\begin{equation}
\begin{aligned}
    [L_z,p^\dg_{m\sg}] &= m \  p^\dg_{m\sg},
    \\
    [L_\pm,p^\dg_{m\sg}] &= \sqrt{2-m(m\pm 1)} \ p^\dg_{m\pm1,\sg},
\end{aligned}
\end{equation}
where $L_\pm = L_x \pm \imu L_y$.
We provide the definition of generic spherical tensors in Appendix~\ref{sec:spherical-tensor}.
As for the annihilation operator, since $p_{m\sg}$ itself does not behave as a spherical tensor, we define
\begin{align}
    \tilde p_{m\sg} &= (-1)^{1-m} (-1)^{\frac 1 2 -\sg} p_{-m, \bar \sg},
    \label{eq:def_pfermiontilde}
\end{align}
which satisfies the definitive relation of spherical tensor.
Here $\sg={\frac 1 2}$ for spin $\ua$ and 
$ \bar \sg= -\sg =-{\frac 1 2}$ for spin $\da$.

The ladder operators, which are also regarded as spherical tensors are defined by
\begin{equation}
\begin{aligned}
    \mathscr L_{\pm 1} &= \mp (L_x \pm \imu L_y)/\sqrt 2,
    \\
    \mathscr L_0 &= L_z.
\end{aligned}
\end{equation}
These are spherical tensors of rank 1:
\begin{equation}
\begin{aligned}
    [L_z,\mathscr L_{m}] &= m \  \mathscr L_{m}
    ,
    \\
    [L_\pm,\mathscr L_{m}] &= \sqrt{2-m(m\pm 1)} \ \mathscr L_{m\pm1}.
\end{aligned}
\end{equation}
The matrix element is given by
\begin{align}
    \la L'M'|\mathscr L_{m}|L M\ra
    &=\delta_{LL'} \sqrt{L(L+1)}  \la L M 1 m |  L M' \ra,
    \label{eq:mate_angular_momentum}
\end{align}
where $\left\langle {Lm1m|LM'} \right\rangle$ are the Clebsch--Gordan (CG) coefficients and only the diagonal component with respect to $L$ becomes nonzero.

\subsection{Rank-2 electron tensors}
\label{sec:am_vs_cart_basis_C}

We define the spherical-basis rank-2 tensors by
\begin{equation}
\begin{aligned}
    \mathscr Q_{\pm 2} &= (Q_3\pm \imu Q_1)/\sqrt 2,
    \\
    \mathscr Q_{\pm 1} &= \mp (Q_4\pm \imu Q_6)/\sqrt 2, 
    \\
    \mathscr Q_{0} &= Q_8,
\end{aligned}
\label{eq:QM_def}
\end{equation}
with which $\mathscr Q_M^\dg = (-1)^M \mathscr Q_{-M}$ is satisfied.
Equation~\eqref{eq:QM_def} originates from the relation in Eq.~\eqref{eq:relat_quadrupoles}.
The left-hand side of Eq.~\eqref{eq:QM_def} is called `complex spherical tensor', and the right-hand side is called `tesseral spherical tensor'.

The operator $\mathscr Q_M$ satisfies the definitive relation of a spherical tensor:
\begin{equation}
\begin{aligned}
    [L_z,\mathscr Q_M] &= M \mathscr Q_M,
    \\
    [L_\pm, \mathscr Q_M] &= \sqrt{6-M(M\pm1)}\ \mathscr Q_{M\pm1}.
\end{aligned}
\end{equation}
The matrix element of the electronic quadrupole is given by
\begin{equation}
\begin{aligned}
    \la 1 m'| \mathscr Q_M | 2 m\ra
    &=  - \sqrt{10}\  \la 2m 2M | 1m' \ra,
    \\
    \la 2 m'| \mathscr Q_M | 1 m\ra
    &=   \sqrt{6}\  \la 1m 2M | 2m' \ra,
\end{aligned}
\end{equation}
and others are zero (see Eq.~\eqref{eq:red_mate_Q} for more details).
As discussed in Sec.~\ref{sec:Cart_Comp_quadru}, the matrix elements of the standard quadrupole operator within the $L=1$ subspace vanish, motivating consideration of the composite quadrupole.

\subsection{Phonon angular momentum and tensors}
\label{sec:am_vs_cart_basis_D}

We define the following boson operators associated with angular momentum tensor:
\begin{equation}
\begin{aligned}
    d^\dg_{\pm 2} &=  (a^\dg_3 \pm \imu a^\dg_1) /\sqrt 2,
    \\
    d^\dg_{\pm 1}
    &=   \mp (a^\dg_4 \pm \imu a^\dg_6) /\sqrt 2,
    \\
    d^\dg_{0}
    &= a_8^\dg.
\end{aligned}
\end{equation}
Here, the label `$d$' is used to emphasize the character of the angular momentum $\ell =2$ (called $d$-boson).
The $d$-bosons satisfy standard commutation relations and are regarded as a rank-2 spherical tensor:
\begin{equation}
\begin{aligned}
    [L_z^{\rm ph},d_M^\dg] &= M d_M^\dg,
    \\
    [L_\pm^{\rm ph},d_M^\dg] &= \sqrt{6-M(M\pm1)}\ d_{M\pm1}^\dg,
\end{aligned}
\end{equation}
where the angular momentum operator for $d$-bosons is given by~\cite{Iachello_book}
\begin{equation}
\begin{aligned}
    L^{\rm ph}_z &= \sum_M M d_M^\dg d_M,
    \\
    L^{\rm ph}_\pm &= \sum_M \sqrt{6 - M(M\pm1)} \  d_{M\pm1}^\dg d_M.
\end{aligned}
\end{equation}
The boson annihilation (creation) operator $d_M$ ($d_M^\dg$) is interpreted as removing (adding) 
the magnetic quantum number $M$.
The $x$ and $y$ components are then also introduced 
as $L^{\rm ph}_x = (L^{\rm ph}_++L^{\rm ph}_-)/2$ and $L^{\rm ph}_y = (L^{\rm ph}_+-L^{\rm ph}_-)/2\imu$,
with which one can confirm that $\bm L^{\rm tot} = \bm L+\bm L^{\rm ph}$ and $(\bm L^{\rm tot})^2$ commute with the total Hamiltonian.

The electron--phonon coupling and the bare phonon Hamiltonian is then rewritten as
\begin{align}
    \mathscr H_{\rm ep}
    &= 
    g\sum_M \qty( \mathscr Q_M^\dg d_M^\dg + \mathscr Q_M d_M ),
    \label{eq:Hep_ang}
    \\
    \mathscr H_{\rm p}
    &=   \omega_0 \sum_{M} d_M^\dg d_{M},
    \label{eq:Hp_ang}
\end{align}
where the constant term is neglected.
Since $d_M$ annihilates a phonon carrying angular momentum projection $M$, 
the tensor $\mathscr Q_M$ may be viewed as an operator that carries (and thus adds) the corresponding angular momentum.
The total magnetic quantum number is thus conserved as a manifestation of the rotational symmetry, which imposes a constraint on the 
Hilbert space relevant to electron--phonon coupling.

The annihilation-operator tensor is also defined as
\begin{align}
    \tilde d_M = (-1)^{M} d_{-M}.
\end{align}
Using this, we further introduce
\begin{align}
    x_M = d_M^\dagger + \tilde d_M,
    \label{eq:def_of_x_op}
\end{align}
which corresponds to the spherical-tensor representation of the displacement operator $\phi_\eta$.
The electron--phonon coupling Hamiltonian now becomes
\begin{align}
    \mathscr H_{\rm ep} &= g\sum_M \mathscr Q_M^\dg x_M = g\sum_M \mathscr Q_M x_M^\dg.
\end{align}
It follows from this expression that $x_M$ acts as an operator that raises the phonon angular momentum by $M$, whereas $x_M^\dagger$ lowers it by the same amount.

\subsection{Composite quadrupoles with spherical tensor}
\label{sec:am_vs_cart_basis_E}

Whereas a careful definition of the composite quadrupole is required in Cartesian representation, as discussed in Sec.~\ref{sec:Cart_Comp_quadru}, it can be introduced in a straightforward manner using spherical tensors.
We consider both the electron and phonon degrees of freedom and define
the following composite quadrupoles:
\begin{align}
    [\mathscr Q\otimes \mathscr Q]_M^{(2)} &= \sum_{M_1M_2} \la 2M_12M_2 | 2M\ra
    \mathscr Q_{M_1} \mathscr Q_{M_2},
    \label{eq:QQdef}
    \\
    [x\otimes x]_M^{(2)} &= \sum_{M_1M_2} \la 2M_12M_2 | 2M\ra,
    x_{M_1} x_{M_2},
    \label{eq:composite_xx_def}
    \\
    [\mathscr Q\otimes x]_M^{(2)} &= \frac 1 2\sum_{M_1M_2} \la 2M_12M_2 | 2M\ra
    (\mathscr Q_{M_1} x_{M_2}+ x_{M_1} \mathscr Q_{M_2}).
\end{align}
Equation~\eqref{eq:QQdef} is a spherical tensor version of the previously defined composite quadrupole in Eq.~\eqref{eq:QQdef_Cartesian}.
The matrix element of these operators are determined by the Clebsch-Gordan coefficients by the Wigner-Eckart theorem in Eq.~\eqref{eq:Wigner_Eckart}: 
\begin{equation}
\begin{aligned}
    \lla 1 m'|\  [\mathscr Q \otimes \mathscr Q]^{(2)}_{M}  \ | 1 m \rra
    &=  A_{QQ} \  \la 1m2M | 1m' \ra,
    \\
    \lla 1 m'|\  [x \otimes x]^{(2)}_{M}\  | 1 m\rra
    &=   A_{xx} \  \la 1m2M | 1m' \ra,
    \\
    \lla 1 m'|\  [\mathscr Q \otimes x]^{(2)}_{M}\  | 1 m\rra
    &=  A_{Qx} \  \la 1m2M | 1m' \ra,
\end{aligned}
\label{eq:reduced_comp_Q}
\end{equation}
where we have introduced the electron-phonon-coupled state $|L_{\rm tot},M\rra$ with a total angular momentum for a general multiphonon configuration.
The coefficients $A$ are the reduced matrix elements of the composite quadrupoles. 
Note that the factor of $1/\sqrt{2\cdot 1 + 1} = 1/\sqrt{3}$ is absorbed into the definition of $A$.
We show the numerical results in Sec.~\ref{sec:numer_composite_quadrupole}.

\subsection{Analytic solution for the restricted Hilbert space}
\label{sec:am_vs_cart_basis_F}

We consider the restricted Hilbert space with boson number $n_d \leq 1$ for $L_{\rm tot}=1$.
We have three categories of the states:
\begin{align}
    |m\rra_1 &= |1m\ra|0\ra_{\rm ph},
    \\
    |m\rra_2 &= \sum_{m_1m_2} \la 2 m_1 1 m_2 | 1 m\ra \  d_{m_1}^\dg | 1 m_2\ra|0\ra_{\rm ph},
    \\
    |m\rra_3 &= \sum_{m_1m_2} \la 2 m_1 2 m_2 | 1 m\ra \  d_{m_1}^\dg | 2 m_2\ra|0\ra_{\rm ph},
\end{align}
for $m=0,\pm 1$, where the vacuum of phonons is defined by $d_M |0\ra_{\rm ph} = 0$. The spin is omitted for simplicity.
The state $|m\rra_2$ does not hybridize with the other states through $\mathscr H_{\rm ep}$, and we therefore focus on $|m\rra_1$ and $|m\rra_3$.
The origin of this selection rule is discussed in Sec.~\ref{sec:quasispin}.
The matrix element of electron--phonon interaction Hamiltonian is given by
\begin{align}
_a\lla m| \mathscr H_{\mathrm{ep}} |m \rra_b &= \begin{pmatrix}
        2J & \sqrt{10} g \\
        \sqrt{10} g & \omega_0
    \end{pmatrix},
    \label{eq:2by2mat}
\end{align}
where $a,b = 1,3$.
The ground state energy at large $g$ is then calculated as
\begin{align}
    E_{\rm gs} &= \frac{2J+ \omega_0}{2}
    - \frac 1 2 \sqrt{( \omega_0-2J)^2 + 40g^2}.
\end{align}
The corresponding wave function is
\begin{align}
    |m\rra_{\rm gs}
    &= C_1 |m\rra_1 + C_3 |m\rra_3,
\end{align}
where
\begin{align}
C_1 &= \frac{\sqrt{10} \, g}{\sqrt{(E_{\rm gs}-2J)^2+10g^2}},\\
C_3 &= \frac{E_{\rm gs}-2J}{\sqrt{(E_{\rm gs}-2J)^2+10g^2}},
\end{align}
which satisfies the normalization $C_1^2+C_3^2 = 1$.

The straightforward calculation gives the reduced matrix elements for composite quadrupoles defined in Eq.~\eqref{eq:reduced_comp_Q} as
\begin{equation}
\begin{aligned}
A_{QQ}
    &=  - \sqrt{35} \qty( 1 - \frac{7}{10} C_3^2 ),
    \\
A_{xx}    &=  - \frac{\sqrt{35}}{10} C_3^2,
    \\
A_{Qx}
    &= - \sqrt{14} \, C_1C_3.
\end{aligned}
\label{eq:nc=1_exact_Acoef}
\end{equation}
The ability to explicitly evaluate these matrix elements is one of the advantages of the spherical-tensor formalism.

\section{Numerical results}
\label{sec:numerical_results}

\subsection{Calculation method}
We perform numerical diagonalization of the total Hamiltonian $\mathscr{H}$ with a cutoff in the total phonon number $n_{\mathrm{c}}$. 
We choose the basis for the three-electron states as the eigenstates of $\mathscr{H}_{\mathrm{ee}}$, listed in Eq.~\eqref{eq:shorthand_notation}.
Below, we focus on the formulation with a Cartesian basis, but it can also be formulated 
with spherical-tensor basis in a similar manner.
We choose the basis for the Jahn-Teller phonons as the eigenstates of $\mathscr{H}_{\mathrm{p}}$. With these choices, we denote the electron-phonon basis as
\begin{align}
|\alpha,n\rra = |\alpha\rangle \big|\{n\}\big\ra_{\rm ph},
\end{align}
where $\alpha = 1,2,\dots,8$ labels the eigenstates of $\mathscr H_{\mathrm{ee}}$ in Eq.~\eqref{eq:shorthand_notation} and $\{n\} = (n_{1},n_{3},n_{4},n_{6},n_{8})$ labels the eigenstates of $\mathscr H_{\mathrm{p}}$. We truncate the Hilbert space by imposing a total-phonon-number cutoff,
\begin{align}
n_{d}\equiv \sum_{\eta} n_\eta \le n_{\rm c}.
\end{align}
The matrix element of $\mathscr{H}$
is expressed as
\begin{align}
\lla \alpha,n| \mathscr{H}|\beta, n'\rra
&= \lla \alpha,n| \left(\mathscr{H}_{\mathrm{ee}} + \mathscr{H}_{\mathrm{p}} + \mathscr{H}_{\mathrm{ep}} \right)|\beta, n'\rra . 
\label{eq:Matrix_element}
\end{align}
The first and second terms in Eq.~\eqref{eq:Matrix_element} are already diagonal in this basis. The matrix element of the first term is
\begin{align}
\lla \alpha,n| \mathscr{H}_{\mathrm{ee}} |\beta, n'\rra
&= E^{\mathrm{ee}}_{\alpha} \, \delta_{\alpha, \beta} \prod_{\eta} \delta_{n_{\eta}, n'_{\eta}},
\end{align}
where $E^{\mathrm{ee}}_{\alpha}$ are the eigenvalues of $\mathscr{H}_{\mathrm{ee}}$,
\begin{align}
E^{\mathrm{ee}}_{\alpha} &= E_F(L) = 
\begin{cases}
2J & (\alpha = 1,2,3; \,L = 1),\\
0 & (\alpha = 4,5,6,7,8; \, L = 2).
\end{cases}
\end{align}
The matrix element of the second term is
\begin{align}
\lla \alpha,n| \mathscr{H}_{\mathrm{p}} |\beta, n'\rra
&= \omega_{0} \, n_d \, \delta_{\alpha, \beta} \prod_{\eta} 
\delta_{n_{\eta}, n'_{\eta}} .
\end{align}
Finally, we find that the matrix element of the third term, 
due to the electron-phonon coupling $\mathscr{H}_{\mathrm{ep}}$, can be written as
\begin{align}
&\lla \alpha,n| \mathscr{H}_{\mathrm{ep}} |\beta, n'\rra \nonumber\\
&= g \, \sum_{\eta} Q^{\eta}_{\alpha \beta}
\left(\sqrt{n_{\eta} + 1} \, \delta_{n_{\eta} + 1, n'_{\eta}}
      + \sqrt{n'_{\eta} + 1} \, \delta_{n_{\eta}, n'_{\eta} + 1}\right),
      \label{eq:Hep_mat_elem}
\end{align}
where the $\eta$-dependent electronic quadrupole matrix element is defined by
$Q^{\eta}_{\alpha \beta} = \langle \alpha|Q_{\eta}|\beta\rangle $.
The effective Hamiltonian in Eq.~\eqref{eq:Hep_mat_elem} can be interpreted, when viewed as a lattice problem, as a five-dimensional semi-infinite tight-binding model describing a particle with ten internal degrees of freedom. This perspective is useful for recognizing the structure of the Hamiltonian, particularly for readers familiar with multicomponent tight-binding models commonly encountered in condensed matter theory.

\subsection{Low-energy states and energetics}

We choose the parameters in the Hamiltonian to the relevant values in fullerides~\cite{Nomura12, Nomura15_2}. We use Hund's coupling constant $J\simeq 0.03$ eV and the Jahn--Teller phonon frequency $  \omega_0 \simeq 0.2$ eV. Here, the effective electron--electron interaction mediated by phonons is estimated as $\lambda = - 2 g^{2}/  \omega_{0} = -0.05$ eV, and hence the electron--phonon coupling constant is estimated to be $g\simeq 0.07$ eV. In order to clarify the effect of electron--phonon coupling in our model, we systematically change the coupling constant $g$ in the numerical calculations.

\begin{figure}[tb]
\centering
\includegraphics[width=0.4\textwidth]{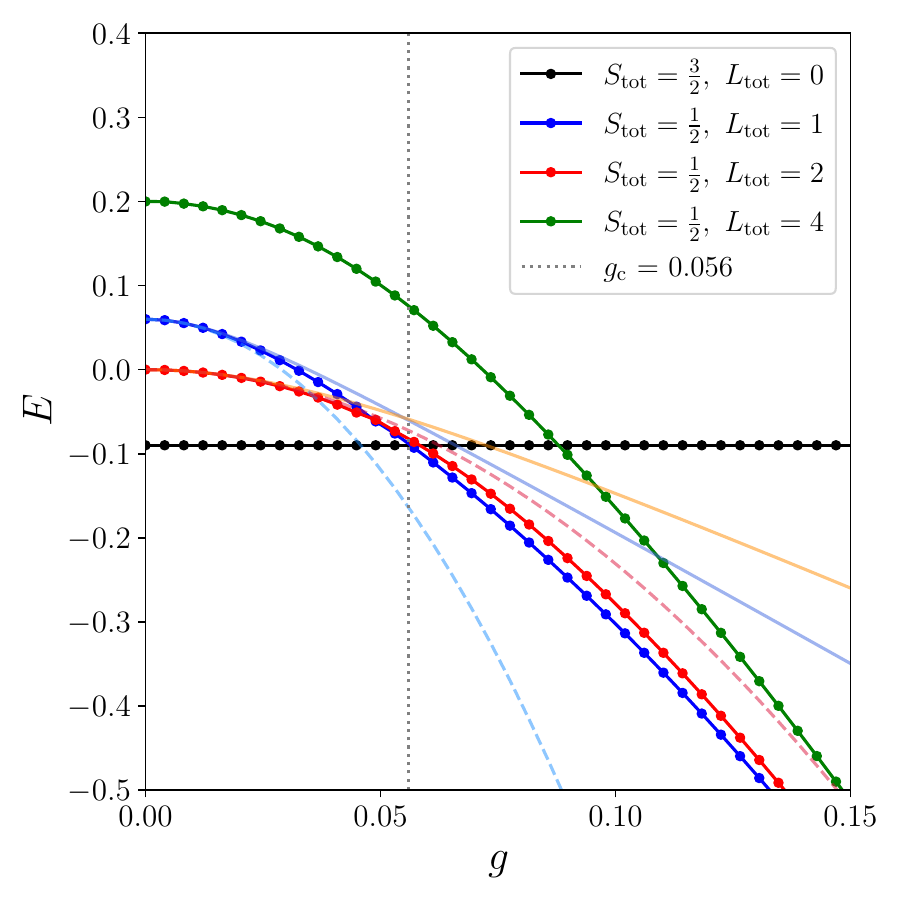}
\caption{Lowest-energy levels $E$ as a function of the electron--phonon coupling $g$ for the four lowest multiplets. The circles connected by the thick solid lines show the exact-diagonalization results of Eq.~\eqref{eq:Matrix_element} with phonon-number cutoff $n_\mathrm{c}=7$. The thin solid lines show the analytic results obtained in the truncated subspace $n_d \le n_\mathrm{c} = 1$, and the dotted lines show the lowest-order perturbation theory; both agree with exact diagonalization in the weak-coupling limit. The vertical dotted line marks the critical coupling $g_{\rm c}=0.056$.
Note that the $L=2$ state at $g=0$ is chosen as the zero-energy reference [see Eq.~\eqref{eq:fermi_system_energy}].
}
    \label{fig:ground_state_energy}
\end{figure}

The energy values of the low-energy states as functions of the electron-phonon coupling constant $g$ are shown in Fig.~\ref{fig:ground_state_energy}. 
Upon coupling the localized electrons to Jahn–Teller phonons, the energies of the $L_{\rm tot} = 1$ and $L_{\rm tot} = 2$ multiplets decrease with increasing electron–phonon coupling $g$, whereas the 
$L_{\rm tot} = 0$ multiplet remains unchanged because it is decoupled from the Jahn–Teller modes. 
The higher-energy states are well separated from these three spherical-tensor multiplets. Only the next higher-energy states $L_{\rm tot} = 4$ are shown as a solid green line.
Although the coupling $g$ preserves the multiplet degeneracies, it lowers the energies of the $L_{\rm tot}=1$ and $L_{\rm tot}=2$ multiplets at different rates. Consequently, the $L_{\rm tot}=1$ multiplet, which lies higher in energy at $g=0$, is driven downward more rapidly, leading to a level crossing and an inversion of the ground-state multiplet at a critical coupling $g_{\rm c} \approx 0.056$. 
At this point, the phonon-induced effective (negative) Hund's exchange overcomes the bare positive $J$.

\begin{figure}[tb]
    \centering
    \includegraphics[width=0.4\textwidth]{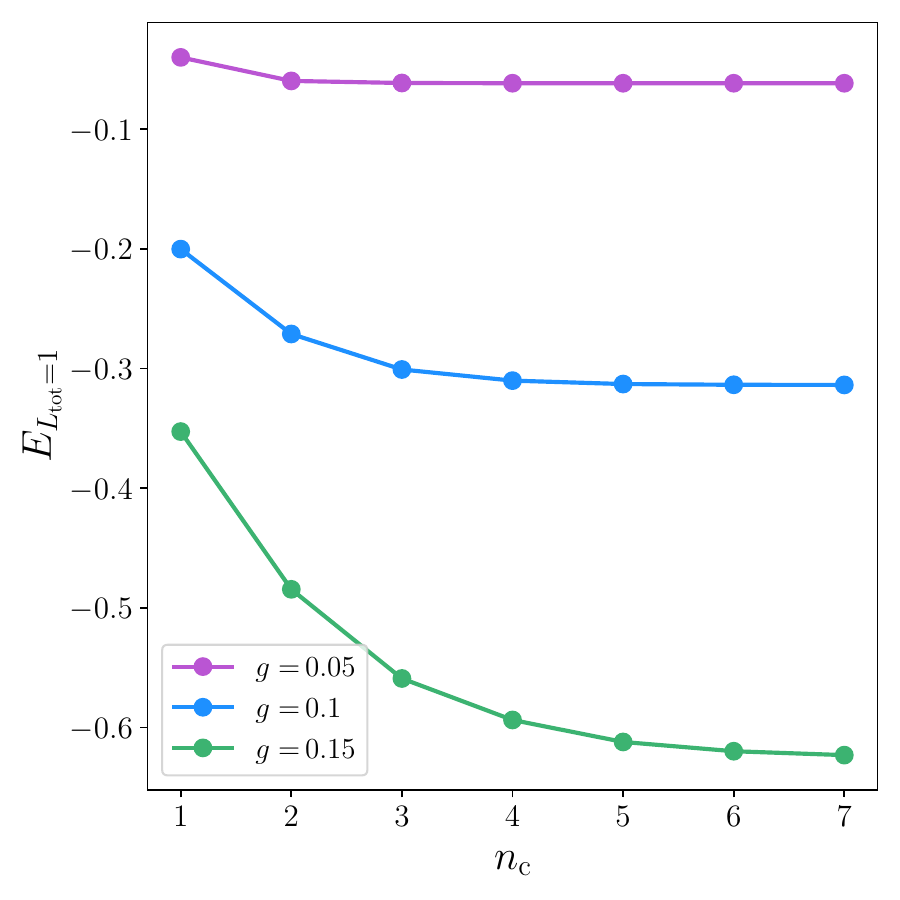}
    \caption{Convergence of the eigenenergies corresponding to $L_{\mathrm{tot}}=1$ with respect to the total-phonon-number cutoff $n_{\mathrm{c}}$ for a representative electron-phonon coupling $g$. The saturation of the eigenenergies for sufficiently large $n_{\mathrm{c}}$ demonstrates that the exact-diagonalization results in Fig.~\ref{fig:ground_state_energy} are numerically converged with respect to the the total phonon-number cutoff at $n_{\mathrm{c}} = 7$.
    }
    \label{fig:convergence}
\end{figure}

The convergence of these eigenenergies with respect to the total-phonon-number cutoff $n_{\mathrm{c}}$ is shown in Fig.~\ref{fig:convergence}. It demonstrates that once $n_{\mathrm{c}}$ is sufficiently large, the energy levels are almost unchanged,
thereby validating our numerical diagonalization treatment.

\subsection{Matrix elements for composite quadrupoles}
\label{sec:numer_composite_quadrupole}

\begin{figure}[tb]
    \centering
    \includegraphics[width=0.42\textwidth]{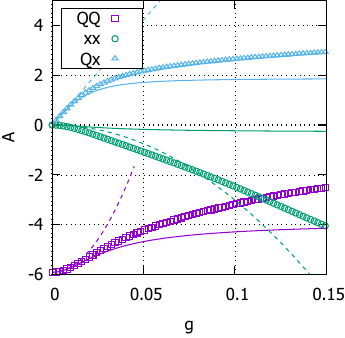}
    \caption{
    The reduced matrix elements $A_{QQ}$, $A_{Qx}$ and $A_{xx}$ of the composite quadrupoles for the lowest $L_{\rm tot}=1$ multiplet defined in Eq.~\eqref{eq:reduced_comp_Q}.
    The solid lines show the analytic results for $n_{\rm c}=1$, and the dotted ones show the lowest-order perturbation theory, both of which coincide with numerical results at small $g$.
    We take the same parameters as those in Fig.~\ref{fig:ground_state_energy}, and the cutoff number $n_{\rm c}$ dependence is shown in Appendix \ref{sec:nc_depend}.
    }
    \label{fig:composite_quadrupole}
\end{figure}

Figure~\ref{fig:composite_quadrupole} shows the reduced matrix elements defined in Eq.~\eqref{eq:reduced_comp_Q} for the lowest-energy $L_{\rm tot}=1$ multiplet. The $[\mathscr Q \otimes \mathscr Q]$-type contribution is nonzero even in the absence of electron--phonon coupling, whereas those involving phonon degrees of freedom, such as $[x \otimes x]$ and $[\mathscr Q \otimes x]$, become finite with increasing coupling $g$. In contrast, non-composite quadrupoles (not shown in Fig.~\ref{fig:composite_quadrupole}) remain zero within the $L_{\rm tot}=1$ subspace even for $g>0$. These results indicate that the composite nature is essential for realizing quadrupolar degrees of freedom in the electron--phonon coupled case.

Figure~\ref{fig:composite_quadrupole} also shows approximate results obtained with $n_{\rm c}=1$ (solid lines) and those from lowest-order perturbation theory in $g$ (dashed lines). At large $g$, the $n_{\rm c}=1$ results exhibit qualitatively better agreement with the numerical data than the lowest-order perturbative results, except for the phonon composite quadrupole $[x \otimes x]$. The large deviation observed for $[x \otimes x]$ reflects the strong sensitivity of phonon observables to the boson number, indicating that the single-phonon truncation is insufficient both quantitatively and qualitatively.

We now discuss the possible relevance of the above results in the context of alkali-doped fullerides.
As shown in Refs.~\cite{Hoshino17,Iwazaki21}, the composite quadrupole order parameter $\langle Q^{\rm C}_\eta \rangle$ emerges spontaneously at low temperatures in the strongly correlated regime, while the conventional quadrupolar order parameter $\langle Q_\eta \rangle$ remains zero.
It creates an effective internal mean-field conjugate to $\langle Q^{\rm C}_\eta \rangle$ as 
\begin{align}
    \mathscr H_{\rm MF} &= - h Q_\eta^{\rm C}
    ,
\end{align}
which is expected to affect the other physical quantities.
In the linear response theory, the physical quantity $\mathscr O$, whose expectation value without $h$ vanishes, is given by
\begin{align}
    \la \mathscr O \ra_h
    &= 
    \chi_{\mathscr O} h + O(h^2),
    \\
    \chi_{\mathscr O} &= \int_0^\beta \diff \tau \la Q_\eta^{\rm C}(\tau)  \mathscr O \ra_0,
\end{align}
where $\beta$ is the inverse temperature, $A(\tau) = \epn^{\tau \mathscr H}A\epn^{-\tau \mathscr H}$ is the Heisenberg representation with imaginary time, and 
\begin{align}
    \la \cdots \ra_h &= \frac{{\rm Tr\,} (\cdots) \epn^{-\beta (\mathscr H + \mathscr H_{\rm MF})}}{{\rm Tr\,} \epn^{-\beta (\mathscr H + \mathscr H_{\rm MF})}}
\end{align}
indicates the quantum-statistical average in the presence of the mean-field $h$.
Using the matrix representation, one obtains
\begin{align}
\chi_{\mathscr O} &=\sum_{nm} 
\frac{\epn^{-\beta E_n}}{Z}
\la n|Q_\eta^{\rm C}|m\ra
\la m| \mathscr O |n \ra
\int_0^\beta \epn^{\tau(E_n - E_m)} \diff \tau,
\label{eq:suscep_def}
\end{align}
where $Z=\sum_n \epn^{-\beta E_n}$ with $\mathscr H| n\ra = E_n |n\ra$.
From this expression, we find that the physical quantity $\langle \mathscr O \rangle$ is induced by the composite quadrupole order $Q^{\rm C}_\eta$ only when the corresponding $m$-$n$ matrix elements are simultaneously nonzero.
However, for the simple quadrupoles $Q_\eta$ and $\phi_\eta$ (or $\mathscr Q_M$ and $x_M$ 
in the spherical-tensor representation), no such matrix elements exist.
We therefore conclude that neither static electronic charge distortion nor lattice distortion is induced by the composite quadrupole order, even though they share the same spatial rotational symmetry.
This result follows from the selection rules inherent to the present electron--phonon coupled system, which are discussed in detail in Sec.~\ref{sec:quasispin}.

\subsection{Entanglement properties of the $L_{\rm tot}=1$ state}
\label{sec:numer_entangle}

\begin{figure*}[t]
    \centering
    \includegraphics[width=1.0\textwidth]{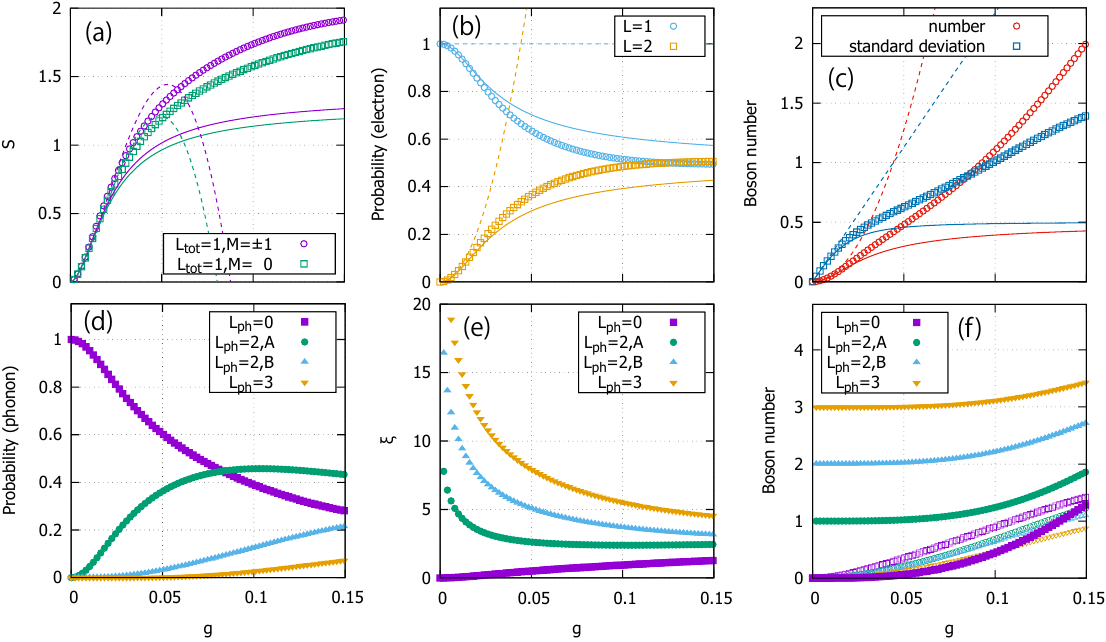}
    \caption{
    Coupling constant ($g$) dependence of 
    (a) the entanglement entropy $S$ [Eq.~\eqref{eq:def_ee}], 
    (b) the probability $(2L+1)\lambda_{\rm el}^{(L)}$ for electrons  [Eq.~\eqref{eq:elec_prob}], 
    (c) the expectation value of the phonon number $\la  n_d  \ra$ and its standard deviation $\varDelta n_d$, 
    (d) the probability for phonons [Eq.~\eqref{eq:ph_prob}], 
    (e) the entanglement spectrum for phonons [Eq.~\eqref{eq:ph_entan}], and 
    (f) the angular momentum decomposition of the phonon numbers [Eq.~\eqref{eq:ph_num_decomp}].
    In panel (f), the standard deviation of the number is plotted as open symbols.
    These results are obtained for the $L_{\rm tot} = 1$ states which become the ground state for $g\gtrsim 0.056$.
    The solid lines show the approximate results obtained 
    for the phonon-number cutoff  $n_{\rm c}=1$, and dotted lines for the lowest-order perturbation theory with respect to $g$.
    }
    \label{fig:EE}
\end{figure*}

Given the eigenenergies and corresponding wave functions, we analyze the entanglement properties that characterize Jahn--Teller phonon-coupled localized electrons. In our model, the Hilbert space factorizes into electronic and phononic sectors, as prescribed by the tensor-product structure of composite quantum systems, which naturally enables a characterization of their quantum correlations. Within the low-energy manifold, we therefore compute the reduced density matrix and the entanglement spectrum between the electronic and phononic degrees of freedom as a function of the coupling constant $g$. While entanglement properties in electron--phonon coupled systems have previously been investigated in one-dimensional Holstein models~\cite{Zhao04,Stojanovic08,Zhang09}, the present study instead focuses on a local multiorbital problem, with particular emphasis on the role of angular momentum.

The eigenstate of the electron-phonon-coupled system is labeled by the total angular momentum as
\begin{align}
    | L_{\rm tot}, M \rra
    &= \sum_{\ell=1}^2 \sum_{m=-\ell}^\ell \sum_{\{n\}} C_{\ell m}^{L_{\rm tot}M}\big( \{n\}\big) \ \big|\ell m\big\ra
    \ \big|\{n\}\big\ra_{\rm ph},
    \label{eq:def_of_ph_state}
\end{align}
where $\{n\} = (n_{-2},n_{-1},n_{0},n_{1},n_{2})$.
Conservation of angular momentum conservation imposes the constraint $M=m + \sum_{m'=-2}^2 m' n_{m'}$.
In this subsection, we focus on the lowest $L_{\rm tot}=1$ multiplet, which becomes the ground state at sufficiently large electron--phonon coupling, and omit the label $L_{\rm tot}$ when no confusion arises. 
We also restrict our attention to the spin sector $\sg=\ua$ and suppress this label in the following.

\subsubsection{Reduced density matrix for electrons}

When the states are classified solely by the total angular momentum, the electronic and phononic components are intrinsically entangled. 
The use of reduced matrix elements, however, allows one to factor out the angular-momentum coupling and thereby access quantities specific to the electron or phonon sectors.

The reduced density matrix for the electronic degrees of freedom is defined by taking the partial trace over the phonon Hilbert space as
\begin{align}
\rho_{{\rm el},M} &=
    {\rm Tr}_{\rm ph}\, |{1M}\rra \lla{1M}|,
\end{align}
where $|L_{\rm tot} M\rra$ is defined in Eq.~\eqref{eq:def_of_ph_state}.
Using this reduced density matrix, we introduce the entanglement entropy for the pure state as
\begin{align}
    S_{M} &= - {\rm Tr}_{\rm el}\, \rho_{{\rm el}, M} \ln \rho_{{\rm el}, M},
    \label{eq:def_ee}
\end{align}
with $M=-1,0,+1$. 
In the quantum Rabi models (see Table~\ref{tab:my_label}, the entanglement entropy has been employed to quantify the correlations between the two-level system and the bosonic mode without internal angular momentum~\cite{Rossatto17,Liu17,Shi22}.

Figure~\ref{fig:EE}(a) shows the $g$ dependence of the entanglement entropy, calculated within a truncated Hilbert space with $n_d \leq n_{\rm{c}} = 7$.
The entanglement entropy depends on the magnetic quantum number $|M|=0,1$ and increases monotonically with increasing electron--phonon coupling.
We also evaluate the entanglement entropy for the $L_{\rm tot}=1$ sector within the minimal truncation $n_d \leq n_{\rm{c}} = 1$, yielding
\begin{align}
    S_M &\simeq 
    - C_1^2\ln C_1^2 - C_3^2\ln C_3^2
    + C_3^2 \tilde S_M.
\end{align}
The deviation among different values of $M$ originates from the entropy $\tilde S_M$, which arises solely from the CG coefficients and is independent of any system parameters such as $\omega_0$ and $g$. It is defined as
\begin{align}
    \tilde S_M &= - \sum_{m_1} A_{m_1}^M \ln A_{m_1}^M,
    \\
    A_{m_1}^M &= \la 2, m_1, 2, M-m_1 | 1M\ra^2.
\end{align}
The solid lines in Fig.~\ref{fig:EE}(a) represent these analytical results, which capture the qualitative behavior of the full numerical calculations.
For comparison, the dotted lines show the lowest-order perturbation-theory results, which deviate significantly for $g \gtrsim 0.05$.
This indicates that the $n_{\rm c}=1$ truncation provides a more reliable approximation in this coupling regime.

It is also of interest to examine the mixed state within the $L_{\rm tot}=1$ subspace, which is relevant to effective quantum statistical properties at low temperatures.
We define the reduced density matrix for the electronic degrees of freedom as
\begin{align}
    \rho_{\rm el} &= \frac{1}{3} \sum_{M=-1}^{1} \rho_{{\rm el},M}
    = \sum_{L=1}^{2} \lambda_{\rm el}^{(L)} \sum_{M=-L}^{L} \ket{LM}\bra{LM}.
    \label{eq:elec_prob}
\end{align}
The probabilities $P_{1}=3\lambda_{\rm el}^{(1)}$ for $L=1$ and $P_{2}=5\lambda_{\rm el}^{(2)}$ for $L=2$, which satisfy $P_{1}+P_{2}=1$, are shown in Fig.~\ref{fig:EE}(b).
At small $g$, the $L_{\rm tot}=1$ state is dominated by the purely electronic $L=1$ component.
With increasing electron--phonon coupling, the electronic $L=2$ component gains weight and exceeds the $L=1$ component for $g \gtrsim 0.12$ due to the coupling to Jahn--Teller phonons.

In our model, the electronic density matrix involves only two energy levels, which allows for an interpretation in terms of an effective `thermal' state. 
The corresponding effective temperature can be defined through the Boltzmann weights associated with the unperturbed energies $E_F(L=1,2)$.
Such a thermal-state interpretation, however, is not generally applicable to three or more levels, and the density matrix should be regarded as an effective `nonequilibrium' state, as in the case of phonons discussed below.

\subsubsection{Reduced density matrix for phonons}

Electron--phonon entanglement induces phonon excitations in the $L_{\rm tot}=1$ ground-state multiplet, leading to an increasing weight of states with larger phonon number. Figure~\ref{fig:EE}(c) shows the $g$ dependence of the expectation value $\langle n_d \rangle$ and its standard deviation $\Delta n_d = \sqrt{\langle n_d^2 \rangle - \langle n_d \rangle^2}$. For the cutoff $n_{\rm c}=1$, these quantities are obtained analytically as
\begin{align}
    \langle n_d \rangle &= C_3^2, \\
    \Delta n_d &= C_1 C_3,
\end{align}
which agree with the numerically exact results only for $g \lesssim 0.02$. 
Both $\langle n_d \rangle$ and $\Delta n_d$ increase approximately linearly with the coupling constant $g$, a behavior not captured by the $n_{\rm c}=1$ approximation.

To gain further insight, we consider the entanglement spectrum for phonons.
While in the context of nuclear physics bosonic degrees of freedom provide an effective description of a deformed shell structure, phonons in condensed matter systems constitute real physical degrees of freedom. We are therefore interested in their intrinsic properties.
We begin with the reduced density matrix for phonons:
\begin{align}
    \rho_{\rm ph} &= \frac{1}{3} {\rm Tr_{el}} \sum_M|1M\rra\lla 1M|
    \\
    &= \sum_{L_{\rm ph}}\sum_{\al} \lambda_{\rm ph}^{(L_{\rm ph},\al)} \sum_{M=-L_{\rm ph}}^{L_{\rm ph}} | L_{\rm ph}\al M \ra_{\rm ph}\,_{\rm ph}\la L_{\rm ph}\al M |.
\end{align}
Note that $|L_{\rm ph} \al M\ra_{\rm ph}$ is introduced here 
and is obtained by diagonalizing the reduced density matrix for phonons.
The index $\al$ is implicitly dependent on $L_{\rm ph}$ and distinguishes the different multiplets with the angular momentum $L_{\rm ph}$.
For example, if there are two $L_{\rm ph}=2$ multiplets, we distinguish them by the label $\al={\rm A}$ and $\al = {\rm B}$.
We suppress the index $\al$ if there is only one $L_{\rm ph}$ multiplet.
The probability for each state labeled by $(L_{\rm ph}, \al)$ is given by 
\begin{align}
    P_{L_{\rm ph},\al} &= (2L_{\rm ph}+1)\  \lambda_{\rm ph}^{(L_{\rm ph},\al)}.
    \label{eq:ph_prob}
\end{align}
We also define the entanglement spectrum by~\cite{Li08}
\begin{align}
    \xi_{L_{\rm ph},\al} = - \ln \lambda_{\rm ph}^{(L_{\rm ph},\al)},
    \label{eq:ph_entan}
\end{align}
which effectively visualizes the density matrix, especially when the probability is small.

Figures.~\ref{fig:EE}(d) and (e) show the probabilities and the entanglement spectrum defined in Eqs.~\eqref{eq:ph_prob} and \eqref{eq:ph_entan}, respectively.
The upper cutoff in Fig.~\ref{fig:EE}(e) corresponds to the probability $\epn^{-20} \sim 10^{-9}$ in Fig.~\ref{fig:EE}(d), which is negligibly small; the relevant contributions are therefore captured by the four spectra shown in Fig.~\ref{fig:EE}(e).
At small $g$, the dominant weight corresponds to $L_{\rm ph}=0$, which is a state close to the phonon vacuum, and hence $L_{\rm tot}=1$ is mainly composed of the electronic wave function with $L=1$.
With increasing electron-phonon coupling $g$, on the other hand, the $(L_{\rm ph},\alpha)=(2,{\rm A})$ sector becomes dominant for $g \gtrsim 0.07$, while subdominant contributions arise from $L_{\rm ph}=0$, $(L_{\rm ph},\alpha)=(2,{\rm B})$, and $L_{\rm ph}=3$, in descending order of weight.

The expectation value of the phonon number can also be decomposed in a manner analogous to the reduced density matrix.
Specifically, we obtain the following expression:
\begin{align}
    \langle n_d \rangle &=
    \sum_{L_{\rm ph},\alpha}
    P_{L_{\rm ph},\alpha}\,
    \langle n_d \rangle_{L_{\rm ph},\alpha}, \\
    \langle \cdots \rangle_{L_{\rm ph},\alpha}
    &= \frac{1}{2L_{\rm ph}\!+\!1}
    \sum_{M}
    {}_{\rm ph}\!\langle L_{\rm ph}\alpha M | (\cdots)
    | L_{\rm ph}\alpha M \rangle_{\rm ph}.
    \label{eq:ph_num_decomp}
\end{align}
Figure~\ref{fig:EE}(f) shows $\langle n_d \rangle_{L_{\rm ph},\alpha}$ together with its standard deviation.
In the weak-coupling limit $g\to 0$, the $L_{\rm ph}=0$, $(L_{\rm ph},\alpha)=(2,{\rm A})$, $(L_{\rm ph},\alpha)=(2,{\rm B})$, and $L_{\rm ph}=3$ multiplets are continuously connected to boson numbers $0$, $1$, $2$, and $3$, respectively.
As the electron-phonon coupling increases, higher phonon-number components are progressively admixed within each multiplet.
In particular, multiplets that are dominated by smaller phonon numbers in the weak-coupling limit exhibit a more pronounced increase in $\langle n_d \rangle$ with increasing $g$.

We now discuss why the phonon angular momentum $L_{\rm ph}=1$ is absent 
for the reduced density matrix of the phonons.
Since the electronic angular momentum takes $L=1$ and $L=2$, the condition for forming a total 
angular momentum $L_{\rm tot}=1$ state requires $L_{\rm ph}\leq 3$.
For bosons, the wave function must satisfy the totally symmetric condition under particle exchange.
As established in Ref.~\cite{Iachello_book}, bosonic states constructed from $d$ bosons are classified according to the quantum numbers of the group chain 
\begin{align}
U(5) \supset SO(5) \supset SO(3) \supset SO(2), 
\end{align}
and the resulting phonon angular momentum never takes the value $L_{\rm ph}=1$.
More explicitly, for the two-boson state with $n_d=2$, the symmetric angular momentum coupling rule yields
\begin{align}
    [2\otimes 2]_{\rm sym} = 4 \oplus 2 \oplus 0 ,
\end{align}
where $[\cdots]_{\rm sym}$ denotes projection onto the totally symmetric subspace appropriate for bosons.
Similarly, for $n_d=3$, the angular momentum coupling gives
\begin{align}
    [2\otimes 2\otimes 2]_{\rm sym} = 6 \oplus 4 \oplus 3 \oplus 2 \oplus 0 ,
\end{align}
none of which contains the $L_{\rm ph}=1$ sector.
The absence of the $L_{\rm ph}=1$ state persists to higher orders, 
reflecting the underlying bosonic angular-momentum coupling constraints.
Although this group-theoretical argument may appear abstract at first sight, it can be understood more explicitly by examining the concrete form of the wave functions, as discussed in detail in Appendix~\ref{sec:Quasispin}.

The above argument is based on purely bosonic systems. Although the $L_{\rm tot}=1$ multiplet in our model consists of electron--phonon entangled states, in the limit $g \to 0$ the expectation value of the boson number approaches $n_d=1$ for the $(L_{\rm ph},\alpha)=(2,{\rm A})$ sector, as shown in Fig.~\ref{fig:EE}(f). Similarly, one finds $n_d=2$ for $(L_{\rm ph},\alpha)=(2,{\rm B})$ and $n_d=3$ for $L_{\rm ph}=3$ as $g \to 0$. These results indicate that each $L_{\rm ph}$ sector continuously connects, in the weak-coupling limit, to a well-defined $n_d$-boson configuration. Therefore, the structure of the $L_{\rm ph}$ states observed in Figs.~\ref{fig:EE}(d)–(f) can be qualitatively understood from the angular-momentum coupling rules of the phonon degrees of freedom in the absence of electron--phonon interaction.

Finally, we discuss the connection to A$_3$C$_{60}$. As discussed above, parameters for fulleride materials correspond to $g \simeq 0.07$.
At this coupling, Figs.~\ref{fig:EE}(d) and \ref{fig:EE}(f) show that the phonon sector is dominated by an approximately equal mixture of the $n_d=0$ ($L_{\rm ph}=0$) and $n_d=1$ ($L_{\rm ph}=2$) states. Subleading contributions arise from the $n_d=2$ ($L_{\rm ph}=2$) component, with a smaller admixture of the $n_d=3$ ($L_{\rm ph}=3$) state. 
It is also notable that, as shown in Fig.~\ref{fig:EE}(f), the expectation value $\langle n_d \rangle_{L_{\rm ph},\alpha}$ at $g \simeq 0.07$ remains nearly unchanged upon inclusion of the electron--phonon coupling, whereas its standard deviation increases more significantly.
Experimentally, if the angular momentum of the phonon states could be identified, for instance, using circularly polarized light or optical vortices, this would provide a direct means to probe the electron--lattice entangled states predicted in this work.

\section{Quasispin selection rules}
\label{sec:quasispin}

\subsection{Quasispin operators for electrons}

The results of Sec.~\ref{sec:numer_composite_quadrupole}, in particular the vanishing of the standard quadrupole in the electron--phonon-coupled $L_{\rm tot}=1$ ground-state multiplet, can be 
understood in terms of the quasispin selection rule~\cite{Judd67,Ceulemans94}.
To elucidate this point, we first define the operators
\begin{align}
    K_+ &\equiv \frac{\sqrt{6}}{2} \sum_{mm'}\sum_{\sg\sg'}
    \la1m1m'|00\ra
    \la\tfrac 1 2 \sg \tfrac 1 2 \sg'|00\ra
    p_{m\sg}^\dg p_{m'\sg'}^\dg .
\end{align}
The prefactor $\sqrt{6}$ originates from the factor $\sqrt{2s+1}\,\sqrt{2\ell+1}$ evaluated at $s=\tfrac{1}{2}$ and $\ell=1$, as adopted in the definition of Ref.~\cite{Ceulemans94}.
We also define
\begin{align}
    K_- &= K_+^\dg = \sum_\gm D_\gm,
 \\
    K_z &= [K_+,K_-]/2  = \frac 1 2 \sum_{m\sg}\qty(p_{m\sg}^\dg p_{m\sg}-\frac 1 2) ,
\end{align}
which satisfies
\begin{align}
    [K_z,K_\pm] &= \pm K_\pm .
\end{align}
We also define $K_{x,y}$ by $K_\pm = K_x \pm \imu K_{y}$.
These operators obey the standard $SU(2)$ algebra.

Using these quasispins, the operators $p^\dagger$ and $\tilde{p}$ are identified as forming a quasispin tensor:
\begin{equation}
\begin{aligned}
    &[K_z, p_{m\sg}^\dg] = \frac 1 2 p_{m\sg}^\dg
    ,\ \ \ 
    [K_z, {\tilde p}_{m\sg}] = - \frac 1 2 \tilde p_{m\sg},
    \\
    &[K_-, p_{m\sg}^\dg] = \tilde p_{m\sg}
    ,\ \ \ 
    [K_+, \tilde p_{m\sg}] =  p_{m\sg}^\dg .
\end{aligned}
\label{eq:p_as_quasispin_tensor}
\end{equation}
Hence the creation and annihilation operators are written in terms of the triple tensor as
\begin{align}
p_{m,\sigma ,{\textstyle{1 \over 2}}}^{\left( {1,{\textstyle{1 \over 2}},{\textstyle{1 \over 2}}} \right)} = p_{m\sigma }^\dag ,~~~p_{m,\sigma , - {\textstyle{1 \over 2}}}^{\left( {1,{\textstyle{1 \over 2}},{\textstyle{1 \over 2}}} \right)} = {\tilde p_{m\sigma }}, 
\end{align}
where the superscripts $(1,\tfrac 1 2, \tfrac 1 2)$ indicate the quantum numbers $(L,S,K)$.
With use of these tensors, one can define another triple tensor operators as
\begin{align}
&T_{M_L,M_S,M_K}^{\left( {L,S,K} \right)} 
\equiv \left[ {{p^{\left( {1,{\textstyle{1 \over 2}},{\textstyle{1 \over 2}}} \right)}} \otimes {p^{\left( {1,{\textstyle{1 \over 2}},{\textstyle{1 \over 2}}} \right)}}} \right]_{M_L,M_S,M_K}^{\left( {L,S,K} \right)}  \nonumber \\
&= \!\! \sum_{mm'\sigma\sigma'\kappa\kappa'} \!\!
\left\langle {1m1m'\left| {LM_L} \right.} 
\right\rangle \left\langle {{\textstyle{1 \over 2}}\sigma {\textstyle{1 \over 2}}\sigma '\left| {S{M_S}} \right.} \right\rangle \left\langle {{\textstyle{1 \over 2}}\kappa {\textstyle{1 \over 2}}\kappa '\left| {K{M_K}} \right.} \right\rangle \nonumber \\
&\hspace{10mm} \times p_{m,\sigma ,\kappa }^{\left( {1,{\textstyle{1 \over 2}},{\textstyle{1 \over 2}}} \right)}p_{m',\sigma ',\kappa '}^{\left( {1,{\textstyle{1 \over 2}},{\textstyle{1 \over 2}}} \right)}.
\end{align}
For example, the angular momentum operator is constructed as
\begin{align}
{\mathscr L_M} 
&=  - 2\sqrt 2 \left[ {{p^{\left( {1,{\textstyle{1 \over 2}},{\textstyle{1 \over 2}}} \right)}} \otimes {p^{\left( {1,{\textstyle{1 \over 2}},{\textstyle{1 \over 2}}} \right)}}} \right]_{M,0,0}^{\left( {1,0,0} \right)}  \nonumber \\
&=  - 2\left[ {{p^\dag } \otimes \tilde p} \right]_{M,0}^{\left( {1,0} \right)},
\end{align}
that has $K=0$ by construction (see Appendix \ref{sec:Quasispin-form} for more details).
Similarly, for the quadrupole operator, one has
\begin{align}
{\mathscr Q_M} 
&=  - \sqrt 2 \left[ {{p^{\left( {1,{\textstyle{1 \over 2}},{\textstyle{1 \over 2}}} \right)}} \otimes {p^{\left( {1,{\textstyle{1 \over 2}},{\textstyle{1 \over 2}}} \right)}}} \right]_{M,0,0}^{\left( {2,0,1} \right)} 
\nonumber \\
&=  - 2\left[ {{p^\dag } \otimes \tilde p} \right]_{M,0}^{\left( {2,0} \right)},
\end{align}
that has $K=1$ by construction.

The quasispin rank $K$ of operators can be more explicitly obtained by considering the eigenvalue of the Casimirian 
\begin{align}
    \mathcal C (\cdots) &= \sum_\mu \big[ K_\mu , [K_\mu, \cdots] \big].
    \label{eq:Casimirian_el}
\end{align}
Here, $\mathcal{C}$ is a superoperator defined through a commutator, analogous in form to the Liouvillian $\mathcal{L}(\cdots) = [\cdots, \mathscr{H}]$.
The eigenvalue of $\mathcal C$ for a tensor operator is given by $K(K+1)$, where $K$ denotes the quasispin rank.
Indeed, for the dipole (angular momentum) and quadrupole operators, there are the relations 
$\mathcal C \mathscr L_M = 0 $ 
and 
$\mathcal C \mathscr Q_M = 2 \mathscr Q_M$ 
each of which corresponds to $K=0$ and $K=1$ tensor operators, respectively.
We can also apply this technique to the composite quadrupole.
By combining the tensor operators, one can construct the composite operator as
\begin{align}
    \mathscr Q_M^{\rm C} &\equiv \qty[ T^{(2,0,1)}\otimes T^{(2,0,1)}]^{(2,0,2)}_{M,0,0},
\end{align}
which has a rank $K=2$ as confirmed by the relation
$\mathcal C \mathscr Q_M^{\rm C} = 6 \mathscr Q_M^{\rm C}$.
The results are summarized in Table~\ref{tab:tensors}(a).

For the purely electronic wave functions, by applying the Casimir operator $\bm K^2$, the quasispin quantum numbers for $|L=1\rangle$ and $|L=2\rangle$ are found to be $(K,M)=(1,0)$ and $(0,0)$, respectively.
According to the Wigner-Eckart theorem, the matrix element $\langle L'|\mathscr Q_M|L\rangle$ is proportional to the CG coefficient 
$\langle K'010|\,K0\rangle$, where $K$ ($K'$) denotes the quasispin of $|L\rangle$ ($|L'\rangle$).
From the properties of the CG coefficients, the condition $K'+1+K=\text{even}$ must be satisfied, implying $K'\neq K$~\cite{Ceulemans94}.
Consequently, the quadrupolar operator $\mathscr Q_M$ has nonzero matrix elements only between states with $L\neq L'$.
The same argument applies to the composite quadrupole $\mathscr Q_M^{\rm C}$, which is our main focus, for which the matrix element is nonzero for $L=L'$ (see Appendix \ref{sec:Quasispin-form} for more discussion).

\begin{table}[t]
    \centering
    \begin{tabular}{c||c|c|c}
    \hline
         (a) Electron\ \  & \ $K_{\rm el}$\  & \ $L_{\rm el}$ \  & \ $S_{\rm el}$ \ \\ \hline
         $p^\dg_{m \sg}$,~$\tilde p_{m \sg}$ &1/2&1&1/2  \\
         $\mathscr L_M$&0&1&0  \\
         $\mathscr Q_M$&1&2&0  \\
         $\mathscr Q^{\rm C}_{M}
         $&2&2&0  \\
         \hline
    \end{tabular}
     \hspace{2mm}
    \begin{tabular}{c||c|c}
    \hline
         (b) Phonon\ \  & \ $K_{\rm ph}$ \ & \ $L_{\rm ph}$\  \\ \hline
         $d_M^\dg$, $\tilde d_M$
         &1/2&2  \\
         $\mathscr L^{\rm ph}_M$&0&1  \\
         $\mathscr Q_{M}^{\rm ph}$
         &1&2  \\ \hline
    \end{tabular}    
    \caption{
    Quantum numbers for the (a) electronic and (b) phononic tensor operators.
    }
    \label{tab:tensors}
\end{table}

\subsection{Electron--phonon coupled case}

We proceed to account for the selection rule in the electron--phonon coupled system.
The above analysis of quasispin operator indicates that the quadrupole operator with $K=1$ mixes the even- and odd-sectors of $L$, which motivates us to define the ``parity'' operator $\Pi_{\rm el}$ for electrons as
\begin{align}
    \Pi_{\rm el} &= \sum_{L=1}^2 \sum_{M=-L}^L (-1)^L |LM\ra \la LM |,
\end{align}
where $|LM\ra$ has been defined in Eqs.~\eqref{eq:L1_def} and \eqref{eq:L2_def}.
Note that this operator is nothing to do with spatial inversion but is just introduced to characterize the three-electron systems.
We also introduce the parity operator $\Pi_{\rm ph}$ for phonons as
\begin{align}
    \Pi_{\rm ph} &= (-1)^{n_d},
\end{align}
where $n_d = \sum_M d_M^\dg d_M$ is the number of bosons.
Then, the total parity $\Pi = \Pi_{\rm el} \Pi_{\rm ph}$ commutes with the Hamiltonian $\mathscr H$ and becomes a conserved quantity.
Hence all the eigenstates of the total Hamiltonian are labeled by $\Pi = \pm$.

While the operators $\mathscr Q_M$ and $x_M$ mix the parity $\Pi$, composite operators such as $[\mathscr Q \otimes \mathscr Q]$ and $[\mathscr Q \otimes x]$ do not. Hence $\mathscr Q$ and $[\mathscr Q \otimes \mathscr Q]$ cannot simultaneously acquire nonzero expectation values. This explains the absence of the susceptibility discussed in Sec.~\ref{sec:numer_composite_quadrupole}. This constraint is not limited to linear-response theory; rather, it is enforced by the parity ($\Pi$) selection rule.

As described in this subsection, the even or odd character of the quasispin quantum number $K$ associated with an operator determines whether its matrix elements connect states of different $\Pi_{\rm el}$ parity sectors. 
An analogous structure arises for phonons, where the integer or half-integer character of $K_{\rm ph}$ (see Sec.~\ref{sec:quasispin_for_phonons}) governs the mixing between $\Pi_{\rm ph}$ sectors.

\subsection{Quasispin for phonons}
\label{sec:quasispin_for_phonons}
Here, we summarize the quasispin properties of tensor operators for phonons.
We begin by defining the following raising operator:
\begin{align}
    K^{\rm ph}_+ &\equiv \frac{\sqrt 5}{2} \sum_{MM'} \la 2M2M'|00\ra d_M^\dg d_{M'}^\dg
    \\
    &= \frac 1 2 \sum_{M}(-1)^M d_M^\dg d_{-M}^\dg.
\end{align}
We then further define the operators
\begin{align}
    K^{\rm ph}_- &= (K^{\rm ph}_+)^\dg,
    \\
    K^{\rm ph}_z &= - [K^{\rm ph}_+,K^{\rm ph}_-]/2 
    = \frac 1 2 \sum_{M} \qty( d_M^\dg d_M + \frac 1 2),
    \label{eq:Qphz_def}
\end{align}
which satisfies
\begin{align}
    [K^{\rm ph}_z,K^{\rm ph}_\pm] &= \pm K^{\rm ph}_\pm.
\end{align}
We also define $K_{x,y}^{\rm ph}$ by $K_\pm^{\rm ph} = K_x^{\rm ph} \pm \imu K^{\rm ph}_{y}$.
The additional minus sign in Eq.~\eqref{eq:Qphz_def} complicates the mathematical structure, 
resulting in an $SU(1,1)$ algebra \cite{Yurke86}.
We have the following relations:
\begin{equation}
\begin{aligned}
    &[K_z^{\rm ph},d_M^\dg] = \frac 1 2 d_M^\dg 
    ,\ \ \ 
    [K_z^{\rm ph},\tilde d_M] = - \frac 1 2 \tilde d_M ,
    \\
    &[K_-^{\rm ph},d_M^\dg] =  \tilde d_M
    ,\ \ \ 
    [K_+^{\rm ph},\tilde d_M] =  - d_M^\dg.
\end{aligned}
\end{equation}
To characterize the quasispin quantum number in analogy with Eq.~\eqref{eq:Casimirian_el}, we define the Casimirian as follows:
\begin{align}
\mathcal C_{\rm ph} (\cdots) = 
    \sum_\mu s_\mu \big[K^{\rm ph}_\mu,[K^{\rm ph}_\mu, \cdots] \big],
\end{align}
where the sign factors are $s_x=s_y=-1$ and $s_z=+1$ originating from $SU(1,1)$ algebra.
Since the relation $\mathcal C_{\rm ph} d_M^\dg = \frac 3 4 d_M^\dg$ holds, the operator $d^\dg_M$ is the $(K_{\rm ph},M) = (1/2,1/2)$ tensor and $\tilde d_M$ is the $(K_{\rm ph},M) = (1/2,-1/2)$ tensor.

We define the composite tensors
\begin{align}
    \mathscr L_M^{\rm ph} &\equiv \sum_{mm'} \la 2m2m'|1M\ra d_m^\dg \tilde d_{m'},
    \\
    \mathscr Q_M^{\rm ph} &\equiv \sum_{mm'} \la 2m2m'|2M\ra d_m^\dg \tilde d_{m'},
\end{align}
which satisfies
\begin{align}
    &\hspace{-3mm}[K^{\rm ph}_z, \mathscr L_M^{\rm ph}] = 0
    ,\ \ 
    \mathcal C_{\rm ph} \mathscr L_M^{\rm ph}
    = 0 ,
    \\
    &\hspace{-3mm}[K^{\rm ph}_z, \mathscr Q_M^{\rm ph}] = 0
    ,\ \ 
    \mathcal C_{\rm ph} \mathscr Q_M^{\rm ph}
    = 2 \mathscr Q_M^{\rm ph}.
\end{align}
Hence $\mathscr L^{\rm ph}_{M}$ is the $(K_{\rm ph},M) = (0,0)$ tensor, and $\mathscr Q^{\rm ph}_{M}$ is the $(K_{\rm ph},M) = (1,0)$ tensor.
The results in this section are summarized in Table~\ref{tab:tensors}(b).
Thus, the even-odd character of the number of creation/annihilation operators, which corresponds to the parity $\Pi_{\rm ph}$, are reflected in the half-integer or integer of the quasispin quantum number $K_{\rm ph}$ (see Appendix~\ref{sec:Quasispin-boson} for more details).

\section{Summary and outlook}
\label{sec:summary}

In this paper, we present a detailed analysis of a single-site model consisting of three electrons confined in a three-orbital system and coupled to Jahn--Teller phonons, which serves as a minimal model for fulleride materials in the Mott-insulating regime. We clarify the relationship between the Cartesian and 
spherical tensor representations, highlighting their complementary advantages: the former provides intuitive physical insight, while the latter enables a systematic algebraic analysis based 
on spherical tensor operators.

In the present system, the ground-state multiplet hosts internal degrees of freedom not captured by conventional quadrupole moments, but rather by composite quadrupoles constructed from products of higher-rank operators. We elucidate their tensorial structure and quantitatively evaluate the associated reduced matrix elements by means of detailed numerical calculations, while the selection rules obeyed by these tensor operators are systematically derived within the quasispin formalism. A detailed analysis of the entanglement induced by the electron--phonon interaction further reveals that, while a single-phonon state carries angular momentum $\ell=2$, the interacting multiphonon sector coupled to the electronic degrees of freedom is dominated by states with $L_{\rm ph}=2$ and $L_{\rm ph}=3$. Thus, this work provides a concrete demonstration of the broad utility of angular-momentum tensor methods in condensed matter theory.

Spherical tensors provide a powerful framework for treating more complex interactions.
In fulleride systems, a small but finite spin--orbit coupling is present~\cite{Tosatti96}, and the resulting fine structure can be described systematically within this formalism.
Moreover, in the context of nuclear physics, more elaborate Hamiltonians beyond simple fermion--boson coupling have been considered.
For example, parity-odd $f$ bosons~\cite{Yoshinaga24} can couple to fermionic degrees of freedom, as well as two‑fermion scattering processes associated with collective $d$‑boson modes~\cite{Teng25}.
Similar extensions are expected to be relevant for fullerides, where parity-odd phonon modes indeed exist as molecular vibrations.
Such generalized electron--phonon couplings in the Hamiltonian can be formulated transparently within the spherical-tensor framework by constructing scalar operators such as $\big[[\mathscr Q \otimes \mathscr Q] \otimes d^\dg\big]^{(0)}$ and $\big[\mathscr Q \otimes [f^\dg  \otimes \tilde f]\big]^{(0)}$, in addition to the $\big[\mathscr Q \otimes d^\dg\big]^{(0)}$ term considered in this work.
Exploring these extensions, including the role of parity-odd phonon modes and higher-rank composite couplings, remains an interesting open problem from both physical and symmetry perspectives.

\section*{Acknowledgment}
We are grateful to Atsushi Umeya, Tatsuya Miki, and Natsuki Okada for fruitful discussions.
This work was supported by KAKENHI Grants No.~24K00578, No.~23K25827, No.~25H01249 and
No.~20K03925.
K.T. was partially supported by a University Block Grant Fellowship from the Department of Physics, University of Illinois Urbana-Champaign.

\appendix

\section{Cartesian and spherical coordinates}
\label{sec:Gell-Mann}

\subsection{Gell-Mann matrices and quadrupoles}
The Gell-Mann matrices are defined by
\begin{equation}
\begin{aligned}
    &\hat \lambda^1 = \begin{pmatrix}
        0 & 1 & 0\\
        1 & 0 & 0\\
        0 & 0 & 0\\
    \end{pmatrix}
    ,\ \ \ 
    \hat \lambda^2 = \begin{pmatrix}
        0 & -\imu & 0\\
        \imu & 0 & 0\\
        0 & 0 & 0\\
    \end{pmatrix}
    ,
    \\
    &\hat \lambda^3 = \begin{pmatrix}
        1 & 0 & 0\\
        0 & -1 & 0\\
        0 & 0 & 0\\
    \end{pmatrix}
    ,\ \ \ 
    \hat \lambda^4 = \begin{pmatrix}
        0 & 0 & 1\\
        0 & 0 & 0\\
        1 & 0 & 0\\
    \end{pmatrix}
    ,
    \\
    &\hat \lambda^5 = \begin{pmatrix}
        0 & 0 & -\imu \\
        0 & 0 & 0\\
        \imu & 0 & 0\\
    \end{pmatrix}
    ,\ \ \ 
    \hat \lambda^6 = \begin{pmatrix}
        0 & 0 & 0\\
        0 & 0 & 1\\
        0 & 1 & 0\\
    \end{pmatrix}
    ,
    \\
    &\hat \lambda^7 = \begin{pmatrix}
        0 & 0 & 0\\
        0 & 0 & -\imu\\
        0 & \imu & 0\\
    \end{pmatrix}
    ,\ \ \ 
    \hat \lambda^8 = \sqrt{\frac{1}{3}} \begin{pmatrix}
        1 & 0 & 0\\
        0 & 1 & 0\\
        0 & 0 & -2\\
    \end{pmatrix}
    ,
\end{aligned}
\end{equation}
where the basis is $\gm=x,y,z$ for electron single-orbitals.
There is a correspondence between the label $\eta = 1,3,4,6,8$ and Cartesian coordinates as
\begin{equation}
\begin{aligned}
    \lambda^1 &\ \ \longleftrightarrow\ \  q_{1} = 2xy,
    \\
    \lambda^3 &\ \ \longleftrightarrow\ \  q_{3} = x^2-y^2,
    \\
    \lambda^4 &\ \ \longleftrightarrow\ \  q_{4} = 2zx,
    \\
    \lambda^6 &\ \ \longleftrightarrow\ \  q_6 = 2yz,
    \\
    \lambda^8 &\ \ \longleftrightarrow\ \  q_8 = (x^2+y^2-2z^2)/\sqrt 3,
\end{aligned}
\end{equation}
which form quadrupoles.
This correspondence table indicates that the Gell-Mann matrices in the left-hand side is transformed as the Cartesian polynomials in the right-hand side under the spatial rotation.

The above polynomials are used for defining the higher-order quadrupoles.
For this purpose, we introduce the scalars
\begin{align}
    r^2 &= x^2+y^2+z^2,
    \\
    r^4 &= \frac{3}{4} \sum_{\eta=1,3,4,6,8} q_\eta^2.
\end{align}
The higher-order quadrupoles are constructed by the following identities:
\begin{equation}
\begin{aligned}
    r^2 q_1 &= \sqrt 3 q_8 q_1 + \frac{3}{2} q_4 q_6,
     \\
    r^2 q_3 &= \sqrt 3 q_8 q_3 + \frac{3}{4} (q_4^2-q_6^2),
    \\
    r^2 q_4 &= \frac{\sqrt 3}{2} (-q_8 + \sqrt 3 q_3) q_4 + \frac 3 2 q_1 q_6,
    \\
    r^2 q_6 &= \frac{\sqrt 3}{2} (-q_8 - \sqrt 3 q_3) q_6 + \frac 3 2 q_1 q_4,
    \\
    r^2 q_8 
    &= \frac{\sqrt 3}{2} \qty[ q_1^2+q_3^2- q_8^2  - \frac 1 2 (q_4^2 + q_6^2) ],
    \label{eq:def_composite_q}
\end{aligned}
\end{equation}
which are used for the definition of $Q_\eta^{\rm C}$ in the main text.

\subsection{Spherical coordinate}

We also define the spherical coordinate by
\begin{equation}
\begin{aligned}
    r^{(1)}_{-1} &= (x - \imu y)/\sqrt 2,
    \\
    r^{(1)}_0 &= z,
    \\
    r^{(1)}_1 &= - (x+\imu y)/\sqrt 2,
\end{aligned}
\end{equation}
which satisfies $r^{(1)*}_m = (-1)^m r^{(1)}_{-m}$ analogous to the spherical harmonics $Y_{1m}$.
In fact, $r^{(1)}_m = r \sqrt{4\pi \over 3} Y_{1m} $.
In terms of the spherical coordinate, we can define the quadrupole in the form
\begin{align}
    q_M^{(2)} &= [r^{(1)}\otimes r^{(1)}]_M^{(2)}
    \\
    &= \sum_{m_1m_2} \la  1m_1 1 m_2 | 2M \ra r^{(1)}_{m_1} r^{(1)}_{m_2}.
\end{align}
It is confirmed that the quadrupole tensor has the following relation:
$q_M^{\dag \left( 2 \right)} = {\left( { - 1} \right)^M}q_{ - M}^{\left( 2 \right)}$.
The transformation from the Cartesian to spherical coordinates
is given by
\begin{equation}
\begin{aligned}
    q_1 &= \qty[q_2^{(2)} - q_{-2}^{(2)}]/\imu,
    \\
    q_3 &= q_2^{(2)} + q_{-2}^{(2)},
    \\
    q_4 &= q_{1}^{(2)} - q_{-1}^{(2)},
    \\
    q_6 &= -\qty[q_1^{(2)} + q_{-1}^{(2)}]/\imu,
    \\
    q_8 &= - \sqrt{2} \, q_0^{(2)},
\end{aligned}
\label{eq:relat_quadrupoles}
\end{equation}
which are used for the definition of the quadrupole tensor operator $\mathscr Q_M$ in the main text.

\section{Cutoff dependence for reduced matrix elements}
\label{sec:nc_depend}

\begin{figure}[t]
    \centering
    \includegraphics[width=0.48\textwidth]{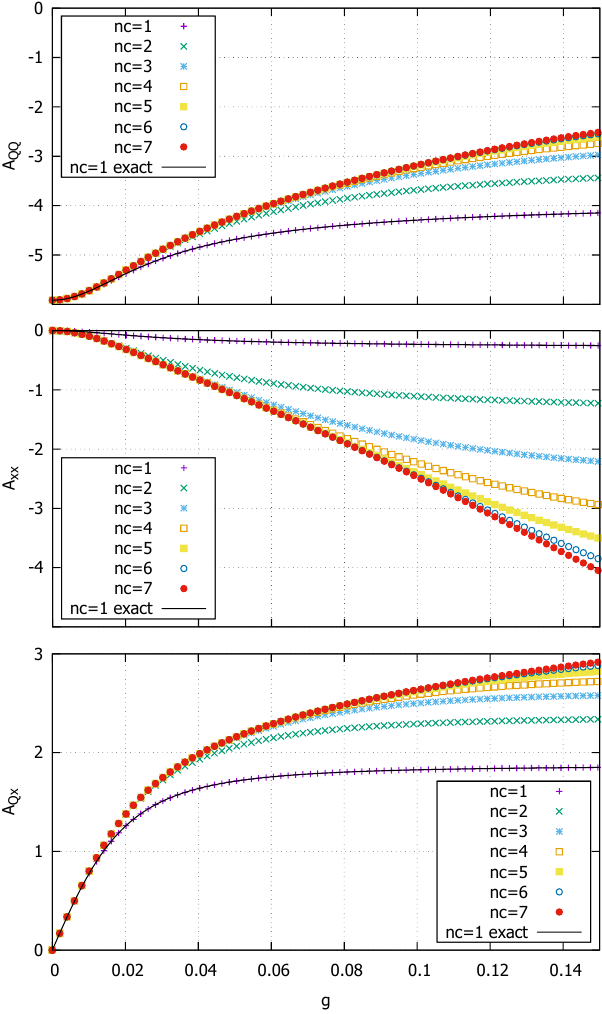}
    \caption{
    Cutoff dependence of the reduced matrix elements shown in Fig.~\ref{fig:composite_quadrupole}.
    The $n_{\rm c}=1$ results shown as solid lines are taken from Eq.~\eqref{eq:nc=1_exact_Acoef}.
    }
    \label{fig:composite_q_cutoff}
\end{figure}

Figure~\ref{fig:composite_q_cutoff} shows the cutoff dependence of the reduced matrix elements $A_{QQ}$, $A_{Qx}$, and $A_{xx}$ defined in Eq.~\eqref{eq:reduced_comp_Q}, calculated for $0 \le n_d \le n_{\rm c}$.
The results are well converged for $g \lesssim 0.15$.
Among them, it is notable that the convergence of $A_{xx}$ with respect 
to the cutoff $n_{\rm c}$ is slower than that of $A_{QQ}$ and $A_{Qx}$.
This is because the operator $[x\otimes x]$ is defined by using phonon operators only.

\section{Spherical tensors and reduced matrix elements}
\label{sec:spherical-tensor}

Here, the definitions of spherical tensors and their useful relations are presented
for $SU(2)$ group.
Angular momentum shift operators $L_{\pm}$ and third component operator $L_0$ are defined
as ${L_ \pm } = {L_x} \pm \mathrm{i} {L_y}$ and ${L_0} = {L_z}$,
which constitute generators of $SU(2)$ group and satisfy the following commutation relations:
$\left[ {{L_i},{L_j}} \right] = \mathrm{i} {\varepsilon _{ijk}}{L_k},$
where $i,j,k$ is either $x$ or $y$ or $z$  and $\varepsilon _{ijk}$ indicates the antisymmetric tensor.

Then a rank-$L$ spherical tensor $T_M^{\left( L \right)}$ 
is defined to satisfy the following commutation relations:
\begin{equation}
\begin{aligned}
\left[ {{L_ \pm },T_M^{\left( L \right)}} \right] 
&= \sqrt {\left( {L \mp M} \right)\left( {L \pm M + 1} \right)}~T_{M \pm 1}^{\left( L \right)},\\
\left[ {{L_0},T_M^{\left( L \right)}} \right] &= M~T_M^{\left( L \right)}.
\label{eq:su(2)}
\end{aligned}
\end{equation}
Defining $\mathscr L_{\pm 1}  =  \mp L_\pm / {\sqrt 2}$ and $\mathscr L_0 \equiv L_0$, 
they themselves satisfy Eq.~\eqref{eq:su(2)} 
as a rank-1 tensor, namely,
\begin{equation}
\begin{aligned}
\left[ {{\mathscr L_{+1} },\mathscr L_{-1}} \right] 
&= - \mathscr L_0,\\
\left[ {{\mathscr L_0},\mathscr L_{\pm 1} } \right] &= \pm \mathscr L_{\pm 1} .
\label{eq:su(2)_2}
\end{aligned}
\end{equation}

One can define a reduced matrix element $ \left\langle {{L_2}\left\| {{T^{\left( L \right)}}} \right\|{L_1}} \right\rangle  $ due to the Wigner-Eckart theorem as
\begin{align}
&
\la L_2M_2|T_M^{(L)}|L_1M_1\ra
= \frac{{\left\langle {{L_1}{M_1}LM\left| {{L_2}{M_2}} \right.} \right\rangle }}{{\sqrt {2{L_2} + 1} }}
\la L_2||T^{(L)}||L_1\ra ,
\label{eq:Wigner_Eckart}
\end{align}
where 
$\left\langle {{L_1}{M_1}LM\left| {{L_2}{M_2}} \right.} \right\rangle $ are the Clebsch--Gordan (CG) coefficients.
The reduced matrix element can be obtained most simply by taking
$M_1 = M_2 = M = 0$ for integer $L_1, L_2$, and $L$.
Note that definition of reduced matrix elements might be different 
in other literature.

Using two spherical tensors of $ T^{\left( {L_1} \right) }_{M_1} $ and
$ U^{\left( {{L_2}} \right)}_{M_2} $, one can construct
their tensor product (rank-$L_{12}$) as
\begin{align}
&\left[ {T^{\left( {{L_1}} \right)} \otimes U^{\left( {{L_2}} \right)}} \right]_{{M_{12}}}^{\left( {{L_{12}}} \right)}  \nonumber \\
&\equiv \sum\limits_{{M_1} =  - {L_1}}^{{L_1}} {} \sum\limits_{{M_2} 
=  - {L_2}}^{{L_2}} {} \left\langle {{L_1}{M_1}{L_2}{M_2}\left| {{L_{12}}{M_{12}}} \right.} \right\rangle  
T_{M_1}^{\left( {{L_1}} \right)} U_{M_2}^{\left( {{L_2}} \right)},
\end{align}
where the symbol $\otimes$ implies the tensor coupling.
In the case of $ {L_{12}} = 0 $, 
one obtains a relation 
\begin{align}
\left[ {T^{\left( L \right)} \otimes U^{\left( L \right)}} \right]_0^{\left( 0 \right)} 
= \frac{{{{\left( { - 1} \right)}^L}}}{{\sqrt {2 L  + 1} }} ~ 
\left ( {T^{\left( L \right)} \cdot U^{\left( L \right)}} \right ),
\end{align}
where the scalar product of $ T^{\left( {{L}} \right)} $ and $ U^{\left( {{L}} \right)} $ 
is defined as
\begin{align}
\left( {T^{\left( L \right)} \cdot U^{\left( L \right)}}  \right)
\equiv \sum\limits_{M =  - L}^L {} {\left( { - 1} \right)^M}T_{M}^{\left( L \right)}~ U_{- M}^{\left( L \right)}.
\label{eq:scalar_product}
\end{align}
Note that a scalar (rank-0) product can be formed only between two tensors of the same rank $L$.

The tensor product of two spherical tensors  $ T^{\left( {{L_1}} \right)} $ 
and $ U^{\left( {{L_2}} \right)} $ has the reduced matrix element~\cite{Bohr_book} 
\begin{align}
&\left\langle {K'\left\| {{{\left[ {{T^{\left( {{L_1}} \right)}} \otimes {U^{\left( {{L_2}} \right)}}} \right]}^{\left( L \right)}}} \right\|K} \right\rangle  \nonumber \\
&= \sqrt {2L + 1} {\left( { - 1} \right)^{K + K' + L}}\sum\limits_{K''} {\left\{ {\begin{array}{*{20}{c}}
{{L_1}}&{{L_2}}&L\\
K&{K'}&{K''}
\end{array}} \right\}} \nonumber \\
& \times \left\langle {K'\left\| {{T^{\left( {{L_1}} \right)}}} \right\|K''} \right\rangle \left\langle {K''\left\| {{U^{\left( {{L_2}} \right)}}} \right\|K} \right\rangle ,
\label{eq:two_product}
\end{align}
where $ \left\{ {\begin{array}{*{20}{c}}
{\begin{array}{*{20}{c}}
{{L_1}}&{{L_2}}&L
\end{array}}\\
{\begin{array}{*{20}{c}}
{{K}}&{{K'}}&K''
\end{array}}
\end{array}} \right\} $ is a Wigner $6j$-symbol.
Here quantum numbers other than 
angular momenta are abbreviated to denote.

The product of two spherical tensors  $ T_1^{\left( {{L_1}} \right)} $ 
and $ T_2^{\left( {{L_2}} \right)} $  that operate on two different spaces, has the reduced matrix element
\begin{align}
&\left\langle {\left( {{K'_1}{K'_2}} \right)K'\left\| {{{\left[ {T_1^{\left( {{L_1}} \right)} \otimes T_2^{\left( {{L_2}} \right)}} \right]}^{\left( {{L_{12}}} \right)}}} \right\|\left( {{K_1}{K_2}} \right)K} \right\rangle   \nonumber \\ 
&=\sqrt {\left( {2K + 1} \right)\left( {2{L_{12}} + 1} \right)\left( {2K' + 1} \right)} \left\{ {\begin{array}{*{20}{c}}
{{K_1}}&{{K_2}}&K\\
{{L_1}}&{{L_2}}&{{L_{12}}}\\
{{K'_1}}&{{K'_2}}&{K'}
\end{array}} 
\right\} \nonumber \\
&\times \left\langle {{K'_1}\left\| {T_1^{\left( {{L_1}} \right)}} \right\|{K_1}} \right\rangle \left\langle {{K'_2}\left\| {T_2^{\left( {{L_2}} \right)}} \right\|{K_2}} \right\rangle ,
\end{align}
where $|(K_1K_2) K\ra$ is the state with the total angular momentum $K$ composed of $K_1$ and $K_2$, and
$ \left\{ {\begin{array}{*{20}{c}}
{{K_1}}&{{K_2}}&K\\
{{L_1}}&{{L_2}}&{{L_{12}}}\\
{{K'_1}}&{{K'_2}}&{K'}
\end{array}} \right\}$ is a Wigner $9j$-symbol. In this notation, quantum numbers other than angular momenta are omitted for simplicity. 
From this formula, 
for a scalar product, one has
\begin{align}
&\left\langle {\left( {{K'_1}{K'_2}} \right)K\left\| 
\left( { ~ {T_1^{\left( L \right)} 
\cdot T_2^{\left( L \right)}~} } \right) \right\|\left( {{K_1}{K_2}} \right)K} \right\rangle  \nonumber \\
&= {\left( { - 1} \right)^{{K_1} + {K'_2} + K}}\sqrt {2K + 1} 
\left\{ {\begin{array}{*{20}{c}}
{\begin{array}{*{20}{c}}
{{K'_1}}&{{K'_2}}&K
\end{array}}\\
{\begin{array}{*{20}{c}}
{{K_2}}&{{K_1}}&L
\end{array}}
\end{array}} 
\right\}
\nonumber \\
&\times \left\langle {{K'_1}\left\| {T_1^{\left( L \right)}} \right\|{K_1}} \right\rangle \left\langle {{K'_2}\left\| {T_2^{\left( L  \right)}} \right\|{K_2}} \right\rangle.
\label{eq:wigner6j}
\end{align}

Especially, in the case that $ T_2^{\left( {{L_2}} \right)} $ 
is taken as the identity operator, i.e., $ T_2^{\left( {{L_2}} \right)} = \hat 1 $, 
one has
\begin{align}
&\left\langle {\left( {{K'_1}{K_2}} \right)K'\left\| {T_1^{\left( {{L_1}} \right)}} \right\|\left( {{K_1}{K_2}} \right)K} \right\rangle  \nonumber \\
&= \sqrt {\left( {2K' + 1} \right)\left( {2K + 1} \right)}~ {\left( { - 1} \right)^{{K_2} + {L_1} + {K'_1} + K}} \nonumber \\
& \times \left\{ 
{\begin{array}{*{20}{c}}
{K'}&{{L_1}}&K\\
{{K_1}}&{{K_2}}&{K'}
\end{array}} \right\}\left\langle {{K'_1}\left\| {T_1^{\left( {{L_1}} \right)}} \right\|{K_1}} \right\rangle.
\label{eq:decoupling}
\end{align}

In some cases, reduced matrix elements are easily obtained
by calculating the left-hand side of Eq.~\eqref{eq:Wigner_Eckart}.
For example, the reduced matrix element of the angular momentum operator 
$\mathscr L$ is given by
\begin{align}
\left\langle {K'\left\| {{\mathscr L}} \right\|K} \right\rangle  = \sqrt {K\left( {K + 1} \right)\left( {2K + 1} \right)} {\delta _{KK'}},
\label{eq:Angular_Momentum}
\end{align}
using Eq.~\eqref{eq:mate_angular_momentum}.

Similarly, in addition to the orbital angular momentum,
one can also include spin degree of freedom to define a
double tensor as $T^{(L,S)}$, which is rank-$L$ tensor for angular momentum 
and rank-$S$ tensor for spin.
One can also include quasispin $K$ to define 
a triple tensor as $T^{(L,S,K)}$.

\section{Spherical tensor formalism for $p$-fermions}
\label{sec:spinorbital}

Below, we summarize the tensorial properties of the spin–orbital system. 
Although some of these results have already been presented in the main text, 
the formulation given here is more closely aligned with approaches commonly used in nuclear physics. 
It thus provides a useful bridge between the theoretical frameworks of Cartesian and spherical tensor representations, and also offers a convenient framework for more complicated cases involving additional orbital degrees of freedom and/or spin--orbit coupling.
Subsections~\ref{sec:spinorbital}~1--4 are devoted to formal description of purely fermionic system, while Sec.~\ref{sec:spinorbital}~5 provides the formulation for the electron--phonon coupled system.

\subsection{Explicit expressions of angular-momentum, spin, and quadrupole operators 
}

Using  $ p_{m\sigma }^\dag $ and  $ {\tilde p_{m\sigma }} $ in Eqs.~\eqref{eq:def_pfermion} and~\eqref{eq:def_pfermiontilde}, 
one can construct any one-body spherical double tensor operator $ \mathscr O^{(L,S)}_{M_L, M_S}$ 
with rank-$ L $ ($0 \le L \le 2$) for the orbital angular momentum and rank-$ S $  ($0 \le S \le 1$) for the spin angular momentum as
\begin{align}
&\mathscr O^{(L,S)}_{M_L,M_S}  
\equiv O_{LS} \left[{p^\dag } \otimes \tilde p\right]_{M_L,M_S}^{\left( {L,S} \right)} \nonumber \\
&\equiv O_{LS}\sum\limits_{mm'} {} \sum\limits_{\sigma \sigma '} {} \left\langle {1m1m'\left| {L{M_L}} \right.} \right\rangle \left\langle {{\textstyle{1 \over 2}}\sigma {\textstyle{1 \over 2}}\sigma '\left| {S{M_S}} \right.} \right\rangle p_{m\sigma }^\dag {\tilde p_{m'\sigma '}},
\end{align}
where $ O_{LS} $ is any real coefficient.
For example, the rank-1 spherical tensor corresponding to the orbital angular momentum operator is given by
\begin{align}
\mathscr L_M =  -2 \left[ {p^\dag } \otimes \tilde p \right]_{M,0}^{\left( {1,0} \right)} .
\label{eq:angular-momentum}
\end{align}
This operator satisfies Eq.~\eqref{eq:su(2)_2}.

Similarly, the rank-1 spherical tensor corresponding to the spin angular momentum operator is given by 
\begin{align}
\mathscr S_M  = - \sqrt {3/2} \left[ {{p^\dag } \otimes \tilde p} \right]_{0,M}^{(0,1)} .
\end{align}
Finally, the quadrupole operator for the orbital part
is given by 
\begin{align}
{\mathscr Q_M}  = -2 \left[ {{p^\dag } \otimes \tilde p} \right]_{M,0}^{\left( {2,0} \right)} , 
\label{eq:quadrupole-moment}
\end{align}
where the factor $ - 2 $ is determined through the equivalence relation $\mathscr Q_0 = Q_8$ in Eq.~\eqref{eq:QM_def}.

\subsection{Explicit tensor expressions of electron states in the $p$-shell}
\label{sec:tensor_expressions}

Here, we consider the wave functions in tensor form. 
The electron states are given explicitly in terms of the creation operators $ p_{m\sigma }^\dag  $. 
The one-body state is given by
$ \left| {{p_{m\sigma }}} \right\rangle  = p_{m\sigma }^\dag \left| 0 \right\rangle  $ 
with $ \left| 0 \right\rangle  $ 
being the closed-shell (or vacuum in our model). 
For two-body states, one has
\begin{align}
&\left| {L{M_L}S{M_S}} \right\rangle
\equiv \frac{1}{{\sqrt 2 }}\left[ {{p^\dag } \otimes {p^\dag }} \right]_{{M_L},{M_S}}^{\left( {L,S} \right)}\left| 0 \right\rangle  \nonumber \\
&\hspace{-2mm}\equiv \frac{1}{{\sqrt 2 }}\sum\limits_{mm'} {} \sum\limits_{\sigma \sigma '} {} \left\langle {1m1m'\left| {L{M_L}} \right.} \right\rangle \left\langle {{\textstyle{1 \over 2}}\sigma {\textstyle{1 \over 2}}\sigma '\left| {S{M_S}} \right.} \right\rangle p_{m\sigma }^\dag p_{m'\sigma '}^\dag \left| 0 \right\rangle
,
\end{align}
where electron orbital angular momenta are coupled to $ L $  ($0 \le L \le 2$) and spin angular momenta, to $ S $  ($0 \le S \le 1$). 
This coupling is denoted as $(L,S)$.

Since  $p_{m\sigma }^\dag p_{m'\sigma '}^\dag  
=  - p_{m'\sigma '}^\dag p_{m\sigma }^\dag $ and 
using properties of CG-coefficients, one has 
\begin{align}
\left[ {{p^\dag } \otimes {p^\dag }} \right]_{{M_L},{M_S}}^{\left( {L,S} \right)} 
&= {\left( { - 1} \right)^{L + S}}\left[ {{p^\dag } \otimes {p^\dag }} \right]_{{M_L},{M_S}}^{\left( {L,S} \right)} .
\end{align}
This indicates that $L + S$ must be an even integer, resulting in $\left( {L,S} \right) = \left( {2,0} \right),\left( {1,1} \right),\left( {0,0} \right)$ combinations.

For three-body states, after considering total anti-symmetry (see Appendix~\ref{sec:su3}),
one has only three independent states:
$ \left( {L,S} \right) = \left( {0,{\textstyle{3 \over 2}}} \right),~\left( {1,{\textstyle{1 \over 2}}} \right),~\left( {2,{\textstyle{1 \over 2}}} \right) $.
The corresponding wave functions are given by
\begin{align}
&\left| {L{M_L}S{M_S}} \right\rangle  \equiv \frac{1}{{\sqrt {{N_L}} }}\left[ {{{\left[ {{p^\dag } \otimes {p^\dag }} \right]}^{\left( {1,1} \right)}} \otimes {p^\dag }} \right]_{{M_L},{M_S}}^{\left( {L,S} \right)}\left| 0 \right\rangle ,
\end{align}
where the normalization factors $N_L$ take the values $N_L = 6,\,3,\,3$ for $L = 0,\,1,\,2$, respectively.
Note that only the $(1,1)$ component
for the intermediate two-body angular momentum
is required to construct independent three-body states in the $p$-shell. 
For this reason, the three-body states are also denoted as
\begin{align}
& \left| {L} \right\rangle \equiv \left| {LS} \right\rangle  \equiv \frac{1}{{\sqrt {{N_L}} }}{\left[ {{{\left[ {{p^\dag } \otimes {p^\dag }} \right]}^{\left( {1,1} \right)}} \otimes {p^\dag }} \right]^{\left( {L,S} \right)}}\left| 0 \right\rangle ,
\label{eq:three_electron_LS}
\end{align}
where the third components of $ L $ and $ S $ (i.e., $M_{L}$ and $M_S$) are abbreviated to denote. 
Specifically for the $ \left( {L,S} \right) = \left( {1,{\textstyle{1 \over 2}}} \right) $ state, 
one has an alternative
expression as
\begin{align}
&\left| {L=1} \right\rangle  = \frac{{\sqrt 3 }}{2} {\left[ {{{\left[ {{p^\dag } \otimes {p^\dag }} \right]}^{\left( {0,0} \right)}} \otimes {p^\dag }} \right]^{\left( {1,1/2} \right)}}\left| 0 \right\rangle .
\label{eq:three_electron_L=1}
\end{align}
As will be discussed in the quasispin formalism (see Appendix~\ref{sec:Quasispin-ope}), this state has $K=1$ in quasispin space, whereas the quantum number $K=0$ is assigned to the $L=0$ and $L=2$ sectors.

The reduced matrix element of the quadrupole operator $\mathscr Q_M$
[Eq.~\eqref{eq:quadrupole-moment}]
is calculated using the three-body wavefunctions by making use of the Wigner-Eckart theorem in Eq.~\eqref{eq:Wigner_Eckart}:
\begin{align}
&\left\langle {{L_2}{M_2}\left| {{{\mathscr Q}_M}} \right|{L_1}{M_1}} \right\rangle  = \frac{{\left\langle {{L_1}{M_1}2M\left| {{L_2}{M_2}} \right.} \right\rangle }}{{\sqrt{2 L_2 + 1} }}\left\langle {{L_2}\left\| {\mathscr Q} \right\|{L_1}} \right\rangle, 
\label{eq:WET}
\end{align}
where the spin degree of freedom is 
abbreviated to denote. 
After a straightforward calculation using the formula Eq.~\eqref{eq:p-overlap}
(see also Appendix~\ref{sec:calc_reduced_Q_matrix_elem}), one obtains 
\begin{align}
\left\langle {L'S'\left\| {\mathscr Q} \right\|LS} \right\rangle  = \sqrt {30} {\left( { - 1} \right)^L}\left( {{\delta _{L,L'}} - 1} \right){\delta _{S,S' = {\textstyle{1 \over 2}}}},
\label{eq:reduced_Q_matrix_elem}
\end{align}
which has nonvanishing matrix element in the case of $S=S'=1/2$.
In the following, 
$ \left\langle {L'S'\left\| {\mathscr Q} \right\|LS} \right\rangle  $ 
is denoted as 
$ \left\langle {L'\left\| {\mathscr Q} \right\|L} \right\rangle  $ in short because the total spin $ S $ is uniquely determined by the total orbital angular momentum $ L $. 
 Therefore, we have
 \begin{equation}
\begin{aligned}
    \la L' m'| \mathscr Q_M | L m\ra
    &= - (-1)^L \sqrt{\frac{{30}}{{2L'\!+\!1}}}\  \la Lm 2M | L'm' \ra,
    \label{eq:red_mate_Q}
\end{aligned}
\end{equation}
for $L\neq L'$ with $s=s'=1/2$.
Then, 
the expectation value of 
$ \left( {\mathscr Q \cdot \mathscr Q} \right) $ 
for any $M$ (third component of angular momentum) with $L=1$ or $L=2$
is given by 
\begin{align}
\left\langle {LM \left| {\left( {\mathscr Q \cdot \mathscr Q} \right)} \right|LM } \right\rangle
= \frac{1}{2L\!+\!1}\sum\limits_I {} {\left\langle {L\left\| {\mathscr Q} \right\|I} \right\rangle ^2} = \frac{{30}}{{2L\!+\!1}} .
\label{eq:QdotQ}
\end{align}
Here, the orthonormalization relation for the CG coefficient
 $\sum\limits_{M_I M} {} \left\langle {IM_I 2M\left| {LM_L } \right.} \right\rangle \left\langle {IM_I 2M\left| {KM_L' } \right.} \right\rangle  = {\delta _{L,K}}{\delta _{M_L ,M_L' }}$ has been used. For $L=0$, one has 
$\left\langle {LM \left| {\left( {\mathscr Q \cdot \mathscr Q} \right)} \right|LM } \right\rangle = 0$.

\subsection{$SU(3)$ structure of electron system}
\label{sec:su3}
By a straightforward calculation, one can show the following commutation relations 
among ${\mathscr L_M}$ and ${\mathscr Q_M}$
[Eq.~\eqref{eq:angular-momentum} and Eq.~\eqref{eq:quadrupole-moment}]:
\begin{equation}
\begin{aligned}
&\left[ {{{\mathscr L}_m},{{\mathscr L}_{m'}}} \right] =  - \sqrt 2 \sum\limits_M \left\langle {1m1m'|1M} \right\rangle {\mathscr L_M},  \\
&\left[ {{{\mathscr Q}_m},{{\mathscr L}_{m'}}} \right] =  - \sqrt 6 \sum\limits_M \left\langle {2m1m'\left| {2M} \right.} \right\rangle {\mathscr Q_M}, \\
&\left[ {{{\mathscr Q}_m},{{\mathscr Q}_{m'}}} \right] = \sqrt {10} \sum\limits_M \left\langle {2m2m'\left| {1M} \right.} \right\rangle {\mathscr L_M}.
\end{aligned}
\end{equation}
The subset of eight operators ${\mathscr L_M}\left( {M =  \pm 1,0} \right)$ and ${\mathscr Q_M}\left( {M =  \pm 2, \pm 1,0} \right)$ 
is associated 
with the group $SU(3)$~\cite{Elliott58_2,Harvey68}. Among them, ${\mathscr L_M}$ are shift operators (or generators) of the group $SU(2)$ [strictly speaking, $SO(3)$].
Thus one has the group chain $SU\left( 3 \right) \supset SO\left( 3 \right)$.
One can therefore construct any electron states in $p$-shell according to the symmetry group ${S_N}$ ($N$ is the number of electrons in $p$-shell), and by taking care of the fermionic anti-symmetry. 

For $U(3)$ or $SU(3)$, states are described by three numbers $\left( {{f_1},{f_2},{f_3}} \right)$ which measure the length of the rows in the Young tableaux $\left[ {{f_1}{f_2}{f_3}} \right]$~\cite{Bohr_book}. However, representations described by tableaux differing only in the number of completed columns (of three squares) all belong to the same representation under $SU(3)$. Thus, only two numbers are needed to classify states according to $SU(3)$, and these can conventionally be defined as $\lambda  = \left( {{f_1} - {f_2}} \right)$ and $\mu  = \left( {{f_2} - {f_3}} \right)$.

For any representation $\left( {\lambda \mu } \right)$ of $SU(3)$, the orbital angular momenta $L$ are given by~\cite{Elliott58}
\begin{align}
L = \left\{ \begin{array}{*{20}{c}}
{K,{\rm{ }}K + 1, \ldots ,\left( {K + \max \left\{ {\lambda ,\mu } \right\}} \right), K \ne 0},\\
{\min \left\{ {\lambda ,\mu } \right\},\min \left\{ {\lambda ,\mu } \right\} - 2,~ \ldots ~,1~{\rm{ or }}~0,~ K = 0},
\end{array} 
\right.
\end{align}
with the integer $K$ taking on the values
\begin{align}
K = \min \left\{ {\lambda ,\mu } \right\},\min \left\{ {\lambda ,\mu } \right\} - 2,~ \ldots ~,1~{\rm{or}}~0.
\end{align}
Note that the indices $\lambda,\mu, K$ here are different from those listed in Table~\ref{tab:notation}.
The three-body electron states have the total anti-symmetry $\left[ f \right] = \left[ {111} \right]$ in the language of Young tableaux. Any state has the orbital inner symmetry $\left[ {{f^{\left( L \right)}}} \right]$ associated with $SU(3)$ and spin inner symmetry $\left[ {{f^{\left( S \right)}}} \right]$ associated with $SU(2)$. These symmetries $\left[ {{f^{\left( L \right)}}} \right]$ and  $\left[ {{f^{\left( S \right)}}} \right]$ should be conjugate to each other to make a completely anti-symmetric  $\left[ f \right] = \left[ {111} \right]$~\cite{Bohr_book}. Since only two rows for the Young tableaux are permissible for $SU(2)$, we have only $\left[ {{f^{\left( S \right)}}} \right] = \left[ 3 \right],\left[ {21} \right]$ whose conjugate representations are $\left[ {{f^{\left( L \right)}}} \right] = \left[ {111} \right],\left[ {21} \right]$, respectively.
$\left[ {{f^{\left( L \right)}}} \right] = \left[ {111} \right]$ corresponds to $\left( {\lambda , \mu } \right) = \left( {0,0} \right)$ and $\left[ {{f^{\left( L \right)}}} \right] = \left[ {21} \right]$ 
to $\left( {\lambda , \mu } \right) = \left( {1,1} \right)$, whereas
$\left[ {{f^{\left( S \right)}}} \right] = \left[ 3 \right],\left[ {21} \right]$ correspond 
to $S = {\textstyle{3 \over 2}},S = {\textstyle{1 \over 2}}$ respectively.
Thus, all three-body electron states have been classified according to the $SU(3)$ representation $\left( {\lambda \mu } \right)$, 
angular momentum $L$, and spin $S$ as shown in Table~\ref{tab:su3-rep}.

\begin{table}[]
    \centering
    \begin{tabular}{c|c|c}
    \hline
       Orbital- and spin-symmetry &   
       $\left| \left( {\lambda,\mu} \right) L,S \right\rangle $ 
       & $\left\langle (\mathscr Q \cdot \mathscr Q) \right\rangle $
       \\
       \hline
        $[{f^{\left( L \right)}}] = [111],~[{f^{\left( S \right)}}] = [3] $ & 
        $\left| {\left( {0,0} \right)L = 0,S = {\textstyle{3 \over 2}}} \right\rangle $  & 0
        \\
        $~[{f^{\left( L \right)}}] = [21],~~~[{f^{\left( S \right)}}] = [21] $ & 
        $\left| {\left( {1,1} \right)L = 1,S = {\textstyle{1 \over 2}}} \right\rangle $  
        & 10
        \\
        $~[{f^{\left( L \right)}}] = [21],~~~[{f^{\left( S \right)}}] = [21] $ &  
        $\left| {\left( {1,1} \right)L = 2,S = {\textstyle{1 \over 2}}} \right\rangle $ 
        & 6
        \\
    \hline
    \end{tabular}
    \caption{Symmetry of three-body fermions in a $p$-shell.
    Table for three-boson system is given in Table~\ref{tab:boson_class}.
    }
    \label{tab:su3-rep}
\end{table}

In terms of creation operators $p^\dagger_{m\sigma}$, any three-body electron states are expressed in tensor representation as
\begin{align}
{\left| {\left( {\lambda \mu } \right)LS\left( {{L_2}{S_2}} \right)} \right\rangle} 
\propto {\left[ 
{{\left[ {{p^\dag } \otimes {p^\dag }} \right]}^{\left( {{L_2},{S_2}} \right)}}  \otimes  {{p^\dag } } 
\right]^{\left( {L,S} \right)}}\left| 0 \right\rangle,
\end{align}
where the first and second  
${p^\dag }$ operators couple to orbital angular momentum ${L_2}$  and spin ${S_2}$, 
and then ${\left[ {{p^\dag } \otimes {p^\dag }} \right]^{\left( {{L_2},{S_2}} \right)}}$ and  the last ${p^\dag }$ operator are coupled to orbital angular momentum $L$  and spin $S$. 
It is directly confirmed that  ${\left\langle {\left( {\lambda \mu } \right)LS\left( {{L'_2}{S'_2}} \right)\left| {\left( {\lambda \mu } \right)LS\left( {{L_2}{S_2}} \right)} \right.} \right\rangle }$ is non-zero. 
Thus, one can set  $\left( {{L_2},{S_2}} \right) = \left( {1,1} \right)$ for any $\left( {\lambda \mu } \right)$ to get independent basis states.
Namely, one has for any three-body electron state
\begin{align}
{\left| \left( {\lambda \mu } \right) LS \right\rangle} 
\propto {\left[ {{\left[ {{p^\dag } \otimes {p^\dag }} \right]}^{\left( {1,1} \right)}} \otimes {{p^\dag }} \right]^{\left( {L,S} \right)}}\left| 0 \right\rangle,
\end{align}
as discussed in Appendix~\ref{sec:tensor_expressions}.

According to Ref.~\cite{Elliott58}, one has 
$\left( {\mathscr Q \cdot \mathscr Q} \right) = \frac{1}{3}\left[ {{{\hat C}_2}\left( {\lambda \mu } \right) - 3\mathscr L \cdot \mathscr L} \right]$ 
in the case that ${\mathscr Q_M} \equiv -2 \left[ {{p^\dag } \otimes \tilde p} \right]_{M,0}^{\left( {2,0} \right)}$
where  ${{\hat C}_2}\left( {\lambda \mu } \right)$
is the two-body Casimir operator of $SU(3)$ group.
Thus, one has
\begin{align}
\left\langle (\mathscr Q \cdot \mathscr Q) \right\rangle 
&\equiv \left\langle {\left( {\lambda \mu } \right)LS} \right|\left( {\mathscr Q \cdot \mathscr Q} \right)\left| {\left( {\lambda \mu } \right)LS} \right\rangle \nonumber \\
&= {\textstyle{1 \over 3}}\left[ {{C_2}\left( {\lambda \mu } \right) - 3L\left( {L + 1} \right)} \right] ,
\label{tab:su3-ene}
\end{align}
where
${C_2}\left( {\lambda \mu } \right) \equiv 4\left( {{\lambda ^2} + {\mu ^2} + \lambda \mu  + 3\lambda  + 3\mu } \right)$.
For each $L$,
$\left\langle (\mathscr Q \cdot \mathscr Q) \right\rangle$ is given in Table~\ref{tab:su3-rep},
which recovers Eq.~\eqref{eq:QdotQ}.

\subsection{Spectroscopic factor and overlap}

Let us also consider the single-particle excitation spectrum as experimentally probed by the photoemission spectroscopy.
We first define the Green's function 
\begin{align}
    G(\imu\omega_n) &= - \int_0^\beta \diff \tau \la p(\tau) p^\dg \ra
    \epn^{\imu\omega_n \tau},
\end{align}
where $\omega_n = (2n+1)\pi T$ is the fermionic Matsubara frequency.
Then the spectrum is given at low $T$ by
\begin{align}
    \rho(\ep) &= - \frac{1}{\pi} {\rm Im\,} G(\ep + \imu 0^+)
    \\
    &= \frac{1}{f(\ep)Z} \sum_{nm} \epn^{-\beta E_n}
    |\la n | p^\dg |m\ra|^2 \delta(\ep + E_m - E_n),
    \label{eq:PES}
\end{align}
where $f(\ep) = 1/ (\epn^{\beta (\ep-\mu)}+1)$ is the Fermi distribution function.
The definitions of $|n\ra$ and $E_n$ are the same as those in Eq.~\eqref{eq:suscep_def}.
This expression contains a matrix element $\langle n | p^\dagger | m \rangle$ that connects a three-electron state $|n\rangle$ with a two-electron state $|m\rangle$.
This matrix element encodes the entanglement between the two-electron subspace and the remaining single electron and is referred to as the spectroscopic factor~\cite{Bohr_book}.
Below, we show the relation between spectroscopic factor and overlap matrix based on our three-electron model.

We define the notation
\begin{align}
{\left| 
{{{
\left[ { {{\left[ {p \otimes p} \right]}^{\left( {{L_2},{S_2}} \right)} \otimes p  }} \right]}
^{\left( {L,S} \right)}}} \right\rangle _U}
\equiv
{\left[ 
{ {{\left[ {{p^\dag } \otimes {p^\dag }} \right]}^{\left( {{L_2},{S_2}} \right)}} {\otimes p^\dag} } \right]^{\left( {L,S} \right)}}\left| 0 \right\rangle ,
\end{align}
where the subscript $U$ indicates an unnormalized state.
One can calculate the overlap between two unnormalized states as
\begin{align}
&O_{{L_2},{S_2},{L'_2},{S'_2} }^{\left( {L,S} \right)} \nonumber \\
&\equiv 
{\left\langle 
{{{\left[ {  {{\left[ {p \otimes p} \right]}^{\left( {{L'_2},{S'_2}}   \right)  }\otimes p}} 
\right]}^{\left( {L,S} 
\right)}} 
\left| 
{{{\left[ { {{\left[ {p \otimes p} \right]}^{\left( {{L_2},{S_2}} \right)}  \otimes p  }} \right]}^{\left( {L,S} \right)}}} \right.} \right\rangle _U}  \nonumber \\
&= 2\left( {{\delta _{{L_2},{L'_2}}}{\delta _{{S_2},{S'_2}}} 
+ 2F_{{L_2},{S_2},{L'_2},{S'_2}}^{\left( {L,S} \right)}} \right) ,
\label{eq:p-overlap}
\end{align}
where 
\begin{align}
F_{{L_2},{S_2},{L'_2},{S'_2}}^{\left( {L,S} \right)}
&\equiv \sqrt {\left( {2{L_2} + 1} \right)\left( {2{L'_2} + 1} \right)\left( {2{S_2} + 1} \right)\left( {2{S'_2}  + 1} \right)}  \nonumber \\
&~~\times \left\{ {\begin{array}{*{20}{c}}
{\begin{array}{*{20}{c}}
1&{{L_2}}&1
\end{array}}\\
{\begin{array}{*{20}{c}}
L&{{L'_2}}&1
\end{array}}
\end{array}} \right\}\left\{ {\begin{array}{*{20}{c}}
{\begin{array}{*{20}{c}}
{{\textstyle{1 \over 2}}}&{{S_2}}&{{\textstyle{1 \over 2}}}
\end{array}}\\
{\begin{array}{*{20}{c}}
S&{{S'_2} }&{{\textstyle{1 \over 2}}}
\end{array}}
\end{array}} \right\},
\label{eq:def_of_F_tensor}
\end{align}
and 
$
 \left\{ {\begin{array}{*{20}{c}}
 {\begin{array}{*{20}{c}}
 a&b&c
 \end{array}}\\
 {\begin{array}{*{20}{c}}
 d&e&f
 \end{array}}
 \end{array}} \right\} $
indicates a Wigner $6j$-symbol.
In the case with ${L'_2} = {L_2},{S'_2} = {S_2}$, 
the norm is given by
\begin{align}
&N_{{L_2},{S_2}}^{\left( {L,S} \right)} \nonumber \\
&\equiv 
{\left\langle 
{{{\left[ {  {{\left[ {p \otimes p} \right]}^{\left( {{L_2},{S_2}}   \right)  }\otimes p}}
\right]}^{\left( {L,S} 
\right)}} 
\left| 
{{{\left[ { {{\left[ {p \otimes p} \right]}^{\left( {{L_2},{S_2}} \right)}  \otimes p  }} \right]}^{\left( {L,S} \right)}}} \right.} \right\rangle _U}  \nonumber \\
&= 2\left( {1 
+ 2F_{{L_2},{S_2},{L_2},{S_2}}^{\left( {L,S} \right)}} \right).
\end{align}
Thus, the normalized overlap is calculated as
\begin{align}
&{\left\langle 
{{{\left[ { {{\left[ {p \otimes p} \right]}^{\left( {{L'_2},{S'_2}} \right)} \otimes p }} \right]}^{\left( {L,S} \right)}}
\left| 
{{{\left[ { {{\left[ {p \otimes p} \right]}^{\left( {{L_2},{S_2}} \right)} \otimes p  }} \right]}^{\left( {L,S} \right)}}} \right.} \right\rangle}  \nonumber \\
&= O_{{L_2},{S_2},{L'_2},{S'_2}}^{\left( {L,S} \right)}{\left( {N_{{L_2},{S_2}}^{\left( {L,S} \right)}N_{{L'_2},{S'_2}}^{\left( {L,S} \right)}} \right)^{ - 1/2}}
\\
&= \sqrt 2 {\left( { - 1} \right)^{L + S - 3/2 }}  {\left( {N_{{L_2},{S_2}}^{\left( {L,S} \right)}\left( {2L + 1} \right)\left( {2S + 1} \right)} \right)^{ - 1/2}} \nonumber \\
& \times \left\langle 
{{{\left[ {{{\left[ {p \otimes p} \right]}^{\left( {{{L'}_2},{{S'}_2}} \right)}} \otimes p} \right]}^{\left( {L,S} \right)}}{{\left\| {{p^\dag }} \right\|}_D}{{\left[ {p \otimes p} \right]}^{\left( {{L_2},{S_2}} \right)}}} \right\rangle.
\end{align}
Here, for a spherical double tensor $\mathscr{D}_{M,\sigma }^{\left( {L,S} \right)}$ 
of orbital angular momentum $L$ (third component $M$) 
and spin angular momentum $S$ (third component $\sigma $), the double-reduced matrix element 
with respect to $L$ and $S$,
$\left\langle {{L_2}{S_2}{{\left\| {{\mathscr{D}^{\left( {L,S} \right)}}} \right\|}_D}{L_1}{S_1}} \right\rangle $ has been defined as: 
\begin{align}
\begin{array}{l}
\left\langle {{L_2}{M_2}{S_2}{\sigma _2}\left| {\mathscr{D}_{M,\sigma }^{\left( {L,S} \right)}} \right|{L_1}{M_1}{S_1}{\sigma _1}} \right\rangle 
\\[2mm]
 = 
 \frac{{\left\langle {{L_1}{M_1}LM\left| {{L_2}{M_2}} \right.} \right\rangle }}{{\sqrt {2{L_2} + 1} }}\frac{{\left\langle {{S_1}{\sigma _1}S\sigma \left| {{S_2}{\sigma _2}} \right.} \right\rangle }}{{\sqrt {2{S_2} + 1} }}\left\langle {{L_2}{S_2}{{\left\| {{\mathscr{D}^{\left( {L,S} \right)}}} \right\|}_D}{L_1}{S_1}} \right\rangle .
 \label{eq:over}
\end{array}
\end{align}
According to Ref.~\cite{Bohr_book}, spectroscopic factor
is defined as
\begin{align}
&S_{{L_2},{S_2},{L'_2},{S'_2}}^{\left( {L,S} \right)} = {\left( {2L + 1} \right)^{ - 1}}{\left( {2S + 1} \right)^{ - 1}} \nonumber \\
& \times {\left| {\left\langle {{{\left[ {{{\left[ {p \otimes p} \right]}^{\left( {{{L_2'}},{{S_2'}}} \right)}} \otimes p} \right]}^{\left( {L,S} \right)}}{{\left\| {{p^\dag }} \right\|}_D}{{\left[ {p \otimes p} \right]}^{\left( {{L_2},{S_2}} \right)}}} \right\rangle } \right|^2}.
\end{align}
From Eq.~\eqref{eq:over}, we obtain
\begin{align}
&S_{{L_2},{S_2},{{L_2'}},{{S_2'}}}^{\left( {L,S} \right)} 
= \frac{{N_{{L_2},{S_2}}^{\left( {L,S} \right)}}}{2}
\nonumber \\
&\times {\left| {\left\langle {{{\left[ {{{\left[ {p \otimes p} \right]}^{\left( {{{L_2'}},{{S_2'}}} \right)}} \otimes p} \right]}^{\left( {L,S} \right)}}\left| {{{\left[ {{{\left[ {p \otimes p} \right]}^{\left( {{L_2},{S_2}} \right)}} \otimes p} \right]}^{\left( {L,S} \right)}}} \right.} \right\rangle } \right|^2}.
\end{align}
This relation shows that the overlap matrix element is expressed in terms of the spectroscopic factor.

\subsection{Matrix elements for the states of three-body fermion system coupled to $d$-bosons
}

At the end of this appendix, we show the matrix elements of the wave functions for the fermion--boson coupled states.
Each term in the total Hamiltonian
$ \mathscr H = 
\mathscr H_{\rm ee}
+ \mathscr H_{\rm p}
+ \mathscr H_{\rm ep}
$
can now be rewritten in terms of the scalar product as
\begin{align}
\mathscr H_{\rm ee} = \frac{J}{2}\left( {\mathscr Q \cdot \mathscr Q} \right),
\end{align}
and
\begin{align}
\mathscr H_{\rm p}
=   \omega_{0} ({d^\dag } \cdot \tilde d) =   \omega_{0} { n_d},
\end{align}
where $ {n_d} $ represents the number operator for the quadrupole phonon. The interaction between electrons and phonons is given by
\begin{align}
\mathscr H_{\rm ep}
&= g\left( {\mathscr Q \cdot x } \right) 
= g\left( {\mathscr Q \cdot {d^\dag }} \right) + {\rm H.c.},
\end{align}
where ${\rm H.c.}$ indicates Hermitian conjugate.

The eigenstate of $\mathscr H_{\rm ee}$ is given by $|LS\ra$.
As for phonons, we have
$
\mathscr H_{\rm p}
\left| {{n_d}\alpha {J_B}} \right\rangle  = \omega_{0} {n_d}\left| {{n_d}\alpha {J_B}} \right\rangle  $, 
where  $ \alpha $ indicates a quantum number that is necessary to uniquely specify a boson state with angular momentum $ {J_B} $ (see Appendix~\ref{sec:Quasispin}). For simplicity, 
$ \left| {{n_d} \alpha {J_B}} \right\rangle  $ is denoted as  $ \left| {{n_d}{J_B}} \right\rangle  $ in the following. Then, the total wavefunction of electrons and phonons is given by  $ \left| {LS{J_F},{n_d}{J_B};{J_T} {M_T}} \right\rangle  $, 
where $ L $ and $ S $ are coupled to the electron angular momentum $ {J_F} $. Next, $ {J_F} $ and $ {J_B} $ are coupled to the total electron--phonon angular momentum $ {J_T} $ 
and the third component $M_T$.
Then, the matrix element of the electron--phonon interaction 
$ \left( {\mathscr Q \cdot {d^\dag }} \right) $ is calculated using Eq.~(\ref{eq:wigner6j}) as follows:
\begin{align}
    &\la J'SJ_F',n_d'J_B';J_T M_T | (\mathscr Q \cdot d^\dg) |LSJ_F,n_dJ_B;J_T M_T\ra
    \nonumber \\
    &=\tfrac{1}{\sqrt{2J_T+1}} \la L'SJ_F',n_d'J_B';J_T || (\mathscr Q \cdot d^\dg) || LSJ_F,n_dJ_B;J_T\ra
    \nonumber \\
    &= (-1)^{J_T+J_F+J_B'} \left\{ \begin{matrix}
        J_F & J_T & J_B \\
        J_B' & 2 & J_F'
        \end{matrix}\right\}
    \nonumber \\
    &\ \ \ \times 
    \la L'SJ_F' || \mathscr Q || LSJ_F \ra
    \la n_d' J_B' || d^\dg || n_d J_B\ra
    \nonumber \\
    &=(-1)^{J_T+S+L+J_B'} \sqrt{(2J_F'+1)(2J_F+1)}
    \nonumber \\
    &\ \ \ \times 
    \left\{ \begin{matrix}
        J_F'& 2& J_F\\
        L& S& L'
    \end{matrix} \right\}
    \left\{ \begin{matrix}
        J_F& J_T& J_B\\
        J_B'& 2& J_F'
    \end{matrix} \right\}
    \nonumber \\
    &\ \ \ \times \la L' || \mathscr Q || L \ra
    \la n_d'J_B'||d^\dg || n_d J_B \ra .
\end{align}
In the above calculation, the following decoupling formula using Eq.~(\ref{eq:decoupling}) has been used:
\begin{align}
&\left\langle {L'S{J'_F}\left\| {\mathscr Q} \right\|LS{J_F}} \right\rangle  
\nonumber \\
&~~= \sqrt {\left( {2{J'_F} + 1} \right)\left( {2{J_F} + 1} \right)} {\left( { - 1} \right)^{S + 2 + L' + {J_F}}}
\nonumber \\
&~~~~~~\times \left\{ 
{\begin{array}{*{20}{c}}
{{J'_F}}&2&{{J_F}}\\
L&S&{L'}
\end{array}} \right\}\left\langle {L'\left\| {\mathscr Q} \right\|L} \right\rangle ,
\end{align}
because the quadrupole operator $ {\mathscr Q_M} $ works only in the orbital angular momentum space. 
The reduced matrix element
$ \left\langle {{n'_d}{J'_B}\left\| {{d^\dag }} \right\|{n_d}{J_B}} \right\rangle  $
has a relation with the $ d $-boson cfp's (coefficients of fractional parentage)
$ \left[ {{d^{{n_d} - 1}}\left( {\alpha L} \right)d,L'\left| {\left. {} \right\}} \right.{d^{{n_d}}}\alpha '} \right] $
as 
\begin{align}
&\left\langle {{n'_d},\alpha 'L'\left\| {{d^\dag }} \right\|{n_d},\alpha L} \right\rangle  
\nonumber \\
&= \sqrt {{n_d} + 1} \sqrt {2L' + 1} \left[ {{d^{{n_d} - 1}}\left( {\alpha L} \right)d,L'\left| {\left. {} \right\}} \right.{d^{{n_d}}}\alpha '} \right], 
\end{align}
where  $ \alpha ,{~} \alpha ' $ are quantum numbers to uniquely specify states 
with the same angular momenta. Especially for $ {n_d} = 0 $, one has 
 $ \left\langle {{n_d} \!=\! 1,{J_B} \!=\! 2\left\| {{d^\dag }} \right\|{n_d} \!=\! 0,{J_B} \!=\! 0} \right\rangle  = \sqrt 5  $, while
cfp’s for $n_d>3$ are usually obtained numerically~\cite{Arima1976}.

\section{Quasispin formalism in fermion systems}
\label{sec:Quasispin-form}
We investigate the quasispin formalism that was first studied 
in nuclear physics by A. K. Kerman~\cite{Kerman1961}.
Here we apply the quasispin formalism to electrons in a $p$-shell (${\ell = 1}$).

\subsection{quasispin operators}
\label{sec:Quasispin-ope}

Using  $ p_{m\sigma }^\dag $ and  $ {\tilde p_{m\sigma }} $ in Eqs.~\eqref{eq:def_pfermion} and~\eqref{eq:def_pfermiontilde}, 
quasispin operators
($K_\pm, K_z)$ are defined as follows:
\begin{align}
{K_+} 
&\equiv \sqrt {\frac{\Omega }{2}} {\left[ {{p^\dag } \otimes {p^\dag }} \right]^{\left( {0,0} \right)}} \nonumber \\
&\equiv \sqrt {\frac{\Omega }{2}} \sum\limits_{m,m',\sigma ,\sigma '} {\left\langle {1m1m' \left| {00} \right.} \right\rangle } \left\langle {{\textstyle{1 \over 2}}\sigma {\textstyle{1 \over 2}}\sigma '\left| {00} \right.} \right\rangle p_{m\sigma }^\dag p_{m'\sigma '}^\dag ,
\\
{K_ - } &\equiv  (K_ +)^\dag = - \sqrt {\frac{\Omega }{2}} {\left[ {\tilde p \otimes \tilde p} \right]^{\left( {0,0} \right)}},\ \ \ 
{K_z} \equiv \frac{1}{2}\left( {n - \Omega } \right),
\end{align}
where $n$ is the number operator and  $\Omega$ (= 3) is 
half the number of the maximally occupied electrons in the $p$-shell.
They satisfy quasispin commutation relations [Eq.~\eqref{eq:su(2)_2}]: 
\begin{align}
\left[ {{K_ + },{K_ - }} \right]{\rm{ = 2}}{K_z},~\left[ {{K_z},{K_ \pm }} \right] = \pm {K_ \pm }.
\end{align}
Defining  ${K_x}$ and ${K_y}$ by  ${K_ \pm } \equiv {K_x} \pm \imu {K_y}$, one has the ordinary commutation relation for angular momentum: $\left[ {{K_k},{K_l}} \right]{\rm{ = }} \imu {\varepsilon _{klm}}{K_m}\left( {k,l,m = x,y,z} \right)$, leading to
the simultaneous eigenstates $\left| {K,{K_0},\alpha } \right\rangle $ 
($\alpha $ is any arbitrary quantum number) of $ {\bm K^2} = K_x^2 + K_y^2 + K_z^2$ and  ${K_z}$. 
Namely, one has ${\bm K^2}\left| {K,{K_0},\alpha } \right\rangle  = K\left( {K + 1} \right)\left| {K,{K_0},\alpha } \right\rangle $ and  ${K_z}\left| {K,{K_0},\alpha } \right\rangle  = {K_0}\left| {K,{K_0},\alpha } \right\rangle $
with  ${K_0} =  - K, - K + 1, \cdots ,K - 1,K$. 
Since the minimum of ${K_0}$ 
is  $ - K$, the relation ${K_ - }\left| {K, - K,\alpha } \right\rangle  = 0$ is satisfied.
For any integer  $p > 0$, one has $\left| {K, - K + p,\alpha } \right\rangle  = {N_{pK}}{\left( {{K_ + }} \right)^p}\left| {K, - K,\alpha } \right\rangle $ with 
${N_{pK}} = \sqrt {{{\left( {2K - p} \right)!} \mathord{\left/
 {\vphantom {{\left( {2K - p} \right)!} {\left( {2K} \right)!p!}}} \right.
 \kern-\nulldelimiterspace} {\left( {2K} \right)!p!}}} $.

The operator $\bm{K}$, although vectorial in form, is a scalar with respect to both the angular momentum $\bm{L}$ and the spin $\bm{S}$, i.e., it is invariant under spatial and spin rotations:
$\left[ {{K_i},{L_j}} \right] = \left[ {{K_i},{S_j}} \right] = 0$.
Therefore, one has the simultaneous eigenstates for ${\bm K^2},{K_z},{\bm L^2},{L_z},{\bm S^2},{S_z}$ denoted
as $\left| {K{K_0},L{L_0},S{S_0},\alpha } \right\rangle $.

\subsection{Construction of states with definite $K$ and ${K_0}$}
In the following we classify the electronic states according to quasispin. For 
one-particle system,
applying ${K_z}$ to 
$p_{m\sigma }^\dag \left| 0 \right\rangle \equiv \left| {K,{K_0},m,\sigma } \right\rangle $, one has
${K_z}\left| {K,{K_0},m,\sigma } \right\rangle  = {K_z}p_{m\sigma }^\dag \left| 0 \right\rangle  = \frac{1}{2}\left( {n - \Omega } \right)p_{m\sigma }^\dag \left| 0 \right\rangle  = \left( { - 1} \right)p_{m\sigma }^\dag \left| 0 \right\rangle $.
Thus ${K_0} =  - 1$. 
Also, by applying $\bm {K^2}$, one obtains the quantum number K as
${\bm K^2}\left| {K,{K_0} \!=\!  - 1,\alpha } \right\rangle  
= 2\left| {K,{K_0},\alpha } \right\rangle .$
Thus, this state has  ${K = 1}$ and is written as $\left| {{p_{m\sigma }}} \right\rangle  = \left| {K \!=\! 1,K_0 \!=\!  - 1,m,\sigma } \right\rangle $.

Next, we consider the three-particle system.
As stated in Eqs.~\eqref{eq:three_electron_LS} and \eqref{eq:three_electron_L=1}, we have the following three orthonormal basis states:
\begin{align}
\left| {L = 0} \right\rangle  \equiv \frac{1}{{\sqrt {2\Omega } }}{\left[ {{{\left[ {{p^\dag } \otimes {p^\dag }} \right]}^{\left( {1,1} \right)}} \otimes {p^\dag }} \right]^{\left( {0,{\textstyle{3 \over 2}}} \right)}}\left| 0 \right\rangle , \\
\left| {L = 2} \right\rangle  \equiv \frac{1}{{\sqrt \Omega  }}{\left[ {{{\left[ {{p^\dag } \otimes {p^\dag }} \right]}^{\left( {1,1} \right)}} \otimes {p^\dag }} \right]^{\left( {2,{\textstyle{1 \over 2}}} \right)}}\left| 0 \right\rangle  , \\
\left| {L = 1} \right\rangle  \equiv \frac{{\sqrt \Omega  }}{2}{\left[ {{{\left[ {{p^\dag } \otimes {p^\dag }} \right]}^{\left( {0,0} \right)}} \otimes {p^\dag }} \right]^{\left( {1,{\textstyle{1 \over 2}}} \right)}}\left| 0 \right\rangle  ,
\end{align}
where the third components of total orbital angular momentum and total spin are abbreviated to denote.

The  $\left| {L = 1} \right\rangle $ state can be constructed by applying ${K_ + }$ to the one-particle state $\left| {{p_{m\sigma }}} \right\rangle  = \left| {K = 1,K =  - 1,m,\sigma } \right\rangle $.
Using the formula $  \left| {K, - K + 1,m,\sigma } \right\rangle
= {N_{p=1,K}}{K_ + }\left| {K, - K,m,\sigma } \right\rangle $ with $K = 1$, 
one has 
\begin{align}
&\sqrt {{{\left( {2K - 1} \right)!} \mathord{\left/
 {\vphantom {{\left( {2K - 1} \right)!} {\left( {2K} \right)!}}} \right.
 \kern-\nulldelimiterspace} {\left( {2K} \right)!}}} {K_ + }\left| {{p_{m\sigma }}} \right\rangle  \nonumber \\
&= \frac{{\sqrt \Omega  }}{2}\left[ {{{\left[ {{p^\dag } \otimes {p^\dag }} \right]}^{\left( {0,0} \right)}} \otimes p^\dag } \right]_{m, \sigma }^{\left( {1,{\textstyle{1 \over 2}}} \right)} \left| 0 \right\rangle \nonumber \\
&= \left| {L = 1} \right\rangle .
\label{eq:exp_of_L=1}
 \end{align}
Thus $\left| {L = 1} \right\rangle $ has $K = 1$ and ${K_0} = 0$, 
while both  $\left| {L = 0} \right\rangle $ and $\left| {L = 2} \right\rangle $ states have 
$K = 0$ and ${K_0} = 0$.

\subsection{Quasispin tensors in terms of creation and annihilation operators}

Since Eq.~\eqref{eq:p_as_quasispin_tensor} is satisfied,
$p_{m\sigma }^\dag $ is a spherical tensor $T_{1/2}^{\left( {1/2} \right)}$ and ${\tilde p_{m\sigma }}$ is a spherical tensor $T_{ - 1/2}^{\left( {1/2} \right)}$ in quasispin space, which
is easily confirmed by Eq.~\eqref{eq:su(2)}.
As a triple tensor in the order of $L,S$ and $K$, we denote
$p_{m,\sigma ,1/2}^{\left( {1,1/2,1/2} \right)} \equiv p_{m\sigma }^\dag$ and $~p_{m,\sigma , - 1/2}^{\left( {1,1/2,1/2} \right)} \equiv {\tilde p_{m\sigma }}$.
Then, the one-body triple tensor operator is defined as
\begin{align}
&T_{{L_0},{S_0},{K_0}}^{\left( {L,S,K} \right)} 
\equiv \left[ {{p^{\left( {1,1/2,1/2} \right)}} \otimes {p^{\left( {1,1/2,1/2} \right)}}} \right]
_{{L_0},{S_0},{K_0}}^{\left( {L,S,K} \right)}  \nonumber \\
&\equiv \sum\limits_{m,m',\sigma,\sigma',\kappa,\kappa'} 
\left\langle {1m1m'\left| {L{L_0}} \right.} \right\rangle \left\langle {{\textstyle{1 \over 2}}\sigma {\textstyle{1 \over 2}}\sigma '\left| {S{S_0}} \right.} \right\rangle \left\langle {{\textstyle{1 \over 2}}\kappa {\textstyle{1 \over 2}}\kappa '\left| {K{K_0}} \right.} \right\rangle \nonumber \\
&~~~~~~~~~~~~~~~~~~~~\times p_{m,\sigma ,\kappa }^{\left( {1,1/2,1/2} \right)}
p_{m',\sigma ',\kappa '}^{\left( {1,1/2,1/2} \right)} .
\end{align}
Upon exchanging the two $p$ operators within $[\cdots]$, one obtains, for $L + S + K \neq 0$,
\begin{align}
&\left[ {{p^{\left( {1,1/2,1/2} \right)}} \otimes {p^{\left( {1,1/2,1/2} \right)}}} \right]
_{{L_0},{S_0},{K_0}}^{\left( {L,S,K} \right)} \nonumber \\
&= {\left( { - 1} \right)^{L + S + K + 1}}\left[ {{p^{\left( {1,1/2,1/2} \right)}} \otimes {p^{\left( {1,1/2,1/2} \right)}}} \right]
_{{L_0},{S_0},{K_0}}^{\left( {L,S,K} \right)} ,
\end{align}
where the factor $(-1)^{L+S+K}$ arises from the phase associated with the exchange of angular momenta $L$, $S$, and $K$, while an additional factor $(-1)$ originates from the exchange of two fermions.
Therefore, one has the selection rule that $L + K + S$ must be an odd integer.

\subsection{Specific cases of triple tensors $T_{{L_0},{S_0},{K_0}}^{\left( {L,S,K} \right)}$}

For concreteness, we list several low-rank triple tensors for $p$-fermions below:
\begin{align}
T_{0,0,1}^{\left( {0,0,1} \right)} 
&= \left[ {{p^{\left( {1,1/2,1/2} \right)}} \otimes {p^{\left( {1,1/2,1/2} \right)}}} \right]_{0,0,1}^{\left( {0,0,1} \right)}  \nonumber \\
&= {\left[ {{p^\dag } \otimes {p^\dag }} \right]^{\left( {0,0} \right)}} 
= \sqrt {\frac{2}{\Omega }} {K_ + },
\\
T_{0,0, - 1}^{\left( {0,0,1} \right)} 
&= \left[ {{p^{\left( {1,1/2,1/2} \right)}} \otimes {p^{\left( {1,1/2,1/2} \right)}}} \right]_{0,0, - 1}^{\left( {0,0,1} \right)}  \nonumber \nonumber \\
&= {\left[ {\tilde p \otimes \tilde p} \right]^{\left( {0,0} \right)}} 
=  - \sqrt {\frac{2}{\Omega }} {K_ - },
\\
T_{0,0, 0}^{\left( {0,0,1} \right)} 
&= \left[ {{p^{\left( {1,1/2,1/2} \right)}} \otimes {p^{\left( {1,1/2,1/2} \right)}}} \right]_{0,0,0}^{\left( {0,0,1} \right)} \nonumber \\
&=  - \frac{2}{{\sqrt \Omega  }}{K_z},
\\
T_{M,0,0}^{\left( {1,0,0} \right)} 
&= \left[ {{p^{\left( {1,1/2,1/2} \right)}} \otimes {p^{\left( {1,1/2,1/2} \right)}}} \right]_{M,0,0}^{\left( {1,0,0} \right)} \nonumber \\
&= \frac{1}{{\sqrt 2 }}\left[ {{p^\dag } \otimes \tilde p} \right]_{M,0}^{\left( {1,0} \right)} 
=  - \frac{{\sqrt 2 }}{4}{\mathscr L_M},
\label{eq:angular-momentum_2}
\\
T_{M,0,0}^{\left( {2,0,1} \right)} 
&= \left[ {{p^{\left( {1,1/2,1/2} \right)}} \otimes {p^{\left( {1,1/2,1/2} \right)}}} \right]_{M,0,0}^{\left( {2,0,1} \right)}  
\nonumber \\
&= \sqrt 2 \left[ {{p^\dag } \otimes \tilde p} \right]_{M,0}^{\left( {2,0} \right)} 
=  - \frac{1}{{\sqrt 2 }}{\mathscr Q_M},
\end{align}
for $S=0$, and 
\begin{align}
T_{0,M,0}^{\left( {0,1,0} \right)} 
 &= \left[ {{p^{\left( {1,1/2,1/2} \right)}} \otimes {p^{\left( {1,1/2,1/2} \right)}}}  \right]_{0,M,0}^{\left( {0,1,0} \right)} 
 =  - \frac{2}{{\sqrt 3 }}{\mathscr S_M},
\end{align}
for $S=1$.

\subsection{Wigner Eckart theorem in quasispin space}
We discuss concrete examples of the matrix elements.
Suppose the one-body operator $Q_0^{(1)}$ has $K = 1$ and $K_0 = 0$ (e.g., a quadrupole operator). 
For diagonal matrix elements, the Wigner--Eckart theorem gives
\begin{align}
&\left\langle {K,{K_z},\alpha \left| {Q_0^{\left( 1 \right)}} \right|K,{K_z},\alpha } \right\rangle  
\nonumber \\
&= \frac{{\left\langle {K,{K_z},1,0\left| {K,{K_z}} \right.} \right\rangle }}{{\sqrt {2K + 1} }}\left\langle {K,\alpha {{\left\| {{Q^{\left( 1 \right)}}} \right\|}_K}K,\alpha } \right\rangle .
\end{align}
Here  
$\left\langle { \cdots {{\left\| {{Q^{\left( 1 \right)}}} \right\|}_K} \cdots } \right\rangle $
indicates a reduced matrix element in quasispin space.
Explicit expression for the CG coefficient 
$\left\langle {K,{K_z},1,0\left| {K,{K_z}} \right.} \right\rangle$ is 
$ \frac{{{K_z}}}{{\sqrt {K\left( {K + 1} \right)} }}$. 
Thus for  $\left| {L = 1} \right\rangle~({K_z} = 0,~K = 1)$  state, the diagonal matrix element vanishes. For  $\left| {L = 2} \right\rangle $ and $\left| {L = 0} \right\rangle $ 
(${K_z} = 0,K = 0$) states, the matrix element vanishes because of $\left\langle {0,0,1,0\left| {0,0} \right.} \right\rangle  = 0$. Thus, for any $\left| L \right\rangle $ state, one has $\left\langle {L\left| {Q_0^{\left( 1 \right)}} \right|L} \right\rangle  = 0$. 

For the off-diagonal matrix element, there is the relation 
${\la L \!=\! 1 | Q_0^{( 1 )} |L \!=\! 0 \ra}  = 0$, because ${Q_0^{\left( 1 \right)}} $ has the rank (angular momentum) $L=2$.
Also, the two states $\left| {L = 2} \right\rangle $ and $\left| {L = 0} \right\rangle $ have 
${K_0} = 0$ and $K = 0$, and hence $\left\langle {L = 2\left| {Q_0^{\left( 1 \right)}} \right|L = 0} \right\rangle  = 0$.
The matrix element between 
$\left| {L = 1} \right\rangle $ and $\left| {L = 2} \right\rangle$ states has 
the only non-vanishing off-diagonal matrix element, that is, 
$\left\langle {L = 2{{\left\| \mathscr Q \right\|}_L}L = 1} \right\rangle  = \sqrt {30} $ by explicit calculation as stated before,
where  $\left\langle \cdots \left\| \mathscr Q \right\|_L \cdots \right\rangle $  indicates reduced matrix element in orbital angular momentum space.

\subsection{Calculation of reduced matrix element 
$\left\langle {L = 2\left\| \mathscr Q \right\|L = 1} \right\rangle $ using quasispin formalism}
\label{sec:calc_reduced_Q_matrix_elem}

Previously, we showed that the diagonal matrix elements 
$\left\langle {L\left\|\mathscr Q \right\|L} \right\rangle $  
for any three-body state $\left| L \right\rangle $  vanish
and also that off-diagonal matrix elements
$\left\langle {L'\left\| \mathscr Q \right\|L} \right\rangle$ vanish except 
for $L = 1$  and $L' = 2$. Here we show its specific value using quasispin formalism. 
We define a triple tensor, which is equal to $\mathscr Q_M$ :
\begin{align}
Q_{M,0,0}^{\left( {2,0,1} \right)} & \equiv  - \sqrt 2 \left[ {{p^{\left( {1,{\textstyle{1 \over 2}},{\textstyle{1 \over 2}}} \right)}} \otimes {p^{\left( {1,{\textstyle{1 \over 2}},{\textstyle{1 \over 2}}} \right)}}} \right]_{M,0,0}^{\left( {2,0,1} \right)} \nonumber \\
&=  - 2{\left[ {{p^\dag } \otimes \tilde p} \right]^{\left( {2,0} \right)}} = {\mathscr Q_M}.
\end{align}
Commutation between  ${K_ + } \equiv \frac{{\sqrt {2\Omega } }}{2}{\left[ {{p^\dag } \otimes {p^\dag }} \right]^{\left( {0,0} \right)}}$ and $Q_{M,0,0}^{\left( {2,0,1} \right)} = \mathscr Q_M $  becomes 
$\left[ {{K_ + },Q_{M,0,0}^{\left( {2,0,1} \right)}} \right] 
= \sqrt 2 Q_{M,0,1}^{\left( {2,0,1} \right)},$
where 
\begin{align}
Q_{M,0,1}^{\left( {2,0,1} \right)} &=  - \sqrt 2 \left[ {{p^{\left( {1,{\textstyle{1 \over 2}},{\textstyle{1 \over 2}}} \right)}} \otimes {p^{\left( {1,{\textstyle{1 \over 2}},{\textstyle{1 \over 2}}} \right)}}} \right]_{M,0,1}^{\left( {2,0,1} \right)} \nonumber \\
&=  - \sqrt 2 \left[ {{p^\dag } \otimes {p^\dag }} \right]_{M,0}^{\left( {2,0} \right)}.
\end{align}
We use the following form of the $L=1$ state:
\begin{align}
\left| {L = 1,{m_1},{\sigma _1}} \right\rangle  
= \frac{1}{{\sqrt 2 }}{K_ + }p_{{m_1}{\sigma _1}}^\dag \left| 0 \right\rangle.
\end{align}
Then one obtains
\begin{align}
&\mathscr Q_M \left| {L = 1,{m_1},{\sigma _1}} 
\right\rangle
\nonumber \\
&= \frac{1}{{\sqrt 2 }}\left\{ { - \left[ {{K_ + },Q_{M,0,0}^{\left( {2,0,1} \right)}} \right] + {K_ + }Q_{M,0,0}^{\left( {2,0,1} \right)}} \right\}p_{{m_1}{\sigma _1}}^\dag \left| 0 \right\rangle.
\end{align}
It should be noted that the second term in $\{\  \}$ vanishes 
for the matrix element between 
$\left| {L = 1} \right\rangle $ and $\left| {L = 2} \right\rangle$ ($K=0$) states
because of  ${K_ - }\left| {L = 2} \right\rangle =0 $ .
Since 
$\left[ {{K_ + },Q_{M,0,0}^{\left( {2,0,1} \right)}} \right] = \sqrt 2 Q_{M,0,1}^{\left( {2,0,1} \right)}$ holds,
one has
\begin{align}
Q_{M,0,1}^{\left( {2,0,1} \right)}\left| {L = 1,{m_1},{\sigma _1}} \right\rangle  
= \sqrt 2 \left[ {{p^\dag } \otimes {p^\dag }} \right]_{M,0}^{\left( {2,0} \right)}p_{{m_1}{\sigma _1}}^\dag \left| 0 \right\rangle  \nonumber \\
= \sqrt 2 \sum\limits_{I,{M_I}} {} \left\langle {2M1{m_1}\left| I \right.{M_I}} \right\rangle \left[ {{{\left[ {{p^\dag } \otimes {p^\dag }} \right]}^{\left( {2,0} \right)}} \otimes {p^\dag }} \right]_{{M_I}{\sigma _1}}^{\left( {I,{\textstyle{1 \over 2}}} \right)}\left| 0 \right\rangle .
\end{align}
Now we use the wave function for $L=1$
\begin{align}
\left| {L = 2,{m_2},{\sigma _2}} \right\rangle  \equiv \frac{1}{{\sqrt \Omega  }}\left[ {{{\left[ {{p^\dag } \otimes {p^\dag }} \right]}^{\left( {1,1} \right)}} \otimes {p^\dag }} \right]_{{m_2}{\sigma _2}}^{\left( {2,{\textstyle{1 \over 2}}} \right)}\left| 0 \right\rangle ,
\end{align}
and then obtain
\begin{align}
\left\langle {L = 2,{m_2}\sigma \left| {{\mathscr Q_M}} \right|L = 1,{m_1}\sigma } \right\rangle  =  - \sqrt 6 \left\langle {2M1{m_1}\left| 2 \right.{m_2}} \right\rangle , 
\end{align}
where the formula Eq.~\eqref{eq:p-overlap} has been used.
Due to Wigner Eckart theorem, one has
\begin{align}
&\left\langle {L = 2,{m_2}\sigma \left| {{\mathscr Q_M}} \right|L = 1,{m_1}\sigma } \right\rangle  \nonumber \\
&= \frac{{\left\langle {1{m_1}2M\left| 2 \right.{m_2}} \right\rangle }}{{\sqrt 5 }}\left\langle {L = 2\left\| \mathscr Q \right\|L = 1} \right\rangle.  \label{eq:Q_matele}
\end{align}
Thus, one finally obtains the desired result :
\begin{align}
\left\langle {L = 2\left\| \mathscr Q \right\|L = 1} \right\rangle  = \sqrt {30},
\end{align}
which is identical to Eq.~\eqref{eq:reduced_Q_matrix_elem}.

\subsection{Explicit construction of many-body operators having non-vanishing diagonal matrix elements}

In terms of quadrupole operator 
$Q_{M,0,0}^{\left( {2,0,1} \right)} 
\equiv  - 2{\left[ {{p^\dag } \otimes \tilde p} \right]^{\left( {2,0} \right)}_{M,0}}
=\mathscr Q_M $, one can construct a two-body quadrupole operator with $K=2$ and $K_0=0$ in a simple manner:
\begin{align}
{\mathscr Q_M^{\rm C}} &\equiv \left[ {Q_{}^{\left( {2,0,1} \right)} \otimes Q_{}^{\left( {2,0,1} \right)}} \right]_{M,0,0}^{\left( {2,0,2} \right)}  \nonumber \\
&= \frac{1}{{\sqrt 6 }}{\left[ {{{\left[ {{p^\dag } \otimes {p^\dag }} \right]}^{\left( {2,0} \right)}} \otimes {{\left[ {\tilde p \otimes \tilde p} \right]}^{\left( {2,0} \right)}}} \right]^{\left( {2,0} \right)}}  \nonumber \\
&\ \ + \frac{4}{{\sqrt 6 }}{\left[ {{{\left[ {{p^\dag } \otimes \tilde p} \right]}^{\left( {2,0} \right)}} \otimes {{\left[ {{p^\dag } \otimes \tilde p} \right]}^{\left( {2,0} \right)}}} \right]^{\left( {2,0} \right)}} \nonumber \\
&\ \ + \frac{1}{{\sqrt 6 }}{\left[ {{{\left[ {\tilde p \otimes \tilde p} \right]}^{\left( {2,0} \right)}} \otimes {{\left[ {{p^\dag } \otimes {p^\dag }} \right]}^{\left( {2,0} \right)}}} \right]^{\left( {2,0} \right)}} .
\end{align}
This   operator does not have diagonal matrix elements in $\left| {L = 0} \right\rangle $ and  $\left| {L = 2} \right\rangle $ states, namely, $\left\langle {L = 0\left| {{\mathscr Q_M^C}} \right|L = 0} \right\rangle  = \left\langle {L = 2\left| {{\mathscr Q_M^C}} \right|L = 2} \right\rangle  = 0$ since both states have 
$K = 0$ and $\left\langle {0,0,2,0\left| {0,0} \right.} \right\rangle  = 0$.

Similarly, the $K = 0$ operator $\mathscr Q_M^{\rm N}$ is constructed, using orbital angular momentum 
$\mathscr L_M$ defined in Eq.~\eqref{eq:angular-momentum} [or in Eq.~\eqref{eq:angular-momentum_2}],
as
\begin{align}
\mathscr Q_M^{\rm N} \equiv 
\left[ {\mathscr L \otimes \mathscr L} \right]_M^{\left( 2 \right)} .
\end{align}
With use of the formula [Eqs.~\eqref{eq:two_product} and \eqref{eq:Angular_Momentum}], 
for this new quadrupole operator $\mathscr Q_M^{\rm N}$ ($K=0,L=2$), reduced matrix elements are easily evaluated as
\begin{align}
\left\langle {L\left\| 
\mathscr Q^{\rm N}
\right\| L} \right\rangle  
= \left\{ {\begin{array}{*{20}{c}}
{0{\rm{     ~~for~~  }}L = 0},\\
{1{\rm{      ~~for~~  }}L = 1},\\
{\sqrt {21} {\rm{ ~~for~~  }}L = 2}.
\end{array}} \right.
\end{align}
These operators, $\mathscr Q_M^{\rm N}$ and $\mathscr Q_M^{\rm C}$, can be regarded as the spherical-tensor versions of the operators $Q_\eta^{\rm C}$ and $\overline{Q}_\eta^{\rm C}$ defined in Eqs.~\eqref{eq:def_QC_1} and \eqref{eq:QQdef_Cartesian} in the main text.

\section{
Spherical tensor formalism in $d$-boson systems
}
\label{sec:Quasispin}
Here we classify the states of phonons with angular momentum $l=2$ ($d$-boson)~\cite{Iachello_book}.
For brevity, we sometimes omit the subscript ``ph''; for example, $L_{\rm ph}$ is written simply as $L$ in the following.

\subsection{$U(5)$ structure of $d$-boson system}
The creation and modified annihilation operators are defined as
$d_m^\dag $ and ${\tilde d_m} \equiv {\left( { - 1} \right)^{2 - m}}{d_{ - m}} = {\left( { - 1} \right)^m}{d_{ - m}}$ ($m =  - 2, - 1,0,1,2$), which behave as spherical tensors.
The following twenty-five operators constitute the generators of $U\left( 5 \right)$ group:
$[ {{d^\dag } \otimes \tilde d} ]_M^{( L )} 
$ $(L = 0,1,2,3,4)$.
Among those, ten operators 
with $L=1$ and $L=3$
constitute the generators of $SO\left( 5 \right)$ group.
We further extract the subgroup by selecting the operators with $L=1$, which
constitute the generators of $SU\left( 2 \right) \cong SO\left( 3 \right)$ group.
Therefore, 
$d$-boson states are classified according to the $U\left( 5 \right) \supset SO\left( 5 \right) \supset SO\left( 3 \right)$ 
subgroup chain.

There are five parameters to characterize the basis states.
The simplest is each boson number $n_{M=-2,-1,0,1,2}$ in spherical representation or $n_{\eta = 1,3,4,6,8}$ in Cartesian representation.
On the other hand,
it is known that $d$-boson states are uniquely classified using five quantum numbers:
${n_d},{n_\beta },{n_\Delta },L,M$ where ${n_\beta }$ is the number of pairs coupled to zero angular momentum, ${n_\Delta }$ counts boson triplets coupled to zero angular momentum~\cite{Kemmer1968, Williams1968}. 
Among the $n_d$ bosons, let $\lambda$ denote the number of bosons that do not form $L=0$ pairs or $L=0$ triplets. One then has the relation
\begin{equation}
n_d = 2n_\beta + 3n_\Delta + \lambda,
\end{equation}
and the allowed values of the total angular momentum for given $n_\beta$ and $n_\Delta$ are expressed in terms of $\lambda$ as
\begin{equation}
L = \lambda,\, \lambda + 1,\, \lambda + 2,\, \cdots,\, 2\lambda - 2,\, 2\lambda,
\end{equation}
with the notable absence of $L = 2\lambda - 1$.
Therefore, any $d$-boson state is uniquely specified by the five integers as $\left| {{n_d},{n_\beta },{n_\Delta },L,M} \right\rangle $.
For more details, see Ref.~\cite{Arima1976}.

The absence of $L = 2\lambda - 1$ can be understood as follows~\cite{Talmi_book}. 
Consider $\lambda$ $d$-bosons ($\ell=2$) that are not coupled to zero angular momentum. 
We first determine the maximum value of the total magnetic quantum number $M$. 
For the highest-weight configuration, each boson carries magnetic quantum number $m=2$, yielding $M_{\max} = 2\lambda$, which corresponds to $L = 2\lambda$. 
The state with $M = 2\lambda - 1$ is obtained by replacing one of the $\lambda$-bosons with $m=1$ while keeping the remaining $\lambda-1$ bosons at $m=2$. 
Because the bosons are identical and the total wave function must be fully symmetric, this configuration is unique. 
Since an $L = 2\lambda$ multiplet already exists, the unique $M = 2\lambda - 1$ state must belong to the $L = 2\lambda$ representation. 
Therefore, the $L = 2\lambda - 1$ multiplet does not exist.
We summarize the $d$-boson states ($n_d \leq 6$) in Table~\ref{tab:boson_class}.

\begin{table}[t]
    \centering
    \begin{tabular}{cccccc}
    \hline
    \ \ $n_d$\ \  & \ \ $n_\beta$\ \  & \ \ $n_\Delta$\ \  & \ \ $\lambda$\ \  & \hspace{13mm} $L$\hspace{13mm} &  $\big(\sum_L (2L+1) \big)$ \\
    \hline
    1 & 0 & 0 & 1 & 2 & (5)\\
    \hline
    2 & 0 & 0 & 2 & 2, 4 & (14)\\
    2 & 1 & 0 & 0 & 0 & (1) \\
    \hline
    3 & 0 & 0 & 3 & 3, 4, 6  & (29)\\
    3 & 0 & 1 & 0 & 0  & (1)\\
    3 & 1 & 0 & 1 & 2  & (5)\\
    \hline
    4 & 0 & 0 & 4 & 4, 5, 6, 8  & (50)\\
    4 & 0 & 1 & 1 & 2  & (5)\\
    4 & 1 & 0 & 2 & 2, 4  & (14)\\
    4 & 2 & 0 & 0 &  0 & (1)\\
    \hline
    5 & 0 & 0 & 5 & 5, 6, 7, 8, 10  & (77)\\
    5 & 0 & 1 & 2 & 2, 4  & (14)\\
    5 & 1 & 0 & 3 & 3, 4, 6  & (29)\\
    5 & 1 & 1 & 0 &  0 & (1)\\
    5 & 2 & 0 & 1 &  2 & (5)\\
    \hline
    6 & 0 & 0 & 6 &  6, 7, 8, 9, 10, 12 & (110)\\
    6 & 0 & 1 & 3 &  3, 4, 6 & (29)\\
    6 & 0 & 2 & 0 &  0 & (1)\\
    6 & 1 & 0 & 4 &  4, 5, 6, 8 & (50)\\
    6 & 1 & 1 & 1 &  2 & (5)\\
    6 & 2 & 0 & 2 &  2, 4 & (14)\\
    6 & 3 & 0 & 0 &  0 & (1)\\
    \hline
    \end{tabular}
    \caption{
Classification of many-body phonon states up to $n_d = 6$.
The rightmost column (in parentheses) shows the number of states in each row.
For a given $n_d$, the total number of states is given by the binomial coefficient $\displaystyle \binom{n_d+4}{4}$.
}
    \label{tab:boson_class}
\end{table}

\subsection{Explicit construction of few-body systems in terms of $d$-boson creation operators}

The group-theoretical analysis in the previous subsection can be concretized by considering the explicit form of the wave functions as follows:
\\
\noindent
(a)	one-body system
\begin{align}
\left| {{n_d} = 1,L = 2,M} \right\rangle  = d_M^\dag \left| 0 \right\rangle,
\end{align}
\noindent
(b)	two-body system
\begin{align}
&\left| {{n_d} = 2,L = 0,2,4,M} \right\rangle  
= \frac{1}{{\sqrt 2 }}\left[ {{d^\dag } \otimes {d^\dag }} \right]_M^{\left( L \right)}\left| 0 \right\rangle  \nonumber \\
&= \frac{1}{{\sqrt 2 }}\sum\limits_{m,m'} {} \left\langle {2m2m'\left| {LM} \right.} \right\rangle d_m^\dag d_{m'}^\dag \left| 0 \right\rangle, 
\end{align}
\noindent
(c) three-body system
\begin{align}
{\left| {{n_d} = 3,LM\left( K \right)} \right\rangle _U} 
= \left[ {{{\left[ {{d^\dag } \otimes {d^\dag }} \right]}^{\left( K \right)}} \otimes {d^\dag }} \right]_M^{\left( L \right)}\left| 0 \right\rangle ,
\end{align}
where the subscript $U$ indicates that this state is unnormalized.

Then, a straightforward calculation shows 
\begin{align}
& O^{(L)}_{KN}  \equiv  {\left\langle {{{\left[ {{{\left[ {d \otimes d} \right]}^{\left( K \right)}} \otimes d} \right]}^{\left( L \right)}}\left| {{{\left[ {{{\left[ {d \otimes d} \right]}^{\left( N \right)}} \otimes d} \right]}^{\left( L \right)}}} \right.} \right\rangle _U} \nonumber \\
&= 2\left( {{\delta _{NK}} + 2\sqrt {\left( {2N + 1} \right)\left( {2K + 1} \right)} \left\{ {\begin{array}{*{20}{c}}
{\begin{array}{*{20}{c}}
2&N&2
\end{array}}\\
{\begin{array}{*{20}{c}}
L&K&2
\end{array}}
\end{array}} \right\}} \right).
\end{align}
Using this formula, we have five normalized basis states for three $d$-boson states:
\begin{align}
&\left|{L\!=\!0,n_\beta\!=\!0,n_\Delta\!\!=\!1,\lambda\!=\!0}\right\rangle
= \frac{1}{{\sqrt 6 }}\left[ {{{\left[ {{d^\dag } \otimes {d^\dag }} \right]}^{\left( 2 \right)}} \!\otimes {d^\dag }} \right]_M^{\left( 0 \right)}\left| 0 \right\rangle ,
\label{eq:nDelta1}
\\
&\left| {L\!=\!2,n_\beta\!=\!1, n_\Delta\!\!=\!0, \lambda\!=\!1} \right\rangle  
= \frac{1}{{\sqrt 6 }} \left[ {{{\left[ {{d^\dag } \otimes {d^\dag }} \right]}^{\left( 0 \right)}} \!\otimes {d^\dag }} \right]_M^{\left( 2 \right)}\left| 0 \right\rangle ,
\label{eq:nbeta1}
\\
&\left| {L\!=\!3, n_\beta\!=\!0, n_\Delta\!\!=\!0, \lambda\!=\!3} \right\rangle  
= \frac{{\sqrt {21} }}{6}\left[ {{{\left[ {{d^\dag } \otimes {d^\dag }} \right]}^{\left( 4 \right)}} \!\otimes {d^\dag }} \right]_M^{\left( 3 \right)}\left| 0 \right\rangle ,
\\
&\left| {L\!=\!4,n_\beta\!=\!0, n_\Delta\!\!=\!0, \lambda\!=\!3} \right\rangle  
= \frac{{\sqrt 7 }}{4}\left[ {{{\left[ {{d^\dag } \otimes {d^\dag }} \right]}^{\left( 4 \right)}} \!\otimes {d^\dag }} \right]_M^{\left( 4 \right)}\left| 0 \right\rangle ,
\\
&\left| {L\!=\!6,n_\beta\!=\!0, n_\Delta\!\!=\!0, \lambda\!=\!3} \right\rangle  
= \frac{1}{{\sqrt 6 }}\left[ {{{\left[ {{d^\dag } \otimes {d^\dag }} \right]}^{\left( 4 \right)}} \!\otimes {d^\dag }} \right]_M^{\left( 6 \right)}\left| 0 \right\rangle .
\end{align}
Here, the superscript $(L)$ denotes a state with total angular momentum $L$. 
In the case of two-particle states, the label $(0)$ corresponds to $n_\beta = 1$ [see Eq.~\eqref{eq:nbeta1}], 
while for three-particle states it represents $n_\Delta = 1$ [see Eq.~\eqref{eq:nDelta1}].

One subtle point concerns the $L=2$ states of three $d$-bosons. 
These can be constructed in two apparently different ways:
$\big[[d \otimes d]^{(0)} \otimes d \big]^{(2)}$ for which it is clear that $n_\beta = 1$, while
$\big[[d \otimes d]^{(2)} \otimes d \big]^{(2)}$
seemingly contains any $L=0$ pair. 
However, these two constructions are not independent: 
one can explicitly verify this by computing the overlap between them, namely,
$O^{(2)}_{02}$ is non-vanishing.
Therefore, another independent $L=2$ state can be constructed. 
A similar reasoning applies to the fermionic $p$-shell system 
when rewriting the $|L=1\rangle$ ground state as
\begin{align}
|L=1\rangle \propto \big[[p \otimes p]^{(0,0)} \otimes p \big]^{(1,1/2)},
\end{align}
which explicitly exhibits that two electrons form a pair with orbital angular momentum $L=0$ and spin $S=0$. 
In contrast, for $|L=0\rangle$ and $|L=2\rangle$, such paired structures do not exist.

We note that the $L=1$ and $L=5$ states are missing because their norms vanish, which is directly confirmed by substituting the value of the Wigner-6j symbol:
\begin{align}
O^{(1)}_{22}
&= 2\left( {1 + 10\left\{ {\begin{array}{*{20}{c}}
{\begin{array}{*{20}{c}}
2&2&2
\end{array}}\\
{\begin{array}{*{20}{c}}
1&2&2
\end{array}}
\end{array}} \right\}} \right)  
= 0,
\\
O^{(5)}_{44}
&= 2\left( {1 + 18\left\{ {\begin{array}{*{20}{c}}
{\begin{array}{*{20}{c}}
2&4&2
\end{array}}\\
{\begin{array}{*{20}{c}}
5&4&2
\end{array}}
\end{array}} \right\}} \right)  
 = 0.
\end{align}

We can also consider $d$-boson states for ${n_d} \ge 4$, but
their explicit construction is tedious. 
Thus, we usually construct those states numerically employing 
coefficients of fractional parentage (cfp) 
$\left[ {{d^{{n_d}}}\left( {\alpha L} \right)d,L'\left| {\left. {} \right\}} \right.{d^{{n_d} + 1}}\alpha '} \right]$ 
where $\alpha $ and $\al'$.
$\left\langle {{n_d} + 1,\alpha 'L'\left\| {{d^\dag }} \right\|{n_d},\alpha L} \right\rangle $ and $\left[ {{d^{{n_d}}}\left( {\alpha L} \right)d,L'\left| {\left. {} \right\}} \right.{d^{{n_d} + 1}}\alpha '} \right]$ are related as
\begin{align}
&\left\langle {{n_d} + 1,\alpha 'L'\left\| {{d^\dag }} \right\|{n_d},\alpha L} \right\rangle  \nonumber \\
&= \sqrt {{n_d} + 1} \sqrt {2L' + 1} \left[ {{d^{{n_d}}}\left( {\alpha L} \right)d,L'\left| {\left. {} \right\}} \right.{d^{{n_d} + 1}}\alpha '} \right] ,
\end{align}
which in fact defines cfp's.
Thus, any ${n_d} + 1$ $d$-boson states are constructed by ${n_d}$ $d$-boson states.

\subsection{Quasispin in $d$-boson space}
\label{sec:Quasispin-boson}

In addition to the group structure of $U\left( 5 \right) \supset SO\left( 5 \right) \supset SO\left( 3 \right)$, the $d$-boson 
system has $SU\left( {1,1} \right)$ structure.
The generators of $SU\left( {1,1} \right)$ group are given in terms of $d$-boson creation and annihilation
operators as
\begin{align}
&K_ + ^{\rm ph} = \frac{1}{2}\sum\limits_m {} {\left( { - 1} \right)^m}d_m^\dag d_{ - m}^\dag , \\
&K_ - ^{\rm ph} = \frac{1}{2}\sum\limits_{m'} {} {\left( { - 1} \right)^{m'}}{d_{ - m'}}{d_{m'}}, \\
&K_z^{\rm ph} = \frac{1}{2}\left( {n + \frac{5}{2}} \right),
\end{align}
where $n = \sum\limits_m {} d_m^\dag {d_m}$ is the number operator. They satisfy
the commutation relations of the $SU\left( {1,1} \right)$ algebra:
\begin{align}
&\left[ {K_ + ^{\rm ph},K_ - ^{\rm ph}} \right] =  - 2K_z^{\rm ph}, \\
&\left[ {K_z^{\rm ph},K_ \pm ^{\rm ph}} \right] =  \pm K_z^{\rm ph},
\end{align}
where a factor of $ - 2$ in front of $K_z^{\rm ph}$ is in contrast with $ + 2$ for $SU\left( 2 \right)$ 
algebra.
A spherical tensor $T_\mu ^{(\lambda)}$ in $SU(1,1)$ group is so defined to satisfy the following 
commutation relations:
\begin{align}
&\left[ {{K_ \pm ^{\rm ph} },T_\mu ^{\left( \lambda  \right)}} \right] =  \mp \sqrt {\left( {\lambda  \mp \mu } \right)\left( {\lambda  \pm \mu  + 1} \right)} T_{\mu  \pm 1}^{\left( \lambda  \right)} , \\
&\left[ {{K_z^{\rm ph}},T_\mu ^{\left( \lambda  \right)}} \right] = \mu T_\mu ^{\left( \lambda  \right)},
\end{align}
where $\mp$ sign in front of 
$\sqrt {\left( {\lambda  \mp \mu } \right)\left( {\lambda  \pm \mu  + 1} \right)}$ 
is in contrast with just $+$ for spherical tensors in $SU(2)$  
[see Eq.~\eqref{eq:su(2)}].

For $d_m^\dag$ and ${\tilde d}_m$, one has the following commutation relations:
\begin{align}
&\left[ {K_z^{\rm ph},d_m^\dag } \right] = \frac{1}{2}d_m^\dag ,\ \ 
\left[ {K_z^{\rm ph},{{\tilde d}_m}} \right] =  - \frac{1}{2}{\tilde d_m}, \\
&\left[ {K_ + ^{\rm ph},{{\tilde d}_m}} \right] =  - d_m^\dag , \ \ 
\left[ {K_ - ^{\rm ph},d_m^\dag } \right] = {\tilde d_m} ,
\end{align}
which imply that $d_m^\dag $ and ${\tilde d_m}$ constitute tensors of $T_{1/2}^{\left( {1/2} \right)}$ and $T_{ - 1/2}^{\left( {1/2} \right)}$ in quasispin space, respectively
and are denoted as
$d_{m,{\textstyle{1 \over 2}}}^{\left( {2,{\textstyle{1 \over 2}}} \right)} \equiv d_m^\dag ,~
d_{m, - {\textstyle{1 \over 2}}}^{\left( {2,{\textstyle{1 \over 2}}} \right)} \equiv {\tilde d_m}$.
Then, one-body spherical double tensors in the order of $L$ and $K$ are defined as
\begin{align}
&D_{{M_0},{K_0}}^{\left( {L,K} \right)} 
\equiv \left[ {d^{(2,{\textstyle{1 \over 2}})} \otimes d^{(2,{\textstyle{1 \over 2}})}} \right]_{{M_0},{K_0}}^{\left( {L,K} \right)} \nonumber \\
&\equiv \sum\limits_{m,m',\kappa ,\kappa'}
{\left\langle {2m2m'\left| {L{M_0}} \right.} \right\rangle } 
{\left\langle {{\textstyle{1 \over 2}}\kappa {\textstyle{1 \over 2}}\kappa '\left| {K{K_0}} \right.} \right\rangle } {d_{m\kappa }^{(2,{\textstyle{1 \over 2}})} }{d_{m'\kappa '}^{(2,{\textstyle{1 \over 2}})}}.
\end{align}
By exchanging two $d's$ in $[~~]$, one has, for $L + K \ne 0$,
\begin{align}
\left[ {d^{(2,{\textstyle{1 \over 2}})} \otimes d^{(2,{\textstyle{1 \over 2}})}} \right]_{{M_0},{K_0}}^{\left( {L,K} \right)} 
\!\!\!\!
= {\left( { - 1} \right)^{L + K + 1}} \!
\left[ {d^{(2,{\textstyle{1 \over 2}})} \otimes d^{(2,{\textstyle{1 \over 2}})}} \right]_{{M_0},{K_0}}^{\left( {L,K} \right)},
\end{align}
where some properties of CG coefficients have been used.
Therefore, one has the selection rule that $L + K $ must be an odd integer.

\subsection{Specific forms of double tensors $D_{{L_0},{K_0}}^{\left( {L,K} \right)}$}

For concreteness, we list several low-rank double tensors for phonons below:
\begin{align}
D_{{0},{1}}^{\left( {0,1} \right)}
&= \frac{2}{{\sqrt 5 }}K_ + ^{\rm ph},
\\
D_{{0},{-1}}^{\left( {0,1} \right)}
&= \frac{2}{{\sqrt 5 }}K_ - ^{\rm ph},
\\
D_{0,0}^{\left( {0,1} \right)} 
&= \frac{2}{5}\sqrt {10} 
K_{z}^{\rm ph},
\\
D_{M,0}^{\left( {1,0} \right)} 
&= \sqrt 2 \left[ {{d^\dag } \otimes \tilde d} \right]_M^{\left( 1 \right)},
\\
D_{M,0}^{\left( {2,1} \right)} 
&= \sqrt 2 \left[ {{d^\dag } \otimes \tilde d} \right]_M^{\left( 2 \right)}.
\end{align}

\subsection{Angular momentum and quadrupole operators}
The bosonic angular momentum operator ($L$=1) and a quadrupole operator ($L$=2) 
can be constructed in a manner similar to the fermions.
The angular momentum in terms of $d$-boson is given as
\begin{align}
\mathscr L_M^{\rm ph} 
= \sqrt {10} \left[ {{d^\dag } \otimes \tilde d} \right]_M^{\left( 1 \right)} 
= \sqrt 5 D_{M,0}^{\left( {1,0} \right)} .
\end{align}
These satisfy the commutation relations of the $SU\left( 2 \right)$ algebra:
\begin{align}
&\left[ {\mathscr L_ {+} ^{\rm ph},\mathscr L_ {-} ^{\rm ph}} \right] =   2 \mathscr L_0^{\rm ph}, \\
&\left[ {\mathscr L_0^{\rm ph},\mathscr L_ \pm ^{\rm ph}} \right] =  \pm \mathscr L_{\pm}^{\rm ph}.
\end{align}
The quadrupole operator is given as
\begin{align}
\mathscr Q_M^{\rm ph} 
= \left[ {{d^\dag } \otimes \tilde d} \right]_M^{\left( 2 \right)} 
=  \sqrt 2 D_{M,0}^{\left( {2,1} \right)}  .
\end{align}
Note that this is a many-body extension of the quadrupole defined in Eq.~\eqref{eq:def_of_x_op}.

Furthermore, using the angular momentum operator in $d$-boson space, 
one can also construct a two-body quadrupole operator
that depends only on the the total angular momentum $L$:
\begin{align}
\mathscr Q_M^{\rm ph, C} \equiv \left[ {{\mathscr L^{\rm ph}} \!\otimes\!\! \, {{\mathscr L^{\rm ph}} }} \right]_M^{\left( 2 \right)} \!=\! 10\!\left[ {{{\left[ {{d^\dag } \otimes \tilde d} \right]}^{\left( 1 \right)}} \!\!\!\otimes\! {{\left[ {{d^\dag } \otimes \tilde d} \right]}^{\left( 1 \right)}}} \right]_M^{\left( 2 \right)}
.
\end{align}
Using the formula [Eqs.~\eqref{eq:two_product} and \eqref{eq:Angular_Momentum}], 
one has
\begin{align}
&\left\langle {n'\alpha 'L'\left\| \mathscr Q^{\rm ph, C} \right\|n\alpha L} \right\rangle  \nonumber \\
&= {\delta _{nn'}}{\delta _{\alpha \alpha '}}\left\langle {n\alpha L\left\| {{{\left[ {{{\mathscr L^{\rm ph}}}\otimes {{\mathscr L^{\rm ph}}}} \right]}^{\left( 2 \right)}}} \right\|{n}\alpha L} \right\rangle \nonumber \\
&= {\delta _{nn'}}{\delta _{\alpha \alpha '}} \sqrt 5 {\left( { - 1} \right)^{L + L + 2}}\left\{ {\begin{array}{*{20}{c}}
1&1&2\\
L&L&L
\end{array}} \right\} \nonumber \\
&\ \ \ \times \left\langle {n\alpha L\left\| {\mathscr L^{\rm ph}} \right\|n\alpha L} \right\rangle 
\left\langle {n\alpha L\left\| {\mathscr L^{\rm ph}} \right\|n\alpha L} \right\rangle  \nonumber \\
&= {\delta _{nn'}}{\delta _{\alpha \alpha '}} \sqrt 5 L\left( {L + 1} \right)\left( {2L + 1} \right)\left\{ {\begin{array}{*{20}{c}}
1&1&2\\
L&L&L
\end{array}} \right\},
\end{align}
where $n,n'$, $\alpha,\alpha'$, and $L,L'$ indicate numbers of $d$-bosons, additional quantum numbers, and angular momenta, respectively.

\nocite{apsrev42control}
\bibliographystyle{apsrev4-2}
\bibliography{fulleride}

\end{document}